\documentclass[longauth]{aa}  
\usepackage{graphicx}
\usepackage{txfonts}
\usepackage{natbib}
\usepackage{longtable}
\usepackage{supertabular}
\bibpunct{(}{)}{;}{a}{}{,}
\newcommand{\tn}[1]{\textnormal{#1}}
\newcommand{\TNT}{T_{90}}
\newcommand{\AV}{\tn{\emph{A}}_{\tn{v}}}

\begin{document}
   \title{The Optical/NIR afterglow of GRB 111209A: Complex yet not Unprecedented\thanks{Partially based on observations obtained under programme 088.A-0051(C), PI: J. P. U. Fynbo}}

   \author{
D. A. Kann		\inst{	1,2,3,4	} \and
P. Schady		\inst{	2	} \and
F. Olivares E.		\inst{	5	} \and
S. Klose		\inst{	1	} \and
A. Rossi		\inst{	6,1	} \and
D. A. Perley		\inst{	7,8,9	} \and
B. Zhang \inst{	10,11	} \and
T. Kr\"uhler		\inst{	2,8,12	} \and
J. Greiner		\inst{	2,3	} \and
A. Nicuesa Guelbenzu		\inst{	1	} \and
J. Elliott		\inst{	2,13	} \and
F. Knust		\inst{	2	} \and
Z. Cano		\inst{	4	} \and
R. Filgas		\inst{	14	} \and
E. Pian		\inst{	5,15	} \and
P. Mazzali		\inst{	16,17	} \and
J. P. U. Fynbo		\inst{	8	} \and
G. Leloudas		\inst{	18,8	} \and
P. M. J. Afonso		\inst{	19	} \and
C. Delvaux		\inst{	2	} \and
J. F. Graham		\inst{	2	} \and
A. Rau		\inst{	2	} \and
S. Schmidl		\inst{	1	} \and
S. Schulze		\inst{	20,21,22	} \and
M. Tanga		\inst{	2	} \and
A. C. Updike		\inst{	23	} \and
K. Varela		\inst{	2	}
 }

   \offprints{D. A. Kann, \email{kann@iaa.es}}

\institute{
Th\"uringer Landessternwarte Tautenburg, Sternwarte 5, 07778 Tautenburg, Germany %Alex, Sylvio, Andrea, Ana, Sebastian
%\email{kann@tls-tautenburg.de}
\and
Max-Planck-Institut f\"ur extraterrestrische Physik, Giessenbachstra\ss e 1, 85748 Garching, Germany %Alex, Patricia, Jonny, Fabian
\and
Universe Cluster, Technische Universit\"at M\"unchen, Boltzmannstra\ss e 2, 85748 Garching, Germany %Alex, Jochen
\and
Instituto de Astrof\'isica de Andaluc\'ia (IAA-CSIC), Glorieta de la Astronom\'ia s/n, 18008 Granada, Spain %Alex
\and
Departamento de Ciencias Fisicas, Universidad Andres Bello, Avda. Republica 252, Santiago, Chile %Felipe
\and
INAF-IASF Bologna, Area della Ricerca CNR, via Gobetti 101, I--40129 Bologna, Italy %Andrea, Elena
\and
Cahill Center for Astrophysics, California Institute of Technology, Pasadena, CA 91125, USA %Dan
\and
Dark Cosmology Centre, Niels Bohr Institute, University of Copenhagen, Juliane Maries Vej 30, 2100 Copenhagen, Denmark %Thomas, Johan, Dan
\and
Astrophysics Research Institute, Liverpool John Moores University, IC2, Liverpool Science Park, 146 Brownlow Hill, Liverpool L3 5RF, UK %Dan
\and
Department of Physics and Astronomy, University of Nevada Las Vegas, Las Vegas, NV 89154, USA
\and
Guangxi Key Laboratory for Relativistic Astrophysics, Department of Physics, Guangxi University, Nanning 530004, China
\and
ESO, Alonso de Cordova 3107, Vitacura, Santiago de Chile, Chile. %Thomas
\and
Harvard-Smithsonian Center for Astrophysics, 60 Garden St., Cambridge, MA 02138, USA %Jonny
\and
Institute of Experimental and Applied Physics, Czech Technical University in Prague, Horsk\'a 3a/22, 12800 Prague, Czech Republic %Robert
\and
Scuola Normale Superiore, Piazza dei Cavalieri 7, 56126 Pisa, Italy %Elena
\and
Astrophysics Research Institute, Liverpool John Moores University, 146 Brownlow Hill, Liverpool L3 5RF, UK %Paolo
\and
%INAF--Osservatorio Astronomico, Vicolo dell'Osservatorio, 5, I-35122 Padova, Italy %Paolo
%\and
Max-Planck Institut f\"ur Astrophysik, Karl-Schwarzschild-Str. 1, 85748 Garching, Germany %Paolo
\and
Department of Particle Physics \& Astrophysics, Weizmann Institute of Science, Rehovot 76100, Israel. %Giorgios
\and
American River College, Physics and Astronomy Dpt., 4700 College Oak Drive, Sacramento, CA 95841, USA %Paulo
\and
Instituto de Astrof\'isica, Facultad de F\'isica, Pontificia Universidad Cat\'olica de Chile, Av. Vicu\~na Mackenna 4860, 306, Santiago 22, Chile. %Steve
\and
Millennium Institute of Astrophysics, Av. Vicu\~na Mackenna 4860, 7820436 Macul, Santiago, Chile. %Steve
\and
Department of Particle Physics and Astrophysics, Faculty of Physics, Weizmann Institute of Science, Rehovot 76100, Israel %Steve
\and
Department of Chemistry and Physics, Roger Williams University, One Old Ferry Road, Bristol, RI 02809, USA %Adria
}

\date{Received 02 June 2017 / Accepted 03 May 2018}

% \abstract{}{}{}{}{} 
% 5 {} token are mandatory

\begin{abstract}
{Afterglows of Gamma-Ray Bursts (GRBs) are simple in the most basic model, but can show many complex features. The ultra-long duration GRB 111209A, {one of the longest GRBs} ever detected, also has the best-monitored afterglow in this rare class of GRBs.}
{We want to address the question whether GRB 111209A was a special event beyond its extreme duration alone, and whether it is a classical GRB or another kind of high-energy transient. The afterglow may yield significant clues.}
{We present afterglow photometry obtained in seven bands with the GROND imager as well as in further seven bands with the UVOT telescope on-board {the Neil Gehrels \emph{Swift} Observatory}. The light curve is analysed by multi-band modelling and joint fitting with power-laws and broken power-laws, and we use the contemporaneous GROND data to study the evolution of the spectral energy distribution. We compare the optical afterglow to a large ensemble we have analysed in earlier works, and especially to that of another ultra-long event, GRB 130925A. We furthermore undertake a photometric study of the host galaxy.}
{We find a strong, chromatic rebrightening event at $\approx0.8$ days after the GRB, during which the spectral slope becomes redder. After this, the light curve decays achromatically, with evidence for a break at about 9 days after the trigger. The afterglow luminosity is found to not be exceptional. We find that a double-jet model is able to explain the chromatic rebrightening. The afterglow features have been detected in other events and are not unique.}
{{The duration aside, the GRB prompt emission} and afterglow parameters of GRB 111209A are in agreement with the known distributions for these parameters. While the central engine of this event may differ from {that of} classical GRBs, there are multiple lines of evidence pointing to GRB 111209A resulting from the core-collapse of a massive star with a stripped envelope.}
\end{abstract}


\authorrunning{D. A. Kann et al.}
\titlerunning{Complex GRB 111209A Afterglow}
\maketitle
%
%________________________________________________________________

\section{Introduction}
\label{SectInt}
\subsection{Gamma-Ray Bursts}

Gamma-Ray Bursts (GRBs) were discovered 50 years ago \citep{Klebesadel1973ApJ} and represent the brightest explosions in the Universe, at cosmological redshifts \citep{Metzger1997Nature}. One class, long GRBs, is associated with the core-collapse of very massive stars (see \citealt{Woosley2006ARAA}, \citealt{HjorthBloom2012Book} and \citealt{Cano2016JAA} for reviews). They are accompanied by afterglow emission \citep{vanParadijs1997Nature, Costa1997Nature, Frail1997Nature} which can be extremely luminous \citep{Akerlof1999Nature,Kann2007AJ,Racusin2008Nature,Bloom2009ApJ,Wozniak2009ApJ,Perley2011AJ,Vestrand2014Science}. In the simplest models, the afterglow is pure synchrotron radiation \citep{Sari1998ApJ,Rossi2011AA} and can be described by a smoothly broken power-law \citep[e.g.,][]{Beuermann1999AA}, the break to a steeper decay stemming from the collimation of the jet \citep{Rhoads1997ApJ,Rhoads1999ApJ,Sari1999ApJ}.

%------------------------------------------------------------------------------------------------------------------------------------------------------------------------------------------------------------------
%------------------------------------------------------------------------------------------------------------------------------------------------------------------------------------------------------------------

\begin{figure*}[!t]
  \centering
 \includegraphics[width=\textwidth]{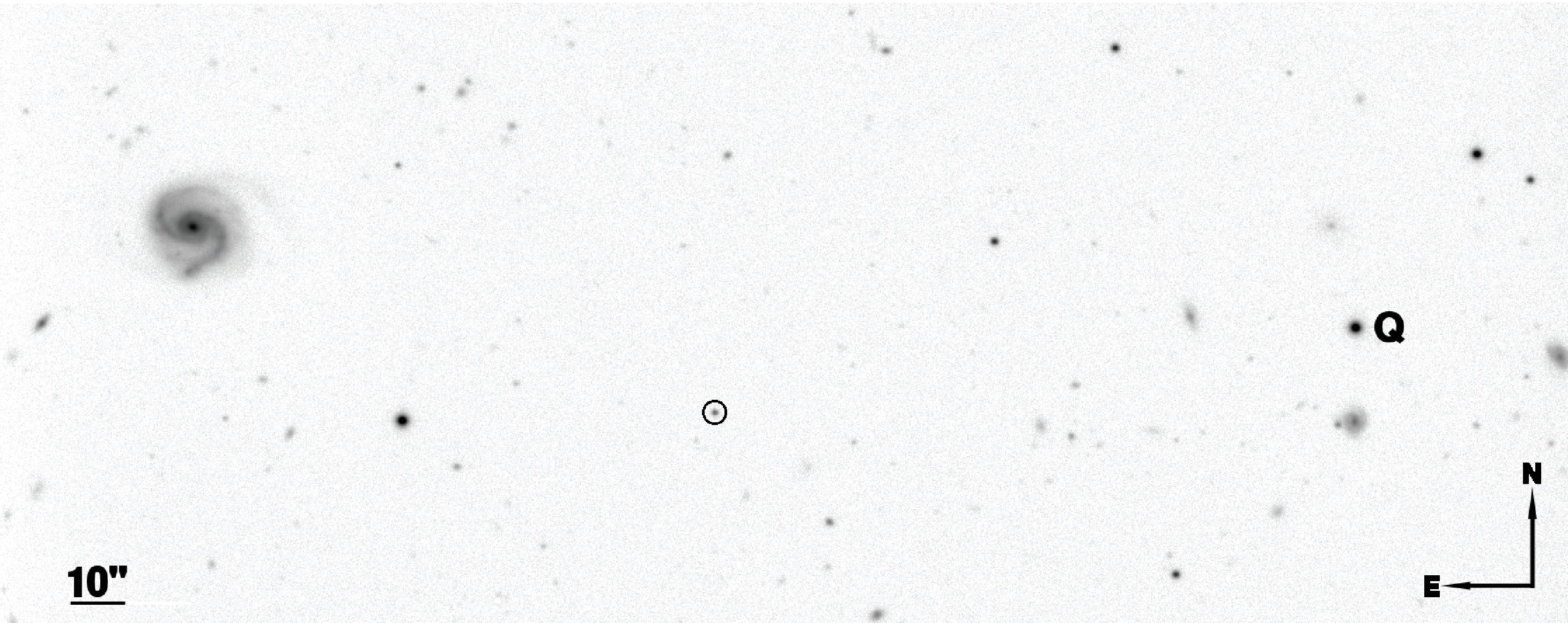}
   \caption{Finding chart of the field of GRB 111209A, the afterglow is circled. This is a GROND $r^\prime$ image at 5.8 days post-burst; the afterglow has $r^\prime=21.2$. Seeing was 0\farcs73 and the limiting magnitude is $r^\prime>25.3$. The field measures 4\farcm9 $\times$ 1\farcm9, pixel scale is 0\farcs158 per pixel. Note the extreme sparsity of stars in the field, most visible sources are distant galaxies. The bright face-on spiral galaxy to the north-east is 6dFGS gJ005732.0-464730 at $z=0.08851$ \citep{Jones2009MNRAS}. The character {Q} marks quasar [VV96] J005711.3-464750 at $z=2.0$ \citep{Maza1993RMxAA}.}
              \label{Field}
    \end{figure*}

\subsection{GRBs of (extremely) long duration}

Much focus in the last years has been on the divide \citep{Mazets1981ApSS,Kouveliotou1993ApJ,Zhang2009ApJ} between long and short GRBs (or, more generally, those not associated with the core-collapse of massive stars), but the other end of the duration distribution is also of great interest, as they pose challenges to the collapsar model of long GRBs \citep{Woosley1993ApJ,MacFadyen1999ApJ}. The duration of a GRB (once corrected for time dilation) generally reflects the duration of the central engine activity after subtracting the time it takes for the jet to break out of the star \citep[e.g.,][and references therein]{Bromberg2012ApJ,Bromberg2013ApJ,Lazzati2013MNRAS,Hamidani2017MNRAS} -- the aforementioned publications find a typical engine activity time of 20 seconds which is far less than the duration of the longest known GRBs. The actual duration measurement is detector-dependent, though, and \cite{Zhang2014ApJ} pointed out that the true duration, which they label $t_{\rm burst}$, is that of any detectable central engine activity \citep{Liang2006ApJ}, which may last much longer than the detectable gamma-rays (but see \citealt{Boer2015ApJ} for a critical take on $t_{\rm burst}$). Such an extreme duration was already discussed by \cite{Zou2006ApJ} in the context of the strong flaring of GRB 050904, which lasts $\gtrsim7500$ s in the rest-frame.

A second aspect which makes such extremely long GRBs interesting is that with modern rapid follow-up observations, both space- and ground-based, GRBs of long duration are still ongoing by the time they become observable by narrow-field instruments. Such observations have yielded the detection of optical flashes that are likely to be directly linked to the prompt emission itself \citep{Blake2005Nature,Vestrand2005Nature,Vestrand2006Nature,Page2007ApJ,Kruehler2009ApJ,Beskin2010ApJ,Thoene2010AA,Guidorzi2011MNRAS,Vestrand2014Science,Greiner2014AA,Elliott2013AA}.

Morphologically, the prompt emission light curves of GRBs are highly diverse, and this is especially true for extremely long GRBs. Different examples may be roughly sorted into five categories, which could indicate a link to different outflow structures from the central engine, or that different processes are responsible for the dominating part of the prompt emission. We give examples of such events, as well as a (rough) phenomenological classification scheme in Appendix \ref{ELDGRBs}. Some further possible extremely long GRBs have been discussed in \cite{Lien2016ApJ}, based on BAT survey mode detections.

\citet[][henceforth L14, see also \citealt{Levan2015Rome}]{Levan2014ApJ} proposed that even among such events of extremely long duration, a few very rare cases exist which they label ``ultra-long duration GRBs'' (ULGRBs), and which may comprise a new class, possibly resulting from novel progenitor channels. Since then, this point has been debated. \cite{Zhang2014ApJ} found more cases of long-lasting central engine activity (though fainter and softer, and therefore not detected in gamma-rays) and no compelling evidence of a separate population in the $t_{\rm burst}$-duration distribution (though conversely also no strong evidence against such a new population). A similar conclusion was reached by \cite{Virgili2013ApJ}, but refuted by \citet[][see also \citealt{Boer2015Rome}b]{Boer2015ApJ} and \cite{Evans2014MNRAS}. \cite{Gao2015ApJ} also found a bimodal distribution and therefore the possibility that ULGRBs may be a different class{, a result they further strengthened in a new paper \citep{Gao2017ApJ}}. The most extreme of the three\footnote{{Two further examples have been detected since then, GRBs 130925A and 170714A}, see Appendix \ref{ELDGRBs}.} events L14 described is GRB 111209A.

%------------------------------------------------------------------------------------------------------------------------------------------------------------------------------------------------------------------
%------------------------------------------------------------------------------------------------------------------------------------------------------------------------------------------------------------------

\subsection{The ULGRB 111209A}

In terms of classical GRBs, even when taking the other events listed in Appendix \ref{ELDGRBs} into account, GRB 111209A is an extreme case, {until recently, it was nothing less than the longest GRB detected so far \citep{Gendre2013ApJ}; a title now held by GRB 170714A (Appendix \ref{ELDGRBs})}. We refer to Appendix \ref{TheGRB} for details on the early multi-wavelength follow-up of this event.

The initial behaviour in the optical and especially the X-rays (Appendix \ref{TheGRB}) led to a comparison with both XRF 060218/SN 2006aj as well as the ``Christmas Burst'' GRB 101225A \citep{Hoversten2011GCN2} and it was also speculated that the event may be a further example of a tidal disruption flare like GRB 110328A/Swift J164449.3+573451 \citep{Gendre2013ApJ, Levan2014ApJ}. The extreme duration of the GRB and its apparent \emph{lack} of a supernova have furthermore fuelled speculation that the progenitor may not have been a stripped-envelope star, but a very-low metallicity Blue SuperGiant (BSG) Progenitor (\citealt{Gendre2013ApJ,Levan2014ApJ,Stratta2013ApJ,Kashiyama2013ApJ,Nakauchi2013ApJ}, see \citealt{Kann2017AA_SN2011kl} for more discussion).

In this paper, we present and discuss observations of the optical/NIR afterglow of GRB 111209A obtained with GROND, the seven-channel Gamma-Ray burst Optical \& Near-infrared Detector \citep{Greiner2008PASP} mounted on the 2.2m MPG/ESO telescope\footnote{The 2.2m MPG telescope starting on the 1st of October 2013.} stationed in La Silla, Chile, as well as UV/optical data obtained with UVOT on-board {the Neil Gehrels \emph{Swift} Observatory} \citep[see also L14,][]{Stratta2013ApJ}. While no observations during the prompt emission could be obtained with GROND due to observing constraints, we discovered strong optical variability even during the following day, and followed the afterglow up regularly until it became unobservable, detecting the multi-wavelength signature of late-time supernova emission.

Results on the supernova discovery, its spectroscopic verification and modelling of the SN, which has been designated SN 2011kl, are given in \citet[][henceforth G15]{Greiner2015Nat}. In this work, we present the entire GROND/UVOT data set, detailed afterglow and host-galaxy modelling, and place the entire event in the context of a large sample of GRB afterglows. In \citet[][henceforth K18A]{Kann2017AA_SN2011kl}, we also place SN 2011kl into the context of large SN samples, and combine all our results to discuss the nature of GRB 111209A.

%The questions we pose are: aside from being the longest GRB ever detected, are there any other features in its afterglow emission and associated supernova which set GRB 111209A apart from other, more typical GRBs? And can the afterglow/supernova help untangle the origin of this event (classical albeit extremely long GRB \textit{or} BSG progenitor GRB \textit{or} Christmas-Burst-like GRB \textit{or} relativistic tidal disruption flare \textit{or} something else again?) and possibly also shed light on the cause of the extreme duration?

The paper is organized as follows: in Sect. \ref{SectObs}, we present the details of our observations and our data analysis. In Sect. \ref{SectRes}, we present the results of fitting the light curve data, both temporally and as spectral energy distributions (SEDs), during the prompt emission and the afterglow phase; we also study the host galaxy and a very similar event, GRB 130925A. In Sect. \ref{SectDisc}, we model the strong chromatic rebrightening, study the energetics of the event and compare GRB 111209A to GRB 130925A, before finally concluding (Sect. \ref{SectConc}). We present additional information on GRBs of extreme duration in the Appendix.

We will follow the convention $F_\nu\propto t^{-\alpha}\nu^{-\beta}$ to describe the temporal and spectral evolution of the afterglow. We use WMAP $\Lambda$CDM concordance cosmology \citep{Spergel2003ApJS} with $H_0=71$km s$^{-1}$ Mpc$^{-1}$, $\Omega_{\rm M}=0.27$, and $\Omega_{\Lambda}=0.73$. Uncertainties are given at 68\% ($1\sigma$) confidence level for one parameter of interest unless stated otherwise, whereas upper limits are given at the $3\sigma$ confidence level.

%------------------------------------------------------------------------------------------------------------------------------------------------------------------------------------------------------------------
%------------------------------------------------------------------------------------------------------------------------------------------------------------------------------------------------------------------
%------------------------------------------------------------------------------------------------------------------------------------------------------------------------------------------------------------------
%------------------------------------------------------------------------------------------------------------------------------------------------------------------------------------------------------------------

\section{Observations}
\label{SectObs}

\onltab{1}{
%\longtab{2}{
\begin{table*}[t]
\tiny
\caption{Local standard stars. Magnitudes are given in the native filter system.}
\label{tab_stand}
\centering                        
\begin{tabular}{l l c c c c c c c}       
\hline\hline   \vspace{1mm}
RA (J2000) & Dec. (J2000) & $g^\prime$ (AB mag) & $r^\prime$ (AB mag) & $i^\prime$ (AB mag) & $z^\prime$ (AB mag) & $J$ (Vega mag) & $H$ (Vega mag) & $K$ (Vega mag) \\\hline   
%\vspace{1mm}
00:57:09.09 & -46:47:16.4 & $19.985\pm0.008$ & $18.440\pm0.009$ & $16.954\pm0.007$ & $16.266\pm0.006$ & $14.743\pm0.015$ & $14.183\pm0.021$ & $13.981\pm0.031$ \\
%00:57:11.24 & -46:47:48.16 & $18.377\pm0.004$ & $17.647\pm0.009$ & $17.211\pm0.007$ & $17.172\pm0.006$ & $16.493\pm0.020$ & $16.012\pm0.026$ & $15.543\pm0.039$ \\
00:57:14.44 & -46:48:33.1 & $21.452\pm0.012$ & $20.039\pm0.011$ & $19.103\pm0.009$ & $18.734\pm0.009$ & $17.448\pm0.027$ & $16.760\pm0.034$ & $16.705\pm0.074$ \\
00:57:15.53 & -46:46:57.0 & $21.191\pm0.012$ & $19.803\pm0.010$ & $18.475\pm0.008$ & $17.922\pm0.007$ & $16.490\pm0.020$ & $15.902\pm0.026$ & $15.653\pm0.040$ \\
00:57:17.67 & -46:47:32.4 & $21.550\pm0.014$ & $20.027\pm0.010$ & $18.479\pm0.008$ & $17.830\pm0.006$ & $16.389\pm0.018$ & $15.861\pm0.024$ & $15.566\pm0.040$ \\
00:57:28.21 & -46:48:05.2 & $18.390\pm0.004$ & $17.736\pm0.009$ & $17.483\pm0.009$ & $17.400\pm0.007$ & $16.389\pm0.018$ & $15.830\pm0.024$ & $15.678\pm0.044$ \\
00:57:30.35 & -46:51:09.6 & $21.661\pm0.016$ & $20.246\pm0.012$ & $19.377\pm0.012$ & $19.086\pm0.011$ & $17.749\pm0.031$ & $17.184\pm0.047$ & $16.818\pm0.094$ \\
00:57:32.34 & -46:51:24.8 & $20.147\pm0.007$ & $19.418\pm0.010$ & $19.096\pm0.012$ & $19.016\pm0.012$ & $17.955\pm0.035$ & $17.415\pm0.047$ & $17.435\pm0.143$ \\
\hline \hline
\end{tabular}
\end{table*}
%}
}

\onltab{2}{
\longtab{2}{
\begin{longtable}{rccrcl}
\caption{\label{obslog} UVOT and GROND observations of the afterglow of GRB 111209A.}\\
\hline
\hline                
$\Delta$t (ks) & $\Delta$t (days) & mag & exp (s) & seeing (arcsec) & filter\\
%\endfirsthead
\hline        \vspace{1mm}
0.7421	&	0.008590	& $	19.09	^{+	0.31	}_{-	0.24	}$ &	19.5	&	$\cdots$	&	$uvw2$	\\\vspace{1mm}
1.1078	&	0.012822	& $	20.03	^{+	0.35	}_{-	0.26	}$ &	38.9	&	$\cdots$	&	$uvw2$	\\\vspace{1mm}
1.4565	&	0.016857	& $	20.15	^{+	0.38	}_{-	0.28	}$ &	38.9	&	$\cdots$	&	$uvw2$	\\\vspace{1mm}
1.8051	&	0.020892	& $	20.25	^{+	0.41	}_{-	0.29	}$ &	38.9	&	$\cdots$	&	$uvw2$	\\\vspace{1mm}
3.7481	&	0.043381	& $	19.69	^{+	0.13	}_{-	0.12	}$ &	201.9	&	$\cdots$	&	$uvw2$	\\\vspace{1mm}
6.7796	&	0.078467	& $	19.76	^{+	0.13	}_{-	0.12	}$ &	196.6	&	$\cdots$	&	$uvw2$	\\\vspace{1mm}
12.3903	&	0.143406	& $	19.73	^{+	0.08	}_{-	0.07	}$ &	885.6	&	$\cdots$	&	$uvw2$	\\\vspace{1mm}
19.0312	&	0.220269	& $	19.97	^{+	0.09	}_{-	0.08	}$ &	694	&	$\cdots$	&	$uvw2$	\\\vspace{1mm}
41.7973	&	0.483765	& $	20.94	^{+	0.12	}_{-	0.11	}$ &	885.6	&	$\cdots$	&	$uvw2$	\\\vspace{1mm}
66.4377	&	0.768955	& $	21.53	^{+	0.13	}_{-	0.11	}$ &	1545.4	&	$\cdots$	&	$uvw2$	\\\vspace{1mm}
98.1442	&	1.135928	& $	21.35	^{+	0.11	}_{-	0.10	}$ &	1771.1	&	$\cdots$	&	$uvw2$	\\\vspace{1mm}
125.9306	&	1.457531	& $	21.29	^{+	0.11	}_{-	0.10	}$ &	1251.9	&	$\cdots$	&	$uvw2$	\\\vspace{1mm}
182.0334	&	2.106868	& $	22.50	^{+	0.40	}_{-	0.29	}$ &	629.6	&	$\cdots$	&	$uvw2$	\\\vspace{1mm}
193.5038	&	2.239628	& $	22.23	^{+	0.33	}_{-	0.25	}$ &	629.4	&	$\cdots$	&	$uvw2$	\\\vspace{1mm}
226.1226	&	2.617160	& $	22.33	^{+	0.52	}_{-	0.35	}$ &	317.4	&	$\cdots$	&	$uvw2$	\\\vspace{1mm}
239.7042	&	2.774354	& $	23.02	^{+	0.64	}_{-	0.40	}$ &	629.5	&	$\cdots$	&	$uvw2$	\\\vspace{1mm}
251.2788	&	2.908319	& $	22.71	^{+	0.48	}_{-	0.33	}$ &	629.4	&	$\cdots$	&	$uvw2$	\\\vspace{1mm}
283.8576	&	3.285389	& $	22.77	^{+	0.47	}_{-	0.33	}$ &	629.4	&	$\cdots$	&	$uvw2$	\\\vspace{1mm}
318.8226	&	3.690077	& $	23.37	^{+	0.72	}_{-	0.43	}$ &	818.8	&	$\cdots$	&	$uvw2$	\\\vspace{1mm}
341.4149	&	3.951561	& $ >	22.47					$ &	794.8	&	$\cdots$	&	$uvw2$	\\\vspace{1mm}
387.4843	&	4.484772	& $	23.20	^{+	0.65	}_{-	0.40	}$ &	814.5	&	$\cdots$	&	$uvw2$	\\\vspace{1mm}
436.5836	&	5.053051	& $	23.10	^{+	0.39	}_{-	0.29	}$ &	1129.4	&	$\cdots$	&	$uvw2$	\\\vspace{1mm}
474.2490	&	5.488994	& $	22.91	^{+	0.69	}_{-	0.42	}$ &	440.5	&	$\cdots$	&	$uvw2$	\\
\hline\vspace{1mm}															
0.6183	&	0.007157	& $	19.03	^{+	0.40	}_{-	0.29	}$ &	19.5	&	$\cdots$	&	$uvm2$	\\\vspace{1mm}
0.7907	&	0.009152	& $	19.21	^{+	0.45	}_{-	0.32	}$ &	19.5	&	$\cdots$	&	$uvm2$	\\\vspace{1mm}
1.1569	&	0.013390	& $	19.82	^{+	0.41	}_{-	0.30	}$ &	38.9	&	$\cdots$	&	$uvm2$	\\\vspace{1mm}
1.5056	&	0.017426	& $	19.49	^{+	0.34	}_{-	0.26	}$ &	38.8	&	$\cdots$	&	$uvm2$	\\\vspace{1mm}
1.8540	&	0.021458	& $	18.97	^{+	0.26	}_{-	0.21	}$ &	38.9	&	$\cdots$	&	$uvm2$	\\\vspace{1mm}
5.7528	&	0.066583	& $	19.82	^{+	0.18	}_{-	0.15	}$ &	196.6	&	$\cdots$	&	$uvm2$	\\\vspace{1mm}
7.1892	&	0.083209	& $	19.11	^{+	0.12	}_{-	0.11	}$ &	196.6	&	$\cdots$	&	$uvm2$	\\\vspace{1mm}
24.0487	&	0.278342	& $	19.78	^{+	0.09	}_{-	0.08	}$ &	885.6	&	$\cdots$	&	$uvm2$	\\\vspace{1mm}
78.9216	&	0.913445	& $	20.96	^{+	0.14	}_{-	0.12	}$ &	1086.6	&	$\cdots$	&	$uvm2$	\\\vspace{1mm}
93.2926	&	1.079776	& $	21.04	^{+	0.12	}_{-	0.11	}$ &	1515	&	$\cdots$	&	$uvm2$	\\\vspace{1mm}
122.9344	&	1.422852	& $	21.49	^{+	0.14	}_{-	0.12	}$ &	1615.7	&	$\cdots$	&	$uvm2$	\\\vspace{1mm}
188.0702	&	2.176739	& $	22.39	^{+	0.76	}_{-	0.44	}$ &	326.7	&	$\cdots$	&	$uvm2$	\\\vspace{1mm}
217.0329	&	2.511955	& $	21.62	^{+	0.32	}_{-	0.25	}$ &	430.3	&	$\cdots$	&	$uvm2$	\\\vspace{1mm}
246.5969	&	2.854131	& $	22.39	^{+	0.20	}_{-	0.17	}$ &	2541.1	&	$\cdots$	&	$uvm2$	\\\vspace{1mm}
258.1727	&	2.988110	& $	22.48	^{+	0.18	}_{-	0.15	}$ &	3498.1	&	$\cdots$	&	$uvm2$	\\\vspace{1mm}
312.9983	&	3.622665	& $ >	20.48					$ &	100.7	&	$\cdots$	&	$uvm2$	\\\vspace{1mm}
341.8239	&	3.956295	& $	23.24	^{+	0.92	}_{-	0.49	}$ &	794.8	&	$\cdots$	&	$uvm2$	\\\vspace{1mm}
387.9034	&	4.489623	& $ >	22.16					$ &	814.5	&	$\cdots$	&	$uvm2$	\\\vspace{1mm}
437.1636	&	5.059764	& $	22.48	^{+	0.31	}_{-	0.24	}$ &	1129.4	&	$\cdots$	&	$uvm2$	\\\vspace{1mm}
474.4777	&	5.491641	& $ >	21.59					$ &	440.5	&	$\cdots$	&	$uvm2$	\\
\hline\vspace{1mm}															
0.6431	&	0.007444	& $	18.61	^{+	0.26	}_{-	0.21	}$ &	19.4	&	$\cdots$	&	$uvw1$	\\\vspace{1mm}
0.8155	&	0.009439	& $	18.56	^{+	0.26	}_{-	0.21	}$ &	19.5	&	$\cdots$	&	$uvw1$	\\\vspace{1mm}
1.1817	&	0.013678	& $	19.02	^{+	0.22	}_{-	0.19	}$ &	38.9	&	$\cdots$	&	$uvw1$	\\\vspace{1mm}
1.5309	&	0.017718	& $	19.32	^{+	0.26	}_{-	0.21	}$ &	38.9	&	$\cdots$	&	$uvw1$	\\\vspace{1mm}
1.8792	&	0.021750	& $	19.09	^{+	0.23	}_{-	0.19	}$ &	38.9	&	$\cdots$	&	$uvw1$	\\\vspace{1mm}
5.9580	&	0.068958	& $	19.63	^{+	0.14	}_{-	0.12	}$ &	196.6	&	$\cdots$	&	$uvw1$	\\\vspace{1mm}
7.3951	&	0.085592	& $	18.81	^{+	0.09	}_{-	0.08	}$ &	196.6	&	$\cdots$	&	$uvw1$	\\\vspace{1mm}
24.8339	&	0.287430	& $	19.66	^{+	0.08	}_{-	0.07	}$ &	642.3	&	$\cdots$	&	$uvw1$	\\\vspace{1mm}
35.6599	&	0.412730	& $	20.33	^{+	0.09	}_{-	0.09	}$ &	885.6	&	$\cdots$	&	$uvw1$	\\\vspace{1mm}
60.5834	&	0.701197	& $	21.05	^{+	0.13	}_{-	0.12	}$ &	1259.8	&	$\cdots$	&	$uvw1$	\\\vspace{1mm}
87.0549	&	1.007580	& $	20.81	^{+	0.09	}_{-	0.08	}$ &	1771.2	&	$\cdots$	&	$uvw1$	\\\vspace{1mm}
138.7108	&	1.605449	& $	20.79	^{+	0.11	}_{-	0.10	}$ &	1042.8	&	$\cdots$	&	$uvw1$	\\\vspace{1mm}
187.3159	&	2.168008	& $ >	21.18					$ &	314.5	&	$\cdots$	&	$uvw1$	\\\vspace{1mm}
199.2553	&	2.306196	& $	21.62	^{+	0.42	}_{-	0.30	}$ &	314.5	&	$\cdots$	&	$uvw1$	\\\vspace{1mm}
211.6930	&	2.450151	& $	21.84	^{+	0.42	}_{-	0.30	}$ &	314.5	&	$\cdots$	&	$uvw1$	\\\vspace{1mm}
223.7850	&	2.590104	& $	21.74	^{+	0.34	}_{-	0.26	}$ &	471.7	&	$\cdots$	&	$uvw1$	\\\vspace{1mm}
233.5397	&	2.703006	& $	21.98	^{+	0.48	}_{-	0.33	}$ &	471.7	&	$\cdots$	&	$uvw1$	\\\vspace{1mm}
241.6857	&	2.797288	& $	22.16	^{+	0.80	}_{-	0.46	}$ &	271.2	&	$\cdots$	&	$uvw1$	\\\vspace{1mm}
250.8648	&	2.903527	& $	21.83	^{+	0.61	}_{-	0.39	}$ &	314.5	&	$\cdots$	&	$uvw1$	\\\vspace{1mm}
262.3900	&	3.036921	& $	21.81	^{+	0.61	}_{-	0.39	}$ &	314.5	&	$\cdots$	&	$uvw1$	\\\vspace{1mm}
312.3202	&	3.614818	& $	22.22	^{+	0.46	}_{-	0.32	}$ &	629	&	$\cdots$	&	$uvw1$	\\\vspace{1mm}
328.1525	&	3.798061	& $	22.47	^{+	0.54	}_{-	0.36	}$ &	546.1	&	$\cdots$	&	$uvw1$	\\\vspace{1mm}
353.2548	&	4.088597	& $ >	22.44					$ &	1100.7	&	$\cdots$	&	$uvw1$	\\\vspace{1mm}
423.4824	&	4.901417	& $	23.00	^{+	0.67	}_{-	0.41	}$ &	748.9	&	$\cdots$	&	$uvw1$	\\\vspace{1mm}
452.1975	&	5.233767	& $	22.86	^{+	0.58	}_{-	0.37	}$ &	675.6	&	$\cdots$	&	$uvw1$	\\
\hline\vspace{1mm}															
0.6675	&	0.007726	& $	18.13	^{+	0.17	}_{-	0.15	}$ &	19.5	&	$\cdots$	&	$u$	\\\vspace{1mm}
0.8400	&	0.009723	& $	18.98	^{+	0.28	}_{-	0.22	}$ &	19.4	&	$\cdots$	&	$u$	\\\vspace{1mm}
1.2065	&	0.013965	& $	18.63	^{+	0.16	}_{-	0.14	}$ &	38.9	&	$\cdots$	&	$u$	\\\vspace{1mm}
1.5560	&	0.018009	& $	18.63	^{+	0.16	}_{-	0.14	}$ &	38.9	&	$\cdots$	&	$u$	\\\vspace{1mm}
1.9036	&	0.022032	& $	18.65	^{+	0.16	}_{-	0.14	}$ &	38.9	&	$\cdots$	&	$u$	\\\vspace{1mm}
6.1636	&	0.071338	& $	19.21	^{+	0.10	}_{-	0.09	}$ &	196.6	&	$\cdots$	&	$u$	\\\vspace{1mm}
7.6001	&	0.087964	& $	18.13	^{+	0.06	}_{-	0.06	}$ &	196.6	&	$\cdots$	&	$u$	\\\vspace{1mm}
29.1265	&	0.337112	& $	19.66	^{+	0.08	}_{-	0.07	}$ &	590.1	&	$\cdots$	&	$u$	\\\vspace{1mm}
32.9249	&	0.381075	& $	19.82	^{+	0.08	}_{-	0.07	}$ &	590.1	&	$\cdots$	&	$u$	\\\vspace{1mm}
45.0843	&	0.521809	& $	20.32	^{+	0.11	}_{-	0.10	}$ &	580.9	&	$\cdots$	&	$u$	\\\vspace{1mm}
61.8953	&	0.716381	& $	20.82	^{+	0.16	}_{-	0.14	}$ &	576.9	&	$\cdots$	&	$u$	\\\vspace{1mm}
76.3139	&	0.883262	& $	20.56	^{+	0.13	}_{-	0.12	}$ &	574.3	&	$\cdots$	&	$u$	\\\vspace{1mm}
87.8171	&	1.016401	& $	20.28	^{+	0.10	}_{-	0.09	}$ &	612.1	&	$\cdots$	&	$u$	\\\vspace{1mm}
96.4025	&	1.115769	& $	20.24	^{+	0.08	}_{-	0.08	}$ &	885.1	&	$\cdots$	&	$u$	\\\vspace{1mm}
139.3566	&	1.612924	& $	20.23	^{+	0.11	}_{-	0.10	}$ &	546	&	$\cdots$	&	$u$	\\\vspace{1mm}
187.4401	&	2.169445	& $	21.14	^{+	0.77	}_{-	0.45	}$ &	157	&	$\cdots$	&	$u$	\\\vspace{1mm}
199.3795	&	2.307633	& $	21.24	^{+	0.58	}_{-	0.38	}$ &	157	&	$\cdots$	&	$u$	\\\vspace{1mm}
211.8172	&	2.451588	& $	21.72	^{+	0.77	}_{-	0.45	}$ &	157	&	$\cdots$	&	$u$	\\\vspace{1mm}
223.9091	&	2.591540	& $	21.25	^{+	0.42	}_{-	0.30	}$ &	235.5	&	$\cdots$	&	$u$	\\\vspace{1mm}
233.6637	&	2.704441	& $	21.75	^{+	0.90	}_{-	0.49	}$ &	235.6	&	$\cdots$	&	$u$	\\\vspace{1mm}
245.1895	&	2.837841	& $	20.82	^{+	0.55	}_{-	0.36	}$ &	156.9	&	$\cdots$	&	$u$	\\\vspace{1mm}
256.7393	&	2.971520	& $	21.74	^{+	2.17	}_{-	0.68	}$ &	157	&	$\cdots$	&	$u$	\\\vspace{1mm}
286.4793	&	3.315733	& $ >	20.66					$ &	84.7	&	$\cdots$	&	$u$	\\\vspace{1mm}
315.6230	&	3.653044	& $	21.84	^{+	0.70	}_{-	0.42	}$ &	314.1	&	$\cdots$	&	$u$	\\\vspace{1mm}
332.6649	&	3.850289	& $	21.98	^{+	0.52	}_{-	0.35	}$ &	382.4	&	$\cdots$	&	$u$	\\\vspace{1mm}
357.8214	&	4.141451	& $	21.78	^{+	0.51	}_{-	0.34	}$ &	844	&	$\cdots$	&	$u$	\\\vspace{1mm}
428.4023	&	4.958360	& $	22.05	^{+	0.44	}_{-	0.31	}$ &	578.3	&	$\cdots$	&	$u$	\\\vspace{1mm}
465.3887	&	5.386443	& $	21.45	^{+	0.42	}_{-	0.30	}$ &	342	&	$\cdots$	&	$u$	\\
\hline\vspace{1mm}															
0.6923	&	0.008013	& $	17.75	^{+	0.18	}_{-	0.16	}$ &	19.5	&	$\cdots$	&	$b$	\\\vspace{1mm}
1.2313	&	0.014252	& $	18.12	^{+	0.16	}_{-	0.14	}$ &	38.9	&	$\cdots$	&	$b$	\\\vspace{1mm}
1.5816	&	0.018305	& $	18.25	^{+	0.17	}_{-	0.15	}$ &	38.9	&	$\cdots$	&	$b$	\\\vspace{1mm}
1.9288	&	0.022324	& $	18.21	^{+	0.17	}_{-	0.15	}$ &	38.9	&	$\cdots$	&	$b$	\\\vspace{1mm}
6.3687	&	0.073712	& $	18.83	^{+	0.11	}_{-	0.10	}$ &	196.6	&	$\cdots$	&	$b$	\\\vspace{1mm}
12.4288	&	0.143852	& $	18.17	^{+	0.06	}_{-	0.06	}$ &	422.9	&	$\cdots$	&	$b$	\\\vspace{1mm}
17.4587	&	0.202068	& $	18.46	^{+	0.06	}_{-	0.05	}$ &	590.1	&	$\cdots$	&	$b$	\\\vspace{1mm}
30.1901	&	0.349422	& $	19.30	^{+	0.07	}_{-	0.07	}$ &	885.1	&	$\cdots$	&	$b$	\\\vspace{1mm}
46.7137	&	0.540668	& $	19.98	^{+	0.21	}_{-	0.18	}$ &	322	&	$\cdots$	&	$b$	\\\vspace{1mm}
55.5361	&	0.642779	& $	20.00	^{+	0.18	}_{-	0.15	}$ &	590.1	&	$\cdots$	&	$b$	\\\vspace{1mm}
78.6596	&	0.910412	& $	20.04	^{+	0.13	}_{-	0.12	}$ &	885.2	&	$\cdots$	&	$b$	\\\vspace{1mm}
113.1247	&	1.309314	& $	20.21	^{+	0.20	}_{-	0.17	}$ &	508.1	&	$\cdots$	&	$b$	\\\vspace{1mm}
181.8280	&	2.104491	& $ >	19.75					$ &	157	&	$\cdots$	&	$b$	\\\vspace{1mm}
193.2988	&	2.237254	& $	20.89	^{+	1.53	}_{-	0.61	}$ &	157	&	$\cdots$	&	$b$	\\\vspace{1mm}
205.7917	&	2.381848	& $ >	19.90					$ &	118.7	&	$\cdots$	&	$b$	\\\vspace{1mm}
217.9545	&	2.522621	& $ >	20.09					$ &	150.1	&	$\cdots$	&	$b$	\\\vspace{1mm}
227.9728	&	2.638574	& $	20.87	^{+	0.93	}_{-	0.49	}$ &	191.1	&	$\cdots$	&	$b$	\\\vspace{1mm}
237.6921	&	2.751066	& $ >	19.86					$ &	235.5	&	$\cdots$	&	$b$	\\\vspace{1mm}
251.0736	&	2.905944	& $	20.76	^{+	1.29	}_{-	0.57	}$ &	157	&	$\cdots$	&	$b$	\\\vspace{1mm}
283.6520	&	3.283009	& $	20.31	^{+	0.51	}_{-	0.34	}$ &	157	&	$\cdots$	&	$b$	\\\vspace{1mm}
315.7073	&	3.654019	& $	21.68	^{+	2.33	}_{-	0.69	}$ &	254.2	&	$\cdots$	&	$b$	\\\vspace{1mm}
332.8640	&	3.852593	& $	21.85	^{+	1.40	}_{-	0.59	}$ &	382.4	&	$\cdots$	&	$b$	\\\vspace{1mm}
358.2559	&	4.146480	& $	21.72	^{+	0.76	}_{-	0.44	}$ &	844	&	$\cdots$	&	$b$	\\\vspace{1mm}
428.7012	&	4.961819	& $	21.57	^{+	0.62	}_{-	0.39	}$ &	578.3	&	$\cdots$	&	$b$	\\\vspace{1mm}
465.5675	&	5.388512	& $ >	20.46					$ &	342.1	&	$\cdots$	&	$b$	\\
\hline\vspace{1mm}															
0.4140	&	0.004791	& $ >	17.60					$ &	9.5	&	$\cdots$	&	$v$	\\\vspace{1mm}
0.5940	&	0.006875	& $	17.20	^{+	0.25	}_{-	0.20	}$ &	19.5	&	$\cdots$	&	$v$	\\\vspace{1mm}
0.7664	&	0.008871	& $	17.47	^{+	0.29	}_{-	0.23	}$ &	19.4	&	$\cdots$	&	$v$	\\\vspace{1mm}
1.1323	&	0.013106	& $	17.77	^{+	0.24	}_{-	0.20	}$ &	38.9	&	$\cdots$	&	$v$	\\\vspace{1mm}
1.4811	&	0.017142	& $	17.86	^{+	0.26	}_{-	0.21	}$ &	38.9	&	$\cdots$	&	$v$	\\\vspace{1mm}
1.8297	&	0.021177	& $	17.73	^{+	0.23	}_{-	0.19	}$ &	38.9	&	$\cdots$	&	$v$	\\\vspace{1mm}
5.5478	&	0.064211	& $	18.04	^{+	0.15	}_{-	0.13	}$ &	196.6	&	$\cdots$	&	$v$	\\\vspace{1mm}
6.9842	&	0.080836	& $	17.70	^{+	0.10	}_{-	0.09	}$ &	196.6	&	$\cdots$	&	$v$	\\\vspace{1mm}
13.0037	&	0.150505	& $	18.06	^{+	0.10	}_{-	0.09	}$ &	295	&	$\cdots$	&	$v$	\\\vspace{1mm}
13.3072	&	0.154018	& $	18.08	^{+	0.10	}_{-	0.09	}$ &	295	&	$\cdots$	&	$v$	\\\vspace{1mm}
18.2242	&	0.210928	& $	18.61	^{+	0.14	}_{-	0.13	}$ &	442.8	&	$\cdots$	&	$v$	\\\vspace{1mm}
23.2940	&	0.269607	& $	18.69	^{+	0.11	}_{-	0.10	}$ &	590.1	&	$\cdots$	&	$v$	\\\vspace{1mm}
51.0158	&	0.590461	& $	19.83	^{+	0.28	}_{-	0.22	}$ &	523.4	&	$\cdots$	&	$v$	\\\vspace{1mm}
67.8238	&	0.784998	& $	19.87	^{+	0.39	}_{-	0.29	}$ &	326.4	&	$\cdots$	&	$v$	\\\vspace{1mm}
81.6539	&	0.945068	& $	20.12	^{+	0.28	}_{-	0.22	}$ &	885.1	&	$\cdots$	&	$v$	\\\vspace{1mm}
98.9052	&	1.144736	& $	19.45	^{+	0.15	}_{-	0.13	}$ &	885.2	&	$\cdots$	&	$v$	\\\vspace{1mm}
119.7450	&	1.385938	& $	19.84	^{+	0.23	}_{-	0.19	}$ &	796.3	&	$\cdots$	&	$v$	\\\vspace{1mm}
182.2380	&	2.109236	& $ >	19.20					$ &	156.2	&	$\cdots$	&	$v$	\\\vspace{1mm}
193.7085	&	2.241996	& $ >	19.14					$ &	157.1	&	$\cdots$	&	$v$	\\\vspace{1mm}
234.1580	&	2.710162	& $	20.32	^{+	1.44	}_{-	0.60	}$ &	132.5	&	$\cdots$	&	$v$	\\\vspace{1mm}
245.6837	&	2.843561	& $ >	19.14					$ &	157	&	$\cdots$	&	$v$	\\\vspace{1mm}
257.2340	&	2.977245	& $	20.61	^{+	2.17	}_{-	0.68	}$ &	157	&	$\cdots$	&	$v$	\\\vspace{1mm}
312.9387	&	3.621976	& $ >	19.20					$ &	157	&	$\cdots$	&	$v$	\\\vspace{1mm}
348.0090	&	4.027882	& $ >	20.20					$ &	713.5	&	$\cdots$	&	$v$	\\\vspace{1mm}
394.1291	&	4.561679	& $	20.97	^{+	0.84	}_{-	0.47	}$ &	565.2	&	$\cdots$	&	$v$	\\\vspace{1mm}
442.6982	&	5.123822	& $	20.88	^{+	0.82	}_{-	0.46	}$ &	551.4	&	$\cdots$	&	$v$	\\
\hline\vspace{1mm}															
0.4323	&	0.005003	& $	18.79	^{+	0.16	}_{-	0.14	}$ &	10	&	$\cdots$	&	$white$	\\\vspace{1mm}
0.4423	&	0.005119	& $	18.80	^{+	0.16	}_{-	0.14	}$ &	10	&	$\cdots$	&	$white$	\\\vspace{1mm}
0.4523	&	0.005235	& $	18.85	^{+	0.16	}_{-	0.14	}$ &	10	&	$\cdots$	&	$white$	\\\vspace{1mm}
0.4623	&	0.005350	& $	18.93	^{+	0.17	}_{-	0.15	}$ &	10	&	$\cdots$	&	$white$	\\\vspace{1mm}
0.4723	&	0.005466	& $	19.06	^{+	0.18	}_{-	0.16	}$ &	10	&	$\cdots$	&	$white$	\\\vspace{1mm}
0.4823	&	0.005582	& $	19.24	^{+	0.21	}_{-	0.17	}$ &	10	&	$\cdots$	&	$white$	\\\vspace{1mm}
0.4923	&	0.005698	& $	19.08	^{+	0.18	}_{-	0.16	}$ &	10	&	$\cdots$	&	$white$	\\\vspace{1mm}
0.5023	&	0.005813	& $	18.85	^{+	0.16	}_{-	0.14	}$ &	10	&	$\cdots$	&	$white$	\\\vspace{1mm}
0.5123	&	0.005929	& $	18.81	^{+	0.16	}_{-	0.14	}$ &	10	&	$\cdots$	&	$white$	\\\vspace{1mm}
0.5223	&	0.006045	& $	19.11	^{+	0.19	}_{-	0.16	}$ &	10	&	$\cdots$	&	$white$	\\\vspace{1mm}
0.5323	&	0.006161	& $	18.79	^{+	0.16	}_{-	0.14	}$ &	10	&	$\cdots$	&	$white$	\\\vspace{1mm}
0.5423	&	0.006276	& $	18.70	^{+	0.15	}_{-	0.13	}$ &	10	&	$\cdots$	&	$white$	\\\vspace{1mm}
0.5523	&	0.006392	& $	18.50	^{+	0.14	}_{-	0.12	}$ &	10	&	$\cdots$	&	$white$	\\\vspace{1mm}
0.5623	&	0.006508	& $	18.70	^{+	0.15	}_{-	0.13	}$ &	10	&	$\cdots$	&	$white$	\\\vspace{1mm}
0.5723	&	0.006624	& $	18.67	^{+	0.15	}_{-	0.13	}$ &	9.8	&	$\cdots$	&	$white$	\\\vspace{1mm}
0.7166	&	0.008294	& $	18.12	^{+	0.09	}_{-	0.08	}$ &	19.5	&	$\cdots$	&	$white$	\\\vspace{1mm}
0.9277	&	0.010737	& $	18.57	^{+	0.05	}_{-	0.05	}$ &	147.4	&	$\cdots$	&	$white$	\\\vspace{1mm}
1.2558	&	0.014535	& $	18.66	^{+	0.08	}_{-	0.07	}$ &	38.9	&	$\cdots$	&	$white$	\\\vspace{1mm}
1.6059	&	0.018587	& $	18.69	^{+	0.08	}_{-	0.07	}$ &	38.9	&	$\cdots$	&	$white$	\\\vspace{1mm}
1.9532	&	0.022607	& $	18.56	^{+	0.07	}_{-	0.07	}$ &	38.9	&	$\cdots$	&	$white$	\\\vspace{1mm}
6.5737	&	0.076084	& $	18.97	^{+	0.05	}_{-	0.05	}$ &	196.6	&	$\cdots$	&	$white$	\\\vspace{1mm}
11.1840	&	0.129445	& $	18.60	^{+	0.05	}_{-	0.04	}$ &	295	&	$\cdots$	&	$white$	\\\vspace{1mm}
11.4876	&	0.132958	& $	18.69	^{+	0.04	}_{-	0.04	}$ &	295	&	$\cdots$	&	$white$	\\\vspace{1mm}
11.7913	&	0.136473	& $	18.72	^{+	0.04	}_{-	0.04	}$ &	295	&	$\cdots$	&	$white$	\\\vspace{1mm}
18.0674	&	0.209113	& $	18.86	^{+	0.03	}_{-	0.03	}$ &	590.1	&	$\cdots$	&	$white$	\\\vspace{1mm}
24.6549	&	0.285358	& $	19.42	^{+	0.04	}_{-	0.04	}$ &	575.8	&	$\cdots$	&	$white$	\\\vspace{1mm}
56.0929	&	0.649223	& $	20.49	^{+	0.07	}_{-	0.07	}$ &	427.1	&	$\cdots$	&	$white$	\\\vspace{1mm}
93.9212	&	1.087051	& $	20.42	^{+	0.07	}_{-	0.07	}$ &	393.4	&	$\cdots$	&	$white$	\\\vspace{1mm}
136.1345	&	1.575631	& $	20.29	^{+	0.05	}_{-	0.05	}$ &	666.6	&	$\cdots$	&	$white$	\\\vspace{1mm}
152.8072	&	1.768602	& $	20.58	^{+	0.06	}_{-	0.06	}$ &	941.7	&	$\cdots$	&	$white$	\\\vspace{1mm}
158.5632	&	1.835222	& $	20.67	^{+	0.06	}_{-	0.06	}$ &	1226.5	&	$\cdots$	&	$white$	\\\vspace{1mm}
164.3322	&	1.901993	& $	20.66	^{+	0.07	}_{-	0.07	}$ &	780.1	&	$\cdots$	&	$white$	\\\vspace{1mm}
170.1173	&	1.968950	& $	20.89	^{+	0.09	}_{-	0.08	}$ &	735.5	&	$\cdots$	&	$white$	\\\vspace{1mm}
219.2282	&	2.537364	& $	21.08	^{+	0.11	}_{-	0.10	}$ &	700.8	&	$\cdots$	&	$white$	\\\vspace{1mm}
269.2423	&	3.116231	& $	21.32	^{+	0.17	}_{-	0.15	}$ &	334.4	&	$\cdots$	&	$white$	\\\vspace{1mm}
275.9110	&	3.193414	& $	21.67	^{+	0.31	}_{-	0.24	}$ &	141.4	&	$\cdots$	&	$white$	\\\vspace{1mm}
310.6214	&	3.595155	& $	22.05	^{+	0.28	}_{-	0.22	}$ &	312	&	$\cdots$	&	$white$	\\\vspace{1mm}
358.6900	&	4.151504	& $	22.34	^{+	0.21	}_{-	0.18	}$ &	844	&	$\cdots$	&	$white$	\\\vspace{1mm}
428.9996	&	4.965273	& $	22.42	^{+	0.28	}_{-	0.22	}$ &	578.3	&	$\cdots$	&	$white$	\\\vspace{1mm}
465.7457	&	5.390575	& $	22.08	^{+	0.32	}_{-	0.25	}$ &	342.1	&	$\cdots$	&	$white$	\\\vspace{1mm}
508.0972	&	5.880754	& $	22.44	^{+	0.12	}_{-	0.11	}$ &	3561.9	&	$\cdots$	&	$white$	\\\vspace{1mm}
519.7252	&	6.015338	& $	22.38	^{+	0.12	}_{-	0.10	}$ &	3422.5	&	$\cdots$	&	$white$	\\\vspace{1mm}
560.4170	&	6.486308	& $	22.43	^{+	0.11	}_{-	0.10	}$ &	3418.5	&	$\cdots$	&	$white$	\\\vspace{1mm}
603.9736	&	6.990435	& $	22.76	^{+	0.22	}_{-	0.19	}$ &	1627.4	&	$\cdots$	&	$white$	\\\vspace{1mm}
619.1009	&	7.165520	& $	22.86	^{+	0.14	}_{-	0.12	}$ &	4709	&	$\cdots$	&	$white$	\\\vspace{1mm}
642.7491	&	7.439226	& $	22.74	^{+	0.15	}_{-	0.13	}$ &	3218.4	&	$\cdots$	&	$white$	\\\vspace{1mm}
668.0231	&	7.731749	& $	22.91	^{+	0.21	}_{-	0.17	}$ &	2451.2	&	$\cdots$	&	$white$	\\\vspace{1mm}
690.6005	&	7.993062	& $	22.54	^{+	0.18	}_{-	0.16	}$ &	1626.3	&	$\cdots$	&	$white$	\\\vspace{1mm}
713.7695	&	8.261221	& $	22.79	^{+	0.13	}_{-	0.11	}$ &	4846.1	&	$\cdots$	&	$white$	\\\vspace{1mm}
739.7536	&	8.561963	& $	22.48	^{+	0.18	}_{-	0.16	}$ &	1485.5	&	$\cdots$	&	$white$	\\\vspace{1mm}
765.6518	&	8.861711	& $	23.09	^{+	0.21	}_{-	0.18	}$ &	3288	&	$\cdots$	&	$white$	\\\vspace{1mm}
782.9717	&	9.062172	& $	23.10	^{+	0.17	}_{-	0.15	}$ &	4948.5	&	$\cdots$	&	$white$	\\\vspace{1mm}
817.6420	&	9.463449	& $	23.03	^{+	0.30	}_{-	0.23	}$ &	1654.1	&	$\cdots$	&	$white$	\\\vspace{1mm}
852.2990	&	9.864572	& $	23.31	^{+	0.26	}_{-	0.21	}$ &	3291.6	&	$\cdots$	&	$white$	\\\vspace{1mm}
870.3247	&	10.07320	& $	23.42	^{+	0.24	}_{-	0.20	}$ &	4641.8	&	$\cdots$	&	$white$	\\\vspace{1mm}
904.9478	&	10.47393	& $	23.72	^{+	0.57	}_{-	0.37	}$ &	1966.5	&	$\cdots$	&	$white$	\\\vspace{1mm}
939.6598	&	10.87569	& $	23.87	^{+	0.41	}_{-	0.29	}$ &	4380.6	&	$\cdots$	&	$white$	\\\vspace{1mm}
982.1179	&	11.36711	& $	23.37	^{+	0.27	}_{-	0.22	}$ &	3550.4	&	$\cdots$	&	$white$	\\\vspace{1mm}
1028.4475	&	11.90333	& $	23.91	^{+	0.69	}_{-	0.42	}$ &	2202.2	&	$\cdots$	&	$white$	\\\vspace{1mm}
1068.6013	&	12.36807	& $	24.07	^{+	0.64	}_{-	0.40	}$ &	3182.9	&	$\cdots$	&	$white$	\\\vspace{1mm}
1103.3277	&	12.77000	& $	24.39	^{+	0.77	}_{-	0.45	}$ &	4852	&	$\cdots$	&	$white$	\\\vspace{1mm}
1149.9545	&	13.30966	& $	24.82	^{+	1.70	}_{-	0.63	}$ &	4368.8	&	$\cdots$	&	$white$	\\\vspace{1mm}
1199.1767	&	13.87936	& $	24.44	^{+	0.84	}_{-	0.47	}$ &	6120.1	&	$\cdots$	&	$white$	\\\vspace{1mm}
1245.3861	&	14.41419	& $	23.63	^{+	0.86	}_{-	0.47	}$ &	893.9	&	$\cdots$	&	$white$	\\\vspace{1mm}
1291.2045	&	14.94450	& $	24.22	^{+	1.16	}_{-	0.55	}$ &	1925.4	&	$\cdots$	&	$white$	\\\vspace{1mm}
1320.0897	&	15.27882	& $	24.10	^{+	0.59	}_{-	0.38	}$ &	5253.9	&	$\cdots$	&	$white$	\\\vspace{1mm}
1351.7119	&	15.64481	& $ >	22.50					$ &	634.4	&	$\cdots$	&	$white$	\\\vspace{1mm}
1400.4694	&	16.20914	& $ >	22.22					$ &	1306.4	&	$\cdots$	&	$white$	\\\vspace{1mm}
1458.1890	&	16.87719	& $ >	20.95					$ &	206.3	&	$\cdots$	&	$white$	\\\vspace{1mm}
1499.3459	&	17.35354	& $	24.09	^{+	0.56	}_{-	0.37	}$ &	5169.2	&	$\cdots$	&	$white$	\\\vspace{1mm}
1542.6334	&	17.85455	& $	24.61	^{+	3.44	}_{-	0.73	}$ &	2189.9	&	$\cdots$	&	$white$	\\\vspace{1mm}
1588.1806	&	18.38172	& $	24.62	^{+	1.85	}_{-	0.65	}$ &	4533.7	&	$\cdots$	&	$white$	\\\vspace{1mm}
1622.8880	&	18.78343	& $ >	22.46					$ &	3294.5	&	$\cdots$	&	$white$	\\\vspace{1mm}
1678.0583	&	19.42197	& $	24.76	^{+	1.83	}_{-	0.65	}$ &	5025.1	&	$\cdots$	&	$white$	\\\vspace{1mm}
1744.7027	&	20.19332	& $	23.74	^{+	0.89	}_{-	0.48	}$ &	1421.7	&	$\cdots$	&	$white$	\\\vspace{1mm}
1794.1713	&	20.76587	& $ >	23.03					$ &	2798.5	&	$\cdots$	&	$white$	\\\vspace{1mm}
1834.5986	&	21.23378	& $	22.99	^{+	0.45	}_{-	0.32	}$ &	838.7	&	$\cdots$	&	$white$	\\\vspace{1mm}
1874.6821	&	21.69771	& $	24.15	^{+	0.65	}_{-	0.40	}$ &	4605.8	&	$\cdots$	&	$white$	\\\vspace{1mm}
1938.2157	&	22.43305	& $ >	22.92					$ &	3090	&	$\cdots$	&	$white$	\\\vspace{1mm}
2001.7916	&	23.16888	& $ >	22.88					$ &	1584.4	&	$\cdots$	&	$white$	\\\vspace{1mm}
2068.2861	&	23.93850	& $ >	22.91					$ &	3074.9	&	$\cdots$	&	$white$	\\\vspace{1mm}
2137.6769	&	24.74163	& $	24.65	^{+	1.97	}_{-	0.66	}$ &	3156.8	&	$\cdots$	&	$white$	\\\vspace{1mm}
2183.9082	&	25.27671	& $	23.62	^{+	0.66	}_{-	0.41	}$ &	1555.5	&	$\cdots$	&	$white$	\\\vspace{1mm}
2235.9254	&	25.87877	& $ >	22.97					$ &	6310.5	&	$\cdots$	&	$white$	\\
\hline															
64.49	&	0.746440	& $	20.426	\pm	0.059	$ &	142	&	0.99	&	$g^\prime$	\\
64.93	&	0.751499	& $	20.478	\pm	0.047	$ &	142	&	0.94	&	$g^\prime$	\\
65.37	&	0.756632	& $	20.507	\pm	0.055	$ &	142	&	0.85	&	$g^\prime$	\\
65.82	&	0.761760	& $	20.493	\pm	0.061	$ &	142	&	0.87	&	$g^\prime$	\\
66.25	&	0.766805	& $	20.476	\pm	0.057	$ &	142	&	0.74	&	$g^\prime$	\\
66.69	&	0.771848	& $	20.464	\pm	0.065	$ &	142	&	0.85	&	$g^\prime$	\\
67.13	&	0.776936	& $	20.462	\pm	0.049	$ &	142	&	0.79	&	$g^\prime$	\\
67.58	&	0.782209	& $	20.461	\pm	0.059	$ &	142	&	0.86	&	$g^\prime$	\\
68.02	&	0.787229	& $	20.502	\pm	0.053	$ &	142	&	0.80	&	$g^\prime$	\\
68.46	&	0.792317	& $	20.492	\pm	0.054	$ &	142	&	0.84	&	$g^\prime$	\\
69.20	&	0.800883	& $	20.462	\pm	0.045	$ &	460	&	0.82	&	$g^\prime$	\\
70.00	&	0.810153	& $	20.469	\pm	0.047	$ &	460	&	0.79	&	$g^\prime$	\\
70.85	&	0.820019	& $	20.429	\pm	0.041	$ &	460	&	0.77	&	$g^\prime$	\\
71.60	&	0.828663	& $	20.444	\pm	0.055	$ &	460	&	0.79	&	$g^\prime$	\\
72.38	&	0.837725	& $	20.430	\pm	0.055	$ &	460	&	0.86	&	$g^\prime$	\\
73.16	&	0.846806	& $	20.456	\pm	0.042	$ &	460	&	0.89	&	$g^\prime$	\\
73.96	&	0.856003	& $	20.432	\pm	0.055	$ &	460	&	1.08	&	$g^\prime$	\\
74.76	&	0.865263	& $	20.382	\pm	0.051	$ &	460	&	1.02	&	$g^\prime$	\\
75.56	&	0.874504	& $	20.364	\pm	0.081	$ &	460	&	1.19	&	$g^\prime$	\\
76.35	&	0.883669	& $	20.303	\pm	0.065	$ &	460	&	1.12	&	$g^\prime$	\\
77.14	&	0.892841	& $	20.304	\pm	0.069	$ &	460	&	1.23	&	$g^\prime$	\\
77.95	&	0.902235	& $	20.262	\pm	0.072	$ &	460	&	1.18	&	$g^\prime$	\\
78.75	&	0.911515	& $	20.170	\pm	0.082	$ &	460	&	1.49	&	$g^\prime$	\\
79.56	&	0.920794	& $	20.172	\pm	0.081	$ &	460	&	1.47	&	$g^\prime$	\\
80.36	&	0.930069	& $	20.166	\pm	0.084	$ &	460	&	1.58	&	$g^\prime$	\\
81.17	&	0.939427	& $	20.080	\pm	0.046	$ &	460	&	1.59	&	$g^\prime$	\\
81.95	&	0.948452	& $	20.002	\pm	0.052	$ &	460	&	2.11	&	$g^\prime$	\\
151.49	&	1.753348	& $	20.047	\pm	0.053	$ &	460	&	1.04	&	$g^\prime$	\\
155.91	&	1.804496	& $	20.073	\pm	0.058	$ &	460	&	1.24	&	$g^\prime$	\\
160.33	&	1.855649	& $	20.048	\pm	0.040	$ &	460	&	1.56	&	$g^\prime$	\\
164.70	&	1.906235	& $	20.138	\pm	0.050	$ &	460	&	1.40	&	$g^\prime$	\\
239.81	&	2.775588	& $	20.810	\pm	0.043	$ &	919	&	0.92	&	$g^\prime$	\\
250.95	&	2.904470	& $	20.852	\pm	0.055	$ &	919	&	1.30	&	$g^\prime$	\\
329.17	&	3.809864	& $	21.156	\pm	0.055	$ &	1133	&	1.10	&	$g^\prime$	\\
415.47	&	4.808631	& $	21.493	\pm	0.048	$ &	1838	&	1.08	&	$g^\prime$	\\
501.08	&	5.799551	& $	21.592	\pm	0.031	$ &	1838	&	0.98	&	$g^\prime$	\\
588.10	&	6.806728	& $	21.851	\pm	0.054	$ &	1838	&	1.18	&	$g^\prime$	\\
669.18	&	7.745094	& $	22.033	\pm	0.053	$ &	919	&	1.24	&	$g^\prime$	\\
843.66	&	9.764627	& $	22.393	\pm	0.027	$ &	1379	&	0.65	&	$g^\prime$	\\
1101.93	&	12.75382	& $	22.855	\pm	0.056	$ &	2420	&	1.07	&	$g^\prime$	\\
1880.55	&	21.76561	& $	23.260	\pm	0.090	$ &	2952	&	1.55	&	$g^\prime$	\\
2401.32	&	27.79309	& $	23.454	\pm	0.186	$ &	4502	&	1.48	&	$g^\prime$	\\
3090.97	&	35.77507	& $	23.803	\pm	0.120	$ &	3630	&	1.51	&	$g^\prime$	\\
4258.44	&	49.28755	& $	24.267	\pm	0.238	$ &	5384	&	1.77	&	$g^\prime$	\\
4732.20	&	54.77078	& $	24.467	\pm	0.352	$ &	5422	&	1.62	&	$g^\prime$	\\
6241.88	&	72.24398	& $ >	24.573			$ &	2758	&	1.66	&	$g^\prime$	\\
24277.46	&	280.9891	& $	25.661	\pm	0.312	$ &	3752	&	1.22	&	$g^\prime$	\\
\hline													
64.49	&	0.746440	& $	20.106	\pm	0.036	$ &	142	&	1.02	&	$r^\prime$	\\
64.93	&	0.751499	& $	20.111	\pm	0.037	$ &	142	&	0.93	&	$r^\prime$	\\
65.37	&	0.756632	& $	20.160	\pm	0.049	$ &	142	&	0.87	&	$r^\prime$	\\
65.82	&	0.761760	& $	20.113	\pm	0.038	$ &	142	&	0.86	&	$r^\prime$	\\
66.25	&	0.766805	& $	20.133	\pm	0.040	$ &	142	&	0.84	&	$r^\prime$	\\
66.69	&	0.771848	& $	20.109	\pm	0.030	$ &	142	&	0.94	&	$r^\prime$	\\
67.13	&	0.776936	& $	20.111	\pm	0.045	$ &	142	&	0.92	&	$r^\prime$	\\
67.58	&	0.782209	& $	20.131	\pm	0.035	$ &	142	&	0.72	&	$r^\prime$	\\
68.02	&	0.787229	& $	20.123	\pm	0.031	$ &	142	&	0.75	&	$r^\prime$	\\
68.46	&	0.792317	& $	20.097	\pm	0.027	$ &	142	&	0.78	&	$r^\prime$	\\
69.20	&	0.800883	& $	20.122	\pm	0.038	$ &	460	&	0.84	&	$r^\prime$	\\
70.00	&	0.810153	& $	20.121	\pm	0.031	$ &	460	&	0.88	&	$r^\prime$	\\
70.85	&	0.820019	& $	20.120	\pm	0.039	$ &	460	&	0.86	&	$r^\prime$	\\
71.60	&	0.828663	& $	20.102	\pm	0.018	$ &	460	&	0.62	&	$r^\prime$	\\
72.38	&	0.837725	& $	20.075	\pm	0.022	$ &	460	&	0.65	&	$r^\prime$	\\
73.16	&	0.846806	& $	20.105	\pm	0.021	$ &	460	&	0.70	&	$r^\prime$	\\
73.96	&	0.856003	& $	20.019	\pm	0.020	$ &	460	&	0.77	&	$r^\prime$	\\
74.76	&	0.865263	& $	19.996	\pm	0.019	$ &	460	&	0.77	&	$r^\prime$	\\
75.56	&	0.874504	& $	19.951	\pm	0.024	$ &	460	&	0.84	&	$r^\prime$	\\
76.35	&	0.883669	& $	19.901	\pm	0.020	$ &	460	&	0.80	&	$r^\prime$	\\
77.14	&	0.892841	& $	19.870	\pm	0.019	$ &	460	&	0.78	&	$r^\prime$	\\
77.95	&	0.902235	& $	19.827	\pm	0.023	$ &	460	&	0.88	&	$r^\prime$	\\
78.75	&	0.911515	& $	19.787	\pm	0.024	$ &	460	&	1.06	&	$r^\prime$	\\
79.56	&	0.920794	& $	19.752	\pm	0.021	$ &	460	&	0.98	&	$r^\prime$	\\
80.36	&	0.930069	& $	19.701	\pm	0.024	$ &	460	&	1.03	&	$r^\prime$	\\
81.17	&	0.939427	& $	19.656	\pm	0.046	$ &	460	&	1.07	&	$r^\prime$	\\
81.95	&	0.948452	& $	19.622	\pm	0.030	$ &	460	&	1.41	&	$r^\prime$	\\
151.49	&	1.753348	& $	19.659	\pm	0.016	$ &	460	&	0.81	&	$r^\prime$	\\
155.91	&	1.804496	& $	19.617	\pm	0.022	$ &	460	&	0.91	&	$r^\prime$	\\
160.33	&	1.855649	& $	19.653	\pm	0.015	$ &	460	&	1.14	&	$r^\prime$	\\
164.70	&	1.906235	& $	19.751	\pm	0.037	$ &	460	&	1.13	&	$r^\prime$	\\
239.81	&	2.775588	& $	20.347	\pm	0.029	$ &	919	&	0.95	&	$r^\prime$	\\
250.95	&	2.904470	& $	20.492	\pm	0.023	$ &	919	&	0.95	&	$r^\prime$	\\
329.17	&	3.809864	& $	20.737	\pm	0.026	$ &	1133	&	0.71	&	$r^\prime$	\\
415.47	&	4.808631	& $	21.084	\pm	0.031	$ &	1838	&	0.86	&	$r^\prime$	\\
501.08	&	5.799551	& $	21.188	\pm	0.024	$ &	1838	&	0.73	&	$r^\prime$	\\
588.10	&	6.806728	& $	21.461	\pm	0.027	$ &	1838	&	0.98	&	$r^\prime$	\\
669.18	&	7.745094	& $	21.668	\pm	0.076	$ &	919	&	1.25	&	$r^\prime$	\\
843.66	&	9.764627	& $	22.009	\pm	0.028	$ &	1379	&	0.69	&	$r^\prime$	\\
1101.93	&	12.75382	& $	22.425	\pm	0.038	$ &	2420	&	0.94	&	$r^\prime$	\\
1880.55	&	21.76561	& $	22.680	\pm	0.046	$ &	2952	&	1.25	&	$r^\prime$	\\
2401.32	&	27.79309	& $	23.005	\pm	0.093	$ &	4502	&	1.24	&	$r^\prime$	\\
3090.97	&	35.77507	& $	23.106	\pm	0.082	$ &	3630	&	1.20	&	$r^\prime$	\\
4258.44	&	49.28755	& $	23.598	\pm	0.134	$ &	5384	&	1.36	&	$r^\prime$	\\
4732.20	&	54.77078	& $	23.923	\pm	0.148	$ &	5422	&	1.19	&	$r^\prime$	\\
6241.88	&	72.24398	& $	24.450	\pm	0.278	$ &	2758	&	1.17	&	$r^\prime$	\\
24277.46	&	280.9891	& $	25.044	\pm	0.184	$ &	3752	&	0.96	&	$r^\prime$	\\
\hline													
64.49	&	0.746440	& $	19.914	\pm	0.043	$ &	142	&	0.90	&	$i^\prime$	\\
64.93	&	0.751499	& $	19.916	\pm	0.051	$ &	142	&	0.84	&	$i^\prime$	\\
65.37	&	0.756632	& $	19.899	\pm	0.032	$ &	142	&	0.80	&	$i^\prime$	\\
65.82	&	0.761760	& $	19.917	\pm	0.037	$ &	142	&	0.75	&	$i^\prime$	\\
66.25	&	0.766805	& $	19.929	\pm	0.036	$ &	142	&	0.71	&	$i^\prime$	\\
66.69	&	0.771848	& $	19.894	\pm	0.036	$ &	142	&	0.79	&	$i^\prime$	\\
67.13	&	0.776936	& $	19.892	\pm	0.039	$ &	142	&	0.76	&	$i^\prime$	\\
67.58	&	0.782209	& $	19.915	\pm	0.037	$ &	142	&	0.66	&	$i^\prime$	\\
68.02	&	0.787229	& $	19.914	\pm	0.038	$ &	142	&	0.67	&	$i^\prime$	\\
68.46	&	0.792317	& $	19.904	\pm	0.040	$ &	142	&	0.69	&	$i^\prime$	\\
69.20	&	0.800883	& $	19.921	\pm	0.022	$ &	460	&	0.78	&	$i^\prime$	\\
70.00	&	0.810153	& $	19.904	\pm	0.021	$ &	460	&	0.76	&	$i^\prime$	\\
70.85	&	0.820019	& $	19.902	\pm	0.023	$ &	460	&	0.74	&	$i^\prime$	\\
71.60	&	0.828663	& $	19.899	\pm	0.020	$ &	460	&	0.57	&	$i^\prime$	\\
72.38	&	0.837725	& $	19.884	\pm	0.022	$ &	460	&	0.61	&	$i^\prime$	\\
73.16	&	0.846806	& $	19.860	\pm	0.023	$ &	460	&	0.64	&	$i^\prime$	\\
73.96	&	0.856003	& $	19.835	\pm	0.024	$ &	460	&	0.72	&	$i^\prime$	\\
74.76	&	0.865263	& $	19.788	\pm	0.024	$ &	460	&	0.72	&	$i^\prime$	\\
75.56	&	0.874504	& $	19.748	\pm	0.025	$ &	460	&	0.78	&	$i^\prime$	\\
76.35	&	0.883669	& $	19.654	\pm	0.019	$ &	460	&	0.73	&	$i^\prime$	\\
77.14	&	0.892841	& $	19.634	\pm	0.017	$ &	460	&	0.75	&	$i^\prime$	\\
77.95	&	0.902235	& $	19.578	\pm	0.022	$ &	460	&	0.80	&	$i^\prime$	\\
78.75	&	0.911515	& $	19.533	\pm	0.028	$ &	460	&	0.97	&	$i^\prime$	\\
79.56	&	0.920794	& $	19.460	\pm	0.029	$ &	460	&	0.89	&	$i^\prime$	\\
80.36	&	0.930069	& $	19.408	\pm	0.025	$ &	460	&	0.94	&	$i^\prime$	\\
81.17	&	0.939427	& $	19.414	\pm	0.036	$ &	460	&	0.98	&	$i^\prime$	\\
81.95	&	0.948452	& $	19.400	\pm	0.024	$ &	460	&	1.28	&	$i^\prime$	\\
151.49	&	1.753348	& $	19.360	\pm	0.030	$ &	460	&	0.81	&	$i^\prime$	\\
155.91	&	1.804496	& $	19.392	\pm	0.033	$ &	460	&	0.95	&	$i^\prime$	\\
160.33	&	1.855649	& $	19.360	\pm	0.021	$ &	460	&	1.10	&	$i^\prime$	\\
164.70	&	1.906235	& $	19.434	\pm	0.029	$ &	460	&	1.12	&	$i^\prime$	\\
239.81	&	2.775588	& $	20.105	\pm	0.024	$ &	919	&	0.85	&	$i^\prime$	\\
250.95	&	2.904470	& $	20.201	\pm	0.029	$ &	919	&	0.92	&	$i^\prime$	\\
329.17	&	3.809864	& $	20.432	\pm	0.031	$ &	1133	&	0.74	&	$i^\prime$	\\
415.47	&	4.808631	& $	20.811	\pm	0.022	$ &	1838	&	0.82	&	$i^\prime$	\\
501.08	&	5.799551	& $	20.902	\pm	0.019	$ &	1838	&	0.72	&	$i^\prime$	\\
588.10	&	6.806728	& $	21.181	\pm	0.038	$ &	1838	&	0.90	&	$i^\prime$	\\
669.18	&	7.745094	& $	21.396	\pm	0.089	$ &	919	&	1.12	&	$i^\prime$	\\
843.66	&	9.764627	& $	21.754	\pm	0.042	$ &	1379	&	0.63	&	$i^\prime$	\\
1101.93	&	12.75382	& $	22.198	\pm	0.067	$ &	2420	&	0.86	&	$i^\prime$	\\
1880.55	&	21.76561	& $	22.398	\pm	0.073	$ &	2952	&	1.14	&	$i^\prime$	\\
2401.32	&	27.79309	& $	22.627	\pm	0.105	$ &	4502	&	1.18	&	$i^\prime$	\\
3090.97	&	35.77507	& $	22.811	\pm	0.102	$ &	3630	&	1.17	&	$i^\prime$	\\
4258.44	&	49.28755	& $	23.262	\pm	0.233	$ &	5384	&	1.33	&	$i^\prime$	\\
4732.20	&	54.77078	& $	23.378	\pm	0.160	$ &	5422	&	1.14	&	$i^\prime$	\\
6241.88	&	72.24398	& $	23.678	\pm	0.320	$ &	2758	&	1.12	&	$i^\prime$	\\
24277.46	&	280.9891	& $	24.364	\pm	0.218	$ &	3752	&	0.94	&	$i^\prime$	\\
\hline													
64.49	&	0.746440	& $	19.716	\pm	0.053	$ &	142	&	0.81	&	$z^\prime$	\\
64.93	&	0.751499	& $	19.745	\pm	0.053	$ &	142	&	0.77	&	$z^\prime$	\\
65.37	&	0.756632	& $	19.759	\pm	0.065	$ &	142	&	0.71	&	$z^\prime$	\\
65.82	&	0.761760	& $	19.734	\pm	0.062	$ &	142	&	0.70	&	$z^\prime$	\\
66.25	&	0.766805	& $	19.717	\pm	0.051	$ &	142	&	0.63	&	$z^\prime$	\\
66.69	&	0.771848	& $	19.721	\pm	0.053	$ &	142	&	0.72	&	$z^\prime$	\\
67.13	&	0.776936	& $	19.767	\pm	0.059	$ &	142	&	0.71	&	$z^\prime$	\\
67.58	&	0.782209	& $	19.692	\pm	0.041	$ &	142	&	0.63	&	$z^\prime$	\\
68.02	&	0.787229	& $	19.723	\pm	0.048	$ &	142	&	0.65	&	$z^\prime$	\\
68.46	&	0.792317	& $	19.742	\pm	0.043	$ &	142	&	0.67	&	$z^\prime$	\\
69.20	&	0.800883	& $	19.675	\pm	0.023	$ &	460	&	0.68	&	$z^\prime$	\\
70.00	&	0.810153	& $	19.663	\pm	0.052	$ &	460	&	0.68	&	$z^\prime$	\\
70.85	&	0.820019	& $	19.670	\pm	0.044	$ &	460	&	0.67	&	$z^\prime$	\\
71.60	&	0.828663	& $	19.681	\pm	0.055	$ &	460	&	0.53	&	$z^\prime$	\\
72.38	&	0.837725	& $	19.642	\pm	0.053	$ &	460	&	0.58	&	$z^\prime$	\\
73.16	&	0.846806	& $	19.634	\pm	0.053	$ &	460	&	0.64	&	$z^\prime$	\\
73.96	&	0.856003	& $	19.537	\pm	0.043	$ &	460	&	0.69	&	$z^\prime$	\\
74.76	&	0.865263	& $	19.547	\pm	0.035	$ &	460	&	0.72	&	$z^\prime$	\\
75.56	&	0.874504	& $	19.534	\pm	0.063	$ &	460	&	0.75	&	$z^\prime$	\\
76.35	&	0.883669	& $	19.416	\pm	0.044	$ &	460	&	0.70	&	$z^\prime$	\\
77.14	&	0.892841	& $	19.395	\pm	0.036	$ &	460	&	0.73	&	$z^\prime$	\\
77.95	&	0.902235	& $	19.323	\pm	0.026	$ &	460	&	0.78	&	$z^\prime$	\\
78.75	&	0.911515	& $	19.277	\pm	0.041	$ &	460	&	0.95	&	$z^\prime$	\\
79.56	&	0.920794	& $	19.237	\pm	0.044	$ &	460	&	0.88	&	$z^\prime$	\\
80.36	&	0.930069	& $	19.166	\pm	0.024	$ &	460	&	0.92	&	$z^\prime$	\\
81.17	&	0.939427	& $	19.136	\pm	0.024	$ &	460	&	0.94	&	$z^\prime$	\\
81.95	&	0.948452	& $	19.098	\pm	0.037	$ &	460	&	1.27	&	$z^\prime$	\\
151.49	&	1.753348	& $	19.132	\pm	0.017	$ &	460	&	0.77	&	$z^\prime$	\\
155.91	&	1.804496	& $	19.128	\pm	0.020	$ &	460	&	0.92	&	$z^\prime$	\\
160.33	&	1.855649	& $	19.149	\pm	0.024	$ &	460	&	1.08	&	$z^\prime$	\\
164.70	&	1.906235	& $	19.189	\pm	0.043	$ &	460	&	1.10	&	$z^\prime$	\\
239.81	&	2.775588	& $	19.889	\pm	0.053	$ &	919	&	0.76	&	$z^\prime$	\\
250.95	&	2.904470	& $	19.950	\pm	0.067	$ &	919	&	0.87	&	$z^\prime$	\\
329.17	&	3.809864	& $	20.221	\pm	0.045	$ &	1133	&	0.72	&	$z^\prime$	\\
415.47	&	4.808631	& $	20.604	\pm	0.041	$ &	1838	&	0.79	&	$z^\prime$	\\
501.08	&	5.799551	& $	20.713	\pm	0.038	$ &	1838	&	0.72	&	$z^\prime$	\\
588.10	&	6.806728	& $	20.941	\pm	0.054	$ &	1838	&	0.87	&	$z^\prime$	\\
669.18	&	7.745094	& $	21.174	\pm	0.077	$ &	919	&	1.04	&	$z^\prime$	\\
843.66	&	9.764627	& $	21.569	\pm	0.064	$ &	1379	&	0.61	&	$z^\prime$	\\
1101.93	&	12.75382	& $	22.028	\pm	0.085	$ &	2420	&	0.82	&	$z^\prime$	\\
1880.55	&	21.76561	& $	22.298	\pm	0.088	$ &	2952	&	1.10	&	$z^\prime$	\\
2401.32	&	27.79309	& $	22.360	\pm	0.138	$ &	4502	&	1.10	&	$z^\prime$	\\
3090.97	&	35.77507	& $	22.459	\pm	0.118	$ &	3630	&	1.12	&	$z^\prime$	\\
4258.44	&	49.28755	& $	23.000	\pm	0.319	$ &	5384	&	1.33	&	$z^\prime$	\\
4732.20	&	54.77078	& $	23.147	\pm	0.214	$ &	5422	&	1.13	&	$z^\prime$	\\
6241.88	&	72.24398	& $	23.470	\pm	0.443	$ &	2758	&	1.06	&	$z^\prime$	\\
24277.46	&	280.9891	& $	24.022	\pm	0.279	$ &	3752	&	0.90	&	$z^\prime$	\\
\hline													
64.52	&	0.746718	& $	19.313	\pm	0.128	$ &	240	&	1.38	&	$J$	\\
64.96	&	0.751794	& $	19.411	\pm	0.128	$ &	240	&	1.49	&	$J$	\\
65.40	&	0.756904	& $	19.323	\pm	0.145	$ &	240	&	1.29	&	$J$	\\
65.84	&	0.762041	& $	19.455	\pm	0.166	$ &	240	&	1.36	&	$J$	\\
66.28	&	0.767086	& $	19.332	\pm	0.167	$ &	240	&	1.32	&	$J$	\\
66.71	&	0.772126	& $	19.314	\pm	0.147	$ &	240	&	1.38	&	$J$	\\
67.15	&	0.777215	& $	19.349	\pm	0.147	$ &	240	&	1.36	&	$J$	\\
67.61	&	0.782483	& $	19.327	\pm	0.110	$ &	240	&	1.24	&	$J$	\\
68.04	&	0.787512	& $	19.313	\pm	0.098	$ &	240	&	1.14	&	$J$	\\
68.48	&	0.792610	& $	19.339	\pm	0.130	$ &	240	&	1.23	&	$J$	\\
69.22	&	0.801194	& $	19.312	\pm	0.137	$ &	480	&	1.31	&	$J$	\\
70.02	&	0.810466	& $	19.324	\pm	0.155	$ &	480	&	1.36	&	$J$	\\
70.82	&	0.819633	& $	19.106	\pm	0.159	$ &	480	&	1.33	&	$J$	\\
71.62	&	0.828972	& $	19.239	\pm	0.172	$ &	480	&	1.08	&	$J$	\\
72.41	&	0.838035	& $	19.353	\pm	0.163	$ &	480	&	1.13	&	$J$	\\
73.19	&	0.847117	& $	19.198	\pm	0.149	$ &	480	&	1.17	&	$J$	\\
73.99	&	0.856314	& $	19.226	\pm	0.099	$ &	480	&	1.18	&	$J$	\\
74.79	&	0.865581	& $	19.121	\pm	0.150	$ &	480	&	1.09	&	$J$	\\
75.58	&	0.874812	& $	19.140	\pm	0.115	$ &	480	&	1.21	&	$J$	\\
76.38	&	0.883978	& $	18.998	\pm	0.136	$ &	480	&	1.14	&	$J$	\\
77.17	&	0.893158	& $	18.910	\pm	0.121	$ &	480	&	1.09	&	$J$	\\
77.98	&	0.902545	& $	19.022	\pm	0.121	$ &	480	&	1.17	&	$J$	\\
78.78	&	0.911825	& $	18.911	\pm	0.106	$ &	480	&	1.26	&	$J$	\\
79.58	&	0.921103	& $	18.910	\pm	0.116	$ &	480	&	1.23	&	$J$	\\
80.38	&	0.930378	& $	18.890	\pm	0.102	$ &	480	&	1.20	&	$J$	\\
81.19	&	0.939736	& $	18.782	\pm	0.115	$ &	480	&	1.20	&	$J$	\\
81.97	&	0.948770	& $	18.719	\pm	0.097	$ &	480	&	1.63	&	$J$	\\
151.52	&	1.753657	& $	18.722	\pm	0.124	$ &	480	&	1.17	&	$J$	\\
155.94	&	1.804812	& $	18.786	\pm	0.077	$ &	480	&	1.18	&	$J$	\\
160.36	&	1.855967	& $	18.746	\pm	0.105	$ &	480	&	1.36	&	$J$	\\
164.73	&	1.906554	& $	18.749	\pm	0.107	$ &	480	&	1.42	&	$J$	\\
239.84	&	2.775900	& $	19.657	\pm	0.133	$ &	960	&	1.51	&	$J$	\\
250.97	&	2.904789	& $	19.647	\pm	0.111	$ &	960	&	1.25	&	$J$	\\
329.20	&	3.810145	& $	19.875	\pm	0.077	$ &	1920	&	1.06	&	$J$	\\
415.49	&	4.808939	& $	20.173	\pm	0.112	$ &	1920	&	1.25	&	$J$	\\
501.11	&	5.799862	& $	20.254	\pm	0.115	$ &	1920	&	1.19	&	$J$	\\
588.13	&	6.807036	& $	20.625	\pm	0.157	$ &	1920	&	1.47	&	$J$	\\
669.20	&	7.745412	& $	20.698	\pm	0.229	$ &	960	&	1.61	&	$J$	\\
843.69	&	9.764943	& $	21.226	\pm	0.212	$ &	1440	&	1.27	&	$J$	\\
1101.95	&	12.75411	& $	21.833	\pm	0.237	$ &	2160	&	1.28	&	$J$	\\
1880.58	&	21.76593	& $	21.792	\pm	0.237	$ &	2400	&	1.36	&	$J$	\\
2401.32	&	27.79309	& $	22.154	\pm	0.321	$ &	3600	&	1.34	&	$J$	\\
3090.99	&	35.77538	& $ >	22.25			$ &	3240	&	1.33	&	$J$	\\
4258.47	&	49.28783	& $ >	21.54			$ &	4560	&	1.33	&	$J$	\\
4732.22	&	54.77109	& $ >	22.06			$ &	4560	&	1.44	&	$J$	\\
6241.91	&	72.24430	& $ >	21.52			$ &	2880	&	1.35	&	$J$	\\
24277.49	&	280.9894	& $ >	22.39			$ &	3600	&	1.25	&	$J$	\\
\hline													
64.52	&	0.746718	& $	18.986	\pm	0.154	$ &	240	&	1.46	&	$H$	\\
64.96	&	0.751794	& $	19.111	\pm	0.164	$ &	240	&	1.33	&	$H$	\\
65.40	&	0.756904	& $	18.995	\pm	0.165	$ &	240	&	1.32	&	$H$	\\
65.84	&	0.762041	& $	19.099	\pm	0.159	$ &	240	&	1.41	&	$H$	\\
66.28	&	0.767086	& $	18.993	\pm	0.166	$ &	240	&	1.42	&	$H$	\\
66.71	&	0.772126	& $	19.135	\pm	0.169	$ &	240	&	1.51	&	$H$	\\
67.15	&	0.777215	& $	18.966	\pm	0.188	$ &	240	&	1.41	&	$H$	\\
67.61	&	0.782483	& $	19.165	\pm	0.160	$ &	240	&	1.24	&	$H$	\\
68.04	&	0.787512	& $	18.994	\pm	0.167	$ &	240	&	1.18	&	$H$	\\
68.48	&	0.792610	& $	18.993	\pm	0.149	$ &	240	&	1.21	&	$H$	\\
69.22	&	0.801194	& $	19.018	\pm	0.132	$ &	480	&	1.38	&	$H$	\\
70.02	&	0.810466	& $	18.985	\pm	0.143	$ &	480	&	1.34	&	$H$	\\
70.82	&	0.819633	& $	18.973	\pm	0.153	$ &	480	&	1.32	&	$H$	\\
71.62	&	0.828972	& $	18.944	\pm	0.130	$ &	480	&	1.12	&	$H$	\\
72.41	&	0.838035	& $	18.887	\pm	0.112	$ &	480	&	1.11	&	$H$	\\
73.19	&	0.847117	& $	18.857	\pm	0.115	$ &	480	&	1.14	&	$H$	\\
73.99	&	0.856314	& $	18.915	\pm	0.108	$ &	480	&	1.14	&	$H$	\\
74.79	&	0.865581	& $	18.919	\pm	0.118	$ &	480	&	1.10	&	$H$	\\
75.58	&	0.874812	& $	18.844	\pm	0.115	$ &	480	&	1.16	&	$H$	\\
76.38	&	0.883978	& $	18.777	\pm	0.106	$ &	480	&	1.15	&	$H$	\\
77.17	&	0.893158	& $	18.704	\pm	0.112	$ &	480	&	1.14	&	$H$	\\
77.98	&	0.902545	& $	18.590	\pm	0.109	$ &	480	&	1.14	&	$H$	\\
78.78	&	0.911825	& $	18.574	\pm	0.106	$ &	480	&	1.23	&	$H$	\\
79.58	&	0.921103	& $	18.571	\pm	0.103	$ &	480	&	1.19	&	$H$	\\
80.38	&	0.930378	& $	18.441	\pm	0.086	$ &	480	&	1.21	&	$H$	\\
81.19	&	0.939736	& $	18.498	\pm	0.098	$ &	480	&	1.15	&	$H$	\\
81.97	&	0.948770	& $	18.374	\pm	0.101	$ &	480	&	1.55	&	$H$	\\
151.52	&	1.753657	& $	18.309	\pm	0.117	$ &	480	&	1.17	&	$H$	\\
155.94	&	1.804812	& $	18.313	\pm	0.114	$ &	480	&	1.16	&	$H$	\\
160.36	&	1.855967	& $	18.351	\pm	0.104	$ &	480	&	1.33	&	$H$	\\
164.73	&	1.906554	& $	18.404	\pm	0.121	$ &	480	&	1.28	&	$H$	\\
239.84	&	2.775900	& $	19.010	\pm	0.137	$ &	960	&	1.52	&	$H$	\\
250.97	&	2.904789	& $	19.106	\pm	0.121	$ &	960	&	1.18	&	$H$	\\
329.20	&	3.810145	& $	19.390	\pm	0.120	$ &	1920	&	1.08	&	$H$	\\
415.49	&	4.808939	& $	19.881	\pm	0.159	$ &	1920	&	1.27	&	$H$	\\
501.11	&	5.799862	& $	19.987	\pm	0.150	$ &	1920	&	1.14	&	$H$	\\
588.13	&	6.807036	& $	20.247	\pm	0.193	$ &	1920	&	1.34	&	$H$	\\
669.20	&	7.745412	& $	20.365	\pm	0.291	$ &	960	&	1.81	&	$H$	\\
843.69	&	9.764943	& $	20.711	\pm	0.405	$ &	1440	&	1.34	&	$H$	\\
1101.95	&	12.75411	& $	20.817	\pm	0.250	$ &	2160	&	1.30	&	$H$	\\
1880.58	&	21.76593	& $	21.762	\pm	0.274	$ &	2400	&	1.27	&	$H$	\\
2401.32	&	27.79309	& $	21.856	\pm	0.358	$ &	3600	&	1.33	&	$H$	\\
3090.99	&	35.77538	& $ >	21.85			$ &	3240	&	1.27	&	$H$	\\
4258.47	&	49.28783	& $ >	21.05			$ &	4560	&	1.34	&	$H$	\\
4732.22	&	54.77109	& $ >	21.62			$ &	4560	&	1.44	&	$H$	\\
6241.91	&	72.24430	& $ >	20.91			$ &	2880	&	1.39	&	$H$	\\
24277.49	&	280.9894	& $ >	21.84			$ &	3600	&	1.34	&	$H$	\\
\hline													
64.52	&	0.746718	& $	18.920	\pm	0.245	$ &	240	&	1.43	&	$K_S$	\\
64.96	&	0.751794	& $	18.547	\pm	0.179	$ &	240	&	1.33	&	$K_S$	\\
65.40	&	0.756904	& $	18.757	\pm	0.255	$ &	240	&	1.34	&	$K_S$	\\
65.84	&	0.762041	& $	18.714	\pm	0.199	$ &	240	&	1.29	&	$K_S$	\\
66.28	&	0.767086	& $	18.896	\pm	0.242	$ &	240	&	1.31	&	$K_S$	\\
66.71	&	0.772126	& $	18.702	\pm	0.203	$ &	240	&	1.32	&	$K_S$	\\
67.15	&	0.777215	& $	18.861	\pm	0.275	$ &	240	&	1.30	&	$K_S$	\\
67.61	&	0.782483	& $	18.643	\pm	0.214	$ &	240	&	1.29	&	$K_S$	\\
68.04	&	0.787512	& $	18.635	\pm	0.217	$ &	240	&	1.25	&	$K_S$	\\
68.48	&	0.792610	& $	18.564	\pm	0.196	$ &	240	&	1.32	&	$K_S$	\\
69.22	&	0.801194	& $	18.680	\pm	0.219	$ &	480	&	1.36	&	$K_S$	\\
70.02	&	0.810466	& $	18.722	\pm	0.206	$ &	480	&	1.34	&	$K_S$	\\
70.82	&	0.819633	& $	18.702	\pm	0.202	$ &	480	&	1.39	&	$K_S$	\\
71.62	&	0.828972	& $	18.500	\pm	0.175	$ &	480	&	1.25	&	$K_S$	\\
72.41	&	0.838035	& $	18.454	\pm	0.171	$ &	480	&	1.33	&	$K_S$	\\
73.19	&	0.847117	& $	18.525	\pm	0.177	$ &	480	&	1.27	&	$K_S$	\\
73.99	&	0.856314	& $	18.515	\pm	0.169	$ &	480	&	1.36	&	$K_S$	\\
74.79	&	0.865581	& $	18.475	\pm	0.170	$ &	480	&	1.30	&	$K_S$	\\
75.58	&	0.874812	& $	18.476	\pm	0.188	$ &	480	&	1.42	&	$K_S$	\\
76.38	&	0.883978	& $	18.431	\pm	0.173	$ &	480	&	1.28	&	$K_S$	\\
77.17	&	0.893158	& $	18.247	\pm	0.191	$ &	480	&	1.30	&	$K_S$	\\
77.98	&	0.902545	& $	18.470	\pm	0.182	$ &	480	&	1.26	&	$K_S$	\\
78.78	&	0.911825	& $	18.261	\pm	0.187	$ &	480	&	1.43	&	$K_S$	\\
79.58	&	0.921103	& $	18.054	\pm	0.187	$ &	480	&	1.39	&	$K_S$	\\
80.38	&	0.930378	& $	17.999	\pm	0.154	$ &	480	&	1.42	&	$K_S$	\\
81.19	&	0.939736	& $	17.965	\pm	0.171	$ &	480	&	1.43	&	$K_S$	\\
81.97	&	0.948770	& $	17.922	\pm	0.211	$ &	480	&	1.62	&	$K_S$	\\
151.52	&	1.753657	& $	17.843	\pm	0.147	$ &	480	&	1.32	&	$K_S$	\\
155.94	&	1.804812	& $	17.866	\pm	0.146	$ &	480	&	1.35	&	$K_S$	\\
160.36	&	1.855967	& $	18.014	\pm	0.164	$ &	480	&	1.45	&	$K_S$	\\
164.73	&	1.906554	& $	18.095	\pm	0.176	$ &	480	&	1.47	&	$K_S$	\\
239.84	&	2.775900	& $	18.713	\pm	0.173	$ &	960	&	1.35	&	$K_S$	\\
250.97	&	2.904789	& $	18.985	\pm	0.208	$ &	960	&	1.31	&	$K_S$	\\
329.20	&	3.810145	& $	19.102	\pm	0.179	$ &	1920	&	1.27	&	$K_S$	\\
415.49	&	4.808939	& $	19.650	\pm	0.259	$ &	1920	&	1.33	&	$K_S$	\\
501.11	&	5.799862	& $	19.937	\pm	0.320	$ &	1920	&	1.30	&	$K_S$	\\
588.13	&	6.807036	& $	19.667	\pm	0.272	$ &	1920	&	1.41	&	$K_S$	\\
669.20	&	7.745412	& $	19.860	\pm	0.351	$ &	960	&	1.49	&	$K_S$	\\
843.69	&	9.764943	& $	20.486	\pm	0.458	$ &	1440	&	1.22	&	$K_S$	\\
1101.95	&	12.75411	& $	20.573	\pm	0.516	$ &	2160	&	1.33	&	$K_S$	\\
1880.58	&	21.76593	& $	20.698	\pm	0.747	$ &	2400	&	1.47	&	$K_S$	\\
2401.32	&	27.79309	& $ >	20.32			$ &	3600	&	1.50	&	$K_S$	\\
3090.99	&	35.77538	& $ >	20.22			$ &	3240	&	1.50	&	$K_S$	\\
4258.47	&	49.28783	& $ >	19.19			$ &	4560	&	2.12	&	$K_S$	\\
4732.22	&	54.77109	& $ >	20.33			$ &	4560	&	1.37	&	$K_S$	\\
6241.91	&	72.24430	& $ >	20.06			$ &	2880	&	1.50	&	$K_S$	\\
24277.49	&	280.9894	& $ >	20.56			$ &	3600	&	1.31	&	$K_S$	\\
%\hline
%\endfoot
\hline									
\hline									
\end{longtable}
\tablefoot{All data are in AB magnitudes and not corrected for Galactic foreground extinction. Midtimes are derived logarithmically, $t=10^{([log(t_1-t_0)+log(t_2-t_0)]/2)}$, hereby $t_{1.2}$ are the absolute start and stop times, and $t_0$ is the \emph{Swift} trigger time. To obtain Vega magnitudes, it is $uvw2_{AB}-uvw2_{Vega}=1.73$ mag, $uvm2_{AB}-uvm2_{Vega}=1.69$ mag, $uvw1_{AB}-uvw1_{Vega}=1.51$ mag, $u_{AB}-u_{Vega}=1.02$ mag, $b_{AB}-b_{Vega}=-0.13$ mag, $v_{AB}-v_{Vega}=-0.01$ mag, $white_{AB}-white_{Vega}=0.80$ mag for UVOT data (as given at http://swift.gsfc.nasa.gov/analysis/uvot\_digest/zeropts.html), and $g^\prime_{AB}-g^\prime_{Vega}=-0.062$ mag, $r^\prime_{AB}-r^\prime_{Vega}=0.178$ mag, $i^\prime_{AB}-i^\prime_{Vega}=0.410$ mag, $z^\prime_{AB}-z^\prime_{Vega}=0.543$ mag, $J_{AB}-J_{Vega}=0.929$ mag, $H_{AB}-H_{Vega}=1.394$ mag, $K_{S.AB}-K_{S.Vega}=1.859$ mag for GROND data. Corrections for Galactic extinction are, using $E_{(B-V)}=0.017$ \citep{Schlegel1998ApJ} and the Galactic extinction curve of \cite{Cardelli1989ApJ}: $A_{uvw2}=0.134$ mag, $A_{uvm2}=0.150$ mag, $A_{uvw1}=0.119$ mag, $A_u=0.085$ mag, $A_b=0.066$ mag. $A_v=0.053$ mag, $A_{white}=0.084$ mag, $A_{g^\prime}=0.066$ mag, $A_{r^\prime}=0.046$ mag, $A_{i^\prime}=0.034$ mag, $A_{z^\prime}=0.025$ mag, $A_J=0.015$ mag, $A_H=0.010$ mag, $A_{K_S}=0.006$ mag.}\\
}
}

\subsection{GROND}

At the time of the \emph{Swift} trigger, GRB 111209A was just 13 degrees above the horizon as seen from La Silla and setting \citep{Klotz2011GCN1,Nysewander2011GCN}, which is under the pointing limit of the 2.2m telescope, therefore no observations could be obtained % in Rapid Response Mode, and we have no observations 
 during the extended prompt phase of the GRB with GROND. Our observations began the following night at the onset of astronomical twilight % (Sun $12^\circ$ below the horizon)
 and continued for 5.5 hours, until the pointing limit of the telescope was reached. % The observations consisted of ten 4-minute OBs\footnote{Observation blocks, the technical term for a fixed observing sequence.} followed by 17 8-minute OBs.
Analysis of the first observations yielded the detection of the optical/NIR afterglow at RA, Dec. (J2000) = 00:57:22.64, -46:48:03.6 (14.34435, -46.80101) with an uncertainty of 0\farcs3 \citep[][see Fig. \ref{Field}]{Kann2011GCN1}. Further analysis of the $J$-band observations of the first night also revealed the afterglow was actually rising in brightness and not falling \citep{Kann2011GCN2}.

Observations continued in the following nights, in such a way to create a dense follow-up on a logarithmic time scale. A total of 16 further epochs were obtained. Our penultimate observations were obtained on the 18th/19th of February 2012, 72 days after the GRB, thereafter the position moved too close to the Sun and became unobservable. A final epoch was obtained after the GRB position had become visible again, 280 days after the GRB, to assess the host galaxy magnitude.

Afterglow magnitudes in the optical were measured against on-chip standard stars calibrated to the SDSS catalogue \citep{Aihara2011ApJS}, obtained from observing a nearby SDSS field immediately before the afterglow observations on the third day after the GRB, in photometric conditions. % The field is at high Galactic latitude and very empty, so in some cases, an astrometric solution had to be derived versus our standard-star catalogue and not versus SDSS or USNOB1.0 field stars.
NIR magnitudes were measured against on-chip comparison stars taken from the 2MASS catalogue \citep{Skrutskie2006AJ}. Reduction and analysis was performed within a custom pipeline calling upon IRAF tasks \citep{Tody1993ASPC}, following the methods described in detail in \cite{Kruehler2008ApJ} and \cite{Yoldas2008AIPC}. %for the last, deepest epoch, then that result catalogue was used to calibrate all earlier NIR observations. Due to the much larger FOV of the NIR detectors \citep{Greiner2008PASP}, an astrometric solution was always found.

\subsection{UVOT}

We expand our photometric database by adding the UVOT observations from \emph{Swift}. UVOT photometry was carried out on pipeline-processed sky images downloaded from the \emph{Swift} data centre\footnote{www.swift.ac.uk/swift\_portal} following the standard UVOT procedure \citep{Poole2008MNRAS}. Source photometric measurements were extracted from the UVOT early-time event data and later imaging data files using the tool {\sc uvotmaghist} (v1.1) with a circular source extraction region of 3\farcs5 to maximise the signal-to-noise. In order to remain compatible with the effective area calibrations, which are based on $5\arcsec$ aperture photometry \citep{Poole2008MNRAS}, an aperture correction was applied. We note independent reductions of the UVOT data set have been used by \cite{Stratta2013ApJ} and L14. Our UVOT data results agree well with those of L14 where they overlap.

\subsection{X-shooter}

Two X-shooter \citep{Vernet2011AA} spectra are included in L14. We downloaded and independently reduced these spectra, for details see \cite{Kruhler2015AA}. The afterglow spectrum, taken 0.74 days after the GRB, yields a precise redshift of $z=0.67702\pm0.00005$. The late spectrum, taken during the rising phase of the SN at 19.82 days after the GRB, is discussed in detail in G15.

%------------------------------------------------------------------------------------------------------------------------------------------------------------------------------------------------------------------
%------------------------------------------------------------------------------------------------------------------------------------------------------------------------------------------------------------------
%------------------------------------------------------------------------------------------------------------------------------------------------------------------------------------------------------------------
%------------------------------------------------------------------------------------------------------------------------------------------------------------------------------------------------------------------

\section{Results}
\label{SectRes}

\begin{figure*}[!t]
  \centering
 \includegraphics[width=0.8\textwidth]{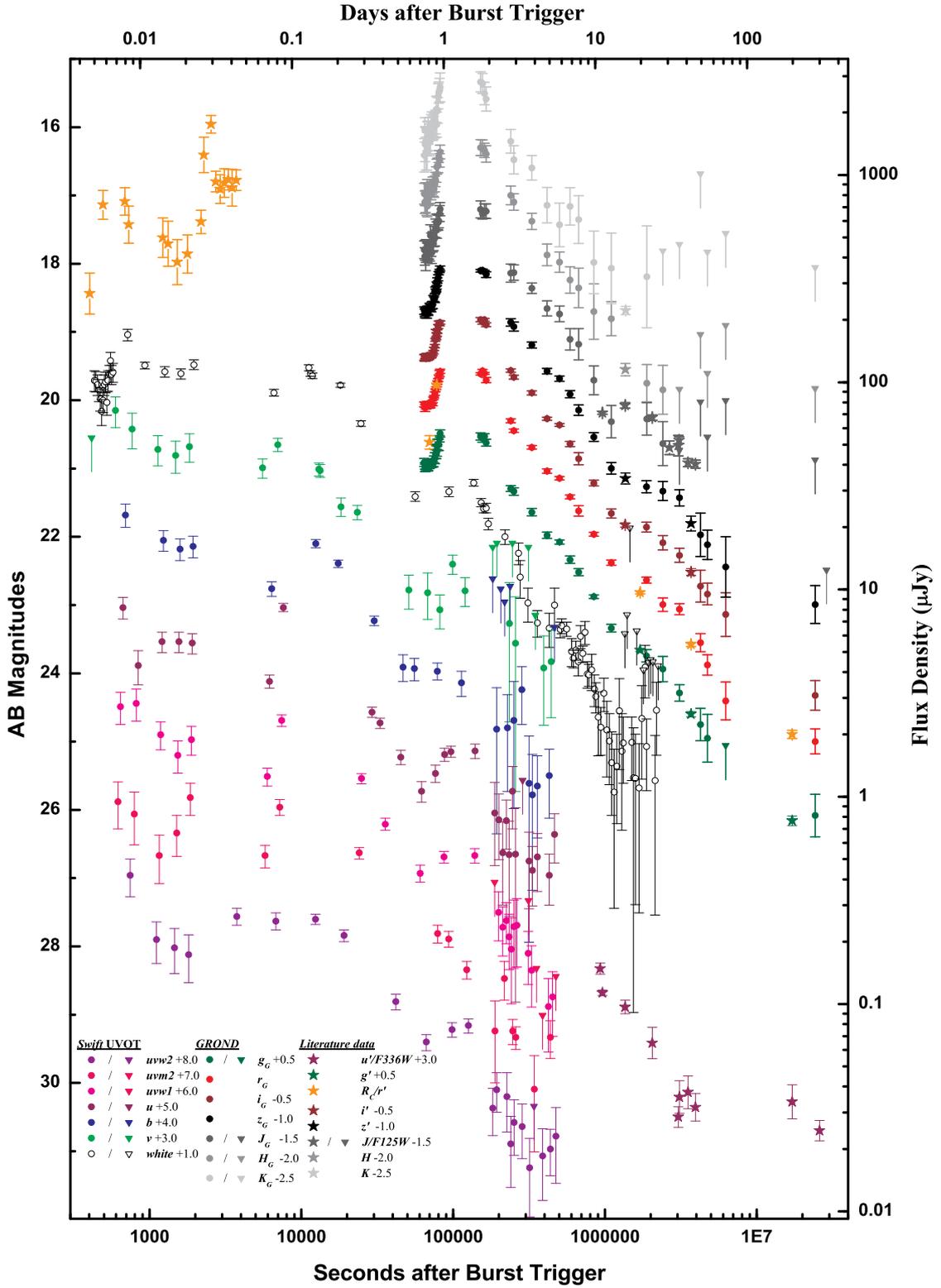}
   \caption{The light curves of the optical counterpart of GRB 111209A as observed by GROND and UVOT (this work), with additional $R_C$ data from \cite{Nysewander2011GCN} and \cite{Stratta2013ApJ}, and $F336Wu^\prime g^\prime r^\prime R_Ci^\prime z^\prime F125WJHK$ data from L14. Downward-pointing triangles are upper limits. The different bands are colour-coded and given in the legend, they are offset for clarity by the amount given. The \emph{G} subscript denotes GROND data. Data in this plot are corrected for the (small) Galactic foreground extinction.}
              \label{Lightcurve}
    \end{figure*}

The GROND and UVOT data of the afterglow of GRB 111209A are given in Table \ref{obslog} (some of the late data were already published in G15) and plotted in their entirety (with additional data from the literature) in Fig. \ref{Lightcurve}. In the $r^\prime i^\prime z^\prime$ bands, the afterglow is detected in all 46 epochs, in the $g^\prime$ band, it is undetected in the last epoch before the host observation. In the NIR, the afterglow is detected in $JH$ until 2.4 Ms/27.79 days (not in the last five epochs), and in the $K_S$ band, it is detected until 1.9 Ms/21.77 days (not in the last six epochs). Multi-colour observations and detections by UVOT extend to $\lesssim500$ ks/6 days, and the afterglow/SN is detected until 2.2 Ms/25.28 days in $white$ light. The data in the table are given in AB magnitudes and have not been corrected for Galactic foreground extinction. For the GRB position, the maps of \cite{Schlegel1998ApJ} yield $E_{(B-V)}=0.020$ mag, using the Milky Way extinction curve of \cite{Cardelli1989ApJ} and the extinction coefficients for the GROND filters\footnote{It is $A_{g^\prime}/A_V=1.255$, $A_{r^\prime}/A_V=0.866$, $A_{i^\prime}/A_V=0.648$, $A_{z^\prime}/A_V=0.482$, $A_J/A_V=0.283$, $A_H/A_V=0.181$, $A_{K_S}/A_V=0.116$.}, we derive small extinction corrections as given in the footnote of Table \ref{obslog}, in other words, the influence of foreground extinction is almost negligible. 

The afterglow light curve (Fig. \ref{Lightcurve}) reveals multiple distinct phases in its evolution. First and foremost, there is strong variability contemporaneous to the long-lasting prompt emission, before the afterglow transits to a constant decay phase, seen in the UVOT data (see also \citealt{Stratta2013ApJ}). Then, during the GROND observations spanning the first night, the afterglow turns from a plateau (more specifically a shallow rise) phase to a very steep rise \citep{Kann2011GCN2,Gendre2013ApJ,Stratta2013ApJ}. A day later, the GROND magnitudes of the afterglow are similar to those at the end of the GROND observations during the first night, they are then seen to decay steadily, implying a turnover must have occurred in between, which the UVOT data confirms. With the exception of two small ``steps'', the afterglow decays smoothly and without exhibiting any visible breaks, until at least 13 days. The afterglow departs from this decay in observations starting three weeks after the trigger, revealing a chromatic bump until the end of our observations. In the $white$ data as well as the $F336W/u^\prime$ data from L14, the afterglow is seen to decay more rapidly.

\subsection{Detailed light-curve fitting of the afterglow}
\label{chap_fitting}
\subsubsection{The prompt emission}
\label{fitting_prompt}
The optical and X-ray emission (which shows evidence for a minor contribution from a blackbody, \citealt{Gendre2013ApJ}, L14) during the prompt phase has been discussed extensively by \cite{Stratta2013ApJ}, therefore we shall just remark on a few results. From 0.008 to 0.025 days (the first \emph{Swift} orbit as well as some TAROT data from \citealt{Stratta2013ApJ}), all data is fit by an achromatic broken power-law:

\begin{equation}
m(t)=-2.5\log\left(10^{-0.4m_k}\left(\left(\frac{t}{t_b}\right)^{\alpha_1n}+\left(\frac{t}{t_b}\right)^{\alpha_2n}\right)^{-1/n}\right).
\end{equation}

Here, $m_k$ is the normalization magnitude, $t$ the time after trigger, $\alpha_{1,2}$ the power-law slopes of the first and second component, $t_b$ the break time, and $n$ the smoothness of the break. We ignore any host galaxy contribution since it lies far below the magnitudes we are dealing with here (Sect. \ref{SectHost}). We find $\alpha_1=1.40\pm0.29$, $\alpha_2=-0.21\pm0.16$, $t_b=1020\pm89$ s, $n=-10$ fixed ($\chi^2=20.1$ for 31 degrees of freedom). This emission is quite red, with a spectral slope $\beta=1.43\pm0.12$, in perfect agreement with \cite{Stratta2013ApJ}, who find $\beta=1.43\pm0.20$.

Our SED exhibits slight curvature, and the addition of NIR data from REM \citep{Fugazza2011GCN} allows an acceptable fit ($\chi^2=5.84$ for 5 degrees of freedom) with a SMC dust model yielding an intrinsic spectral slope $\beta_0=0.63\pm0.24$ and rest-frame extinction $\AV=0.25\pm0.11$ mag. \cite{Stratta2013ApJ} note that the SED is redder than at later times (and also redder than typical pure GRB afterglows) and invoke possible dust destruction. We caution that while we find higher extinction at early times, the increase is only at the $1.2\sigma$ level (see Sect. \ref{SectBeta}), and it is also not clear if the emission truly is an underlying power-law. Furthermore, as the GRB had already been on-going for several ks \citep{Golenetskii2011GCN}, it would be peculiar to still detect further dust destruction and de-reddening, this process is expected to happen on much shorter timescales \citep{Waxman2000ApJ,Starling2008AA,Morgan2014MNRAS}. Also note that the temporal slopes we derive are arbitrary values dependent on the trigger time of \emph{Swift}. If $t_0$ is set significantly earlier, to the beginning of the Konus-\emph{WIND} detection, the slopes become significantly steeper.

The afterglow is further observed by TAROT \citep{Stratta2013ApJ}, who find a strong flare {(see \citealt{Lin2018ApJ} for an interpretation of the X-ray-to-optical delay of this flare)}. The UVOT data from the second \emph{Swift} orbit can be described by the transition from a decay to another steep flare assuming the colours are identical to those we find from the first orbit, a similar result is also derived by \cite{Stratta2013ApJ}. The following orbit may also show a decay followed by a rise, \cite{Stratta2013ApJ} interpret this behaviour as the onset of the forward-shock afterglow. At this time, the X-ray emission begins a steeper decay, but it does not yet transition to the very steep decay dominated by high-latitude emission which marks the end of the prompt emission.

\subsubsection{The early afterglow}
\label{fitting_AG1}

The UVOT data from 0.15 to 0.75 days can be fit by an achromatic power-law decay. We find $\alpha=1.23\pm0.05$  ($\chi^2=13.57$ for 16 degrees of freedom), and a six-colour SED with a small amount of scatter and no evidence for curvature, adequately  ($\chi^2=4.64$ for 4 degrees of freedom) fit by a simple power-law with slope $\beta=0.96\pm0.14$. Note that these values differ quite strongly from those presented in \cite{Stratta2013ApJ}, who find $\alpha=1.6\pm0.1$, $\beta=1.33\pm0.01$. Our results indicate a transition to a bluer spectrum compared to the first orbits, and the smooth decay is indicative of this being a forward-shock afterglow. During this phase, the X-ray afterglow behaves differently, transitioning to a steep decay which is seen both by \emph{Swift}/XRT\footnote{http://www.swift.ac.uk/xrt\_live\_cat/00509336/} ($\alpha_{X1,XRT}=6.45^{+0.13}_{-0.15}$) and \emph{XMM-Newton} \citep[$\alpha_{X1,XMM}=2.23\pm0.10$,][]{Stratta2013ApJ}. This differing behaviour may indicate that the optical emission associated with the prompt GRB emission is significantly fainter than the afterglow, whereas the tail of the prompt emission still dominates the X-ray regime.

\subsubsection{The first day}
\label{fitting_1}
GROND observations begin 0.75 days after the trigger, and we find that the first night of data is described by a (slightly rising) plateau phase followed by a steep rise, first reported by \cite{Kann2011GCN2}. In this case, the values of $\alpha_{1,2}$ will be negative. Furthermore, since the broken power-law is a convex function in this case, the smoothness $n$ must be negative as well, we are not able to let it be a free parameter \citep[see, e.g.,][on $n$ as a free parameter in light curve fits or not]{Zeh2006ApJ}, and we fix it to a hard value, $n=-1000$, as we find no evidence for a smoother rollover and a hard $n$ minimizes the errors of the temporal slopes. Usually, $n=(-)10$ is already considered a hard transition \citep{Zeh2006ApJ}, but in this case, we are working with {very short baselines in $\log t$ space, i.e., the data points do not extend far enough before and after the break. Using $n=-10$ would result in power-law indices that are not yet very close to their asymptotic values, leading to much larger variations and errors, this is mitigated by using $n=-1000$, an abrupt transition between the two slopes}. The addition of UVOT data during this phase does not lead to any improvement of the results, as the data is sparse and suffers from scatter.

\begin{figure}
  \centering
 \includegraphics[width=\columnwidth]{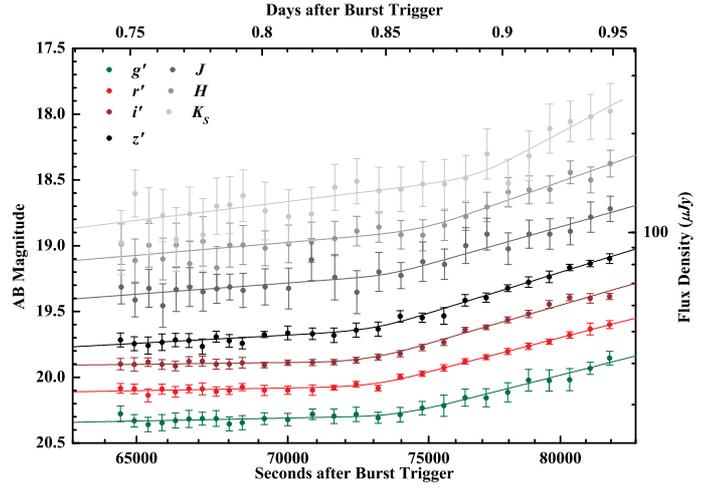}
   \caption{Fit to the light curves of the afterglow of GRB 111209A during the first night of observations. The slopes are free, and different, parameters for all bands. See Table \ref{tab_fit1} for values.}
              \label{Lightcurve_rise}
    \end{figure}

\begin{table}[t]
\caption{Slopes of the plateau/rebrightening phase of the optical/NIR afterglow of GRB 111209A (see Fig. \ref{Lightcurve_rise}). The data spans from 64 ks -- 82 ks. The degrees of freedom for each {single-filter} fit are 23. {For the joint fit, the degrees of freedom are 179.}}
\label{tab_fit1}    
\centering                        
\begin{tabular}{l c c l c}       
\hline\hline   
Filter	&	$\alpha_1$			&	$\alpha_2$			&	break time (ks)			&	$\chi^2$	\\\hline   
$g^\prime$	& $	-0.29	\pm	0.28	$ & $	-3.65	\pm	0.56	$ & $	74.0	\pm	1.0	$ & $	4.16	$ \\
$r^\prime$	& $	-0.26	\pm	0.19	$ & $	-3.73	\pm	0.20	$ & $	73.0	\pm	0.4	$ & $	5.96	$ \\
$i^\prime$	& $	-0.17	\pm	0.20	$ & $	-4.16	\pm	0.19	$ & $	72.7	\pm	0.3	$ & $	12.05	$ \\
$z^\prime$	& $	-0.78	\pm	0.35	$ & $	-4.43	\pm	0.30	$ & $	73.0	\pm	0.7	$ & $	6.26	$ \\
$J$	& $	-1.07	\pm	0.89	$ & $	-4.06	\pm	0.94	$ & $	73.5	\pm	2.5	$ & $	4.39	$ \\
$H$	& $	-1.22	\pm	0.69	$ & $	-5.02	\pm	1.21	$ & $	74.7	\pm	1.9	$ & $	3.89	$ \\
$K_S$	& $	-1.78	\pm	0.77	$ & $	-7.59	\pm	3.28	$ & $	76.7	\pm	2.1	$ & $	5.80	$ \\
All	& $	-0.34	\pm	0.12	$ & $	-4.07	\pm	0.12	$ & $	72.98	\pm	0.23	$ & $	94.54	$ \\
\hline \hline
\end{tabular}
\end{table}

{We performed two different fits. In the first case, we performed a joint fit with all seven bands, leaving the slopes and the break time free to vary, but using only one shared value between all seven bands. This fit yields a good result as given in Table \ref{tab_fit1}. Motivated by our SED results (Sect. \ref{SectBeta}), we also performed broken power-law fits for each band separately, these results are also given in Table \ref{tab_fit1}, and the fit is shown in Fig. \ref{Lightcurve_rise}. The fit finds different values for each of the bands, a detailed look at the differences between filters reveals a span between essentially no differences and a $2\sigma$ offsets. This only weak evidence for chromatic evolution is in agreement with the good fit derived by using shared parameters.}
%On the other hand, the values for the rising slopes, especially the steep slope after the break, show a mostly linear progression toward larger values The fit is shown in Fig. \ref{Lightcurve_rise} and the results are given in Table \ref{tab_fit1}. From these results, we can already see that a spectral evolution must be occurring, as the slopes during the rise are not similar within errors in all bands (which is why we did not perform a joint fit with all slopes left free to vary but fixed between bands, which would be standard procedure in case of an achromatic evolution).
The initial evolution is close to flat in $g^\prime r^\prime i^\prime $, but becomes steeper for redder bands, and this is true as well for the very steep rise starting at $\approx0.86$ days. Indeed, the slope difference $\Delta \alpha$ remains constant within errors over all bands, though we caution the errors especially in the $K_S$ band are very large. The break time $t_b$ is constant in all bands within $2\sigma$. {While the evidence for this chromatic change is not strong from these fits, the SED results (Sect. \ref{SectBeta}) show stronger evidence ($3.4\sigma$) of an intrinsic slope change over this rebrightening phase, indicating that the per-band fit is physically preferred.}

\subsubsection{The transition from steep rise to standard decay}
\label{fitting_2}

\begin{figure}
  \centering
 \includegraphics[width=\columnwidth]{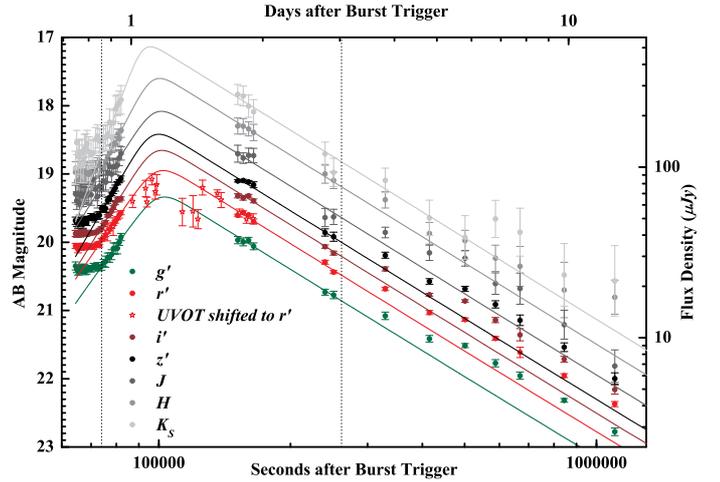}
   \caption{Fit to the light curves of the afterglow of GRB 111209A from the beginning of the rebrightening at $\approx74$ ks to before the ``step'' at about three days (between the vertical dotted lines). This fit uses GROND data only and is the simplest possible transition between the rising and decaying regimes. The stars represent UVOT data (multiple colours) shifted to the $r^\prime$ band, this shows that the evolution over the peak is possibly more complicated than our simple model. The decay slope after the peak is a free but shared parameter for all seven bands, whereas the index of the rise is different for each band. See Table \ref{tab_fit2} for values.}
              \label{Lightcurve_peak}
    \end{figure}

\begin{table}[t]
\caption{Slopes of the rebrightening and first decay phase of the optical/NIR afterglow of GRB 111209A, as well as the peak time and magnitudes. It is $\alpha_3=1.59\pm0.03$ for all bands, and $\chi^2=70$ for 90 degrees of freedom.}
\label{tab_fit2}    
\centering                        
\begin{tabular}{l c r c}       
\hline\hline   \vspace{1mm}
Filter	&	$\alpha_2$			&	$t_{peak}$	(ks)				&	Peak magnitude (AB)	\\
\hline   
\vspace{1mm}
$g^\prime$	& $	-3.62	\pm	0.48	$ & $	102.7	^{+	3.4	}_{-	3.2	}$ & $	19.33	\pm	0.04	$  \\\vspace{1mm}
$r^\prime$	& $	-3.74	\pm	0.18	$  & $	102.1	^{+	1.4	}_{-	1.3	}$ & $	18.94	\pm	0.02	$  \\\vspace{1mm}
$i^\prime$	& $	-4.17	\pm	0.19	$  & $	101.2	\pm1.2 $ & $	18.64	\pm	0.02	$  \\\vspace{1mm}
$z^\prime$	& $	-4.45	\pm	0.28	$  & $	99.6	^{+	1.5	}_{-	1.4	}$ & $	18.42	\pm	0.02	$  \\\vspace{1mm}
$J$	& $	-4.08	\pm	0.79	$ &  $	101.1	^{+	4.8	}_{-	4.5	}$ & $	18.08	\pm	0.07	$  \\\vspace{1mm}
$H$	& $	-5.01	\pm	0.90	$ &  $	100.1	^{+	4.3	}_{-	4.0	}$ & $	17.59	\pm	0.06	$  \\\vspace{1mm}
$K_S$	& $	-6.98	\pm	2.16	$  & $	95.2	^{+	5.9	}_{-	5.1	}$ & $	17.13	\pm	0.09	$  \\
\hline \hline
\end{tabular}
\end{table}

The following night, the afterglow (as measured by GROND)  was at a similar magnitude as at the end of the first night's observations, but had started to decay, which became clear over the following nights. Therefore, a transition must have occurred in between for which we have no GROND data (but UVOT data, see below). The simplest solution is that the afterglow rises to a peak and rolls over into the later decay phase, implying only a single break. Such a behaviour is well-known from early afterglow observations as the ``rising of the forward shock'' \citep[e.g.,][]{Vestrand2006Nature,Molinari2007AA,Kruehler2008ApJ,Ferrero2009AA,Oates2009MNRAS}, but is rarely seen at such late times after the trigger. We find that the decay following the possible peak is achromatic (though there is some scatter), therefore we declare $\alpha_3$ as a shared but otherwise free parameter for all bands. The slope of the steep rise as well as the break time are left free and individual for all bands. We fix the smoothness to $n=5$, as such transitions are usually not sharp, but note that since we only fit data which lies rather far from the peak, the actual result is independent of the chosen smoothness. Following \cite{Molinari2007AA}, the break time is only identical to the peak time for a symmetric rise and decay, which we clearly do not see here, therefore the actual peak times are determined by (in our case, we label the rise $\alpha_2$ and the decay $\alpha_3$):

\begin{equation}
t_{peak}=t_b\left((-\alpha_2/\alpha_3)^{1/[n(\alpha_3-\alpha_2)]}\right).
\end{equation}

The fit is shown in Fig. \ref{Lightcurve_peak} and the results are given in Table \ref{tab_fit2}. The $\alpha_2$ values we find from this fit are in excellent agreement with those given in Table \ref{tab_fit1}. The peak times are identical within errors, the only slight outlier is the $K_S$ band, but here again the errors are larger. We only fit the data until $\approx3$ days after the GRB as afterwards, there is a clear ``step'', the first of two (see below).

This phase between the two GROND epochs is covered by UVOT, though. Again, the data shows large errors, and scatter, but shifting the higher S/N points to the $r^\prime$ band reveals that our simple model may not represent reality. A cluster of data points lies beneath our model fit (see Fig. \ref{Lightcurve_peak}), indicating that there may be an earlier turnover followed by another strong rebrightening that has ended by the time GROND starts observing in the second night. The sparsity of the data does not allow further conclusions, though.

This period is mostly covered by high S/N {XMM}-Newton X-ray observations, as has been mentioned \citep{Gendre2013ApJ,Stratta2013ApJ}. These data show no rebrightening as seen in the optical, instead, after the end of the steep decay (Sect. \ref{fitting_AG1}), the X-rays show a plateau ($\alpha_{X2,XMM}=0.18\pm0.05$), which breaks into a regular decay ($\alpha_{X3,XMM}=1.52\pm0.06$, note this value agrees excellently with our late-time slope in the optical/NIR). A blackbody component as in the early prompt emission is not detected anymore \citep{Gendre2013ApJ}, but \cite{Stratta2013ApJ} report an additional hard spectral component.

Fig. \ref{Lightcurve_peak} shows that at $\approx4$ days, the magnitude of the afterglow increases, though we do not observe the actual transition. Fitting only epochs 34/35 (at $\approx4-5$ days) yields $\alpha_{3.2}=1.44\pm0.09$ and $\chi^2=1.5$ for 8 degrees of freedom. Thereafter, the afterglow ``steps up'' yet again. In the $r^\prime$ band, each of the steps represents a brightening by $\approx0.2$ mag. Note that our slope values are in agreement with those found by \citet[][$\approx1.5-1.6$]{Stratta2013ApJ}.

\subsubsection{The late afterglow (and the supernova)}
\label{fitting_3}

Data beyond six days show no more deviation from a smooth decay {(i.e., another ``step'' rebrightening)} until the supernova becomes dominant {in the optical/NIR bands}. The fit to the entire optical/NIR afterglow (after the rebrightening; taking the ``steps'' into account), the supernova component as well as the host galaxy (Sect. \ref{SectHost}) was first given in G15, and is discussed in detail in K18A. {To summarize, we find from the ultraviolet data, to which the supernova contributes negligibly,} that the afterglow in this segment reveals a light-curve break, and we find $\alpha_{late,1}=1.55\pm0.01$, $\alpha_{late,2}=2.33\pm0.16$, $t_b=9.12\pm0.47$ days; $n=10$ has been fixed. We have also performed a $u^\prime$-band-only fit, and recover the broken power-law fit, $\alpha_{late,u,1}=1.61\pm0.18$, $\alpha_{late,u,2}=2.30\pm0.27$, $t_b=6.45\pm2.71$ days ($n=10$ fixed). The earlier break time is likely caused by lack of data from 5.4 to 10.8 days, but in the joint fit, the unbroken $g^\prime r^\prime i^\prime z^\prime$ data impl{y} a later break. {An attempt to find this break also in the UVOT \emph{white}-band data does not yield conclusive results. While there seems to be a steepening (Fig. \ref{Lightcurve}), the data at $>10$ days are affected by strong scatter. We note that as the \emph{white} bandpass encompasses the entire CCD response of UVOT, it must also contain SN light, as it extends to the $g^\prime V$ filters, where the SN is clearly detected. Therefore, the late \emph{white} light curve has an unknown SN contribution which we can not take into account, but which will somewhat flatten the light curve again.}

{The steepening of the decay, as shown in Figs. 1 of K18A and G15, is not easy to decipher visually. Two main reasons for this are: For one, the magnitude difference between the afterglow at break time and the host galaxy is not that large, $\approx2$ mag, implying the host contributes strongly to the post-break brightness of the transient (i.e., afterglow plus host). G15 show the light-curve decomposition in the $u$ band, where the steep decay becomes evident. Secondly, the slope change $\Delta\alpha=0.78\pm0.16$ is relatively shallow, as a comparison with the sample presented in \citet[][their Fig. 3]{Zeh2006ApJ} shows (barely in the second-lowest bin). GRB afterglows with comparable decay slopes presented in that work are GRBs 000926 and 020124 (\#6 and \#10 in Fig. 1 of \citealt{Zeh2006ApJ}, respectively), in both cases, the slope change is subtle and not easily visible (both GRB afterglows evince a larger $m_{break}$ to $m_{host}$ contrast, though).}

%------------------------------------------------------------------------------------------------------------------------------------------------------------------------------------------------------------------
%------------------------------------------------------------------------------------------------------------------------------------------------------------------------------------------------------------------

\subsection{The evolution of the spectral slope}
\label{SectBeta}
\onltab{5}{
\longtab{5}{
\begin{longtable}{rcclc}
\caption{\label{betatab} The spectral slope $\beta$ of the optical/NIR afterglow of GRB 111209A over time.}\\
\hline
\hline                
$\Delta$t (s) & $\beta$ & $A_V$ & Data set & $\chi^2_\nu$\\\hline   
%\endfirsthead
$572-2176$	& $	1.176	\pm	0.078	$ & $	\cdots			$ &	UVOT+TAROT+REM &	2.24	\\
$572-2176$	& $	0.634	\pm	0.250	$ & $	0.253	\pm	0.105	$ &	UVOT+TAROT+REM &	1.17	\\
$12428-67823$	& $	0.956	\pm	0.140	$ & $	\cdots			$ &	UVOT &	1.16	\\
64492	& $	0.924	\pm	0.074	$ & $	\cdots			$ &	GROND &	0.26	\\
64929	& $	0.962	\pm	0.066	$ & $	\cdots			$ &	GROND &	0.41	\\
65373	& $	1.015	\pm	0.076	$ & $	\cdots			$ &	GROND &	0.13	\\
65816	& $	0.950	\pm	0.075	$ & $	\cdots			$ &	GROND &	0.17	\\
66252	& $	0.960	\pm	0.075	$ & $	\cdots			$ &	GROND &	0.19	\\
66688	& $	0.950	\pm	0.074	$ & $	\cdots			$ &	GROND &	0.11	\\
67127	& $	0.936	\pm	0.074	$ & $	\cdots			$ &	GROND &	0.23	\\
67583	& $	0.977	\pm	0.069	$ & $	\cdots			$ &	GROND &	0.30	\\
68017	& $	1.016	\pm	0.066	$ & $	\cdots			$ &	GROND &	0.16	\\
68456	& $	0.980	\pm	0.066	$ & $	\cdots			$ &	GROND &	0.54	\\
69196	& $	1.017	\pm	0.057	$ & $	\cdots			$ &	GROND &	0.77	\\
69997	& $	0.998	\pm	0.064	$ & $	\cdots			$ &	GROND &	0.25	\\
70850	& $	0.977	\pm	0.062	$ & $	\cdots			$ &	GROND &	0.77	\\
71596	& $	0.995	\pm	0.063	$ & $	\cdots			$ &	GROND &	0.72	\\
72379	& $	1.009	\pm	0.063	$ & $	\cdots			$ &	GROND &	1.03	\\
73164	& $	1.082	\pm	0.057	$ & $	\cdots			$ &	GROND &	0.27	\\
73959	& $	1.038	\pm	0.056	$ & $	\cdots			$ &	GROND &	1.31	\\
74759	& $	1.034	\pm	0.055	$ & $	\cdots			$ &	GROND &	0.51	\\
75557	& $	1.016	\pm	0.067	$ & $	\cdots			$ &	GROND &	0.32	\\
76349	& $	1.077	\pm	0.057	$ & $	\cdots			$ &	GROND &	0.25	\\
77141	& $	1.117	\pm	0.058	$ & $	\cdots			$ &	GROND &	0.50	\\
77953	& $	1.123	\pm	0.055	$ & $	\cdots			$ &	GROND &	0.86	\\
78755	& $	1.123	\pm	0.062	$ & $	\cdots			$ &	GROND &	0.21	\\
79557	& $	1.140	\pm	0.060	$ & $	\cdots			$ &	GROND &	0.39	\\
80358	& $	1.199	\pm	0.052	$ & $	\cdots			$ &	GROND &	0.77	\\
81167	& $	1.169	\pm	0.052	$ & $	\cdots			$ &	GROND &	0.84	\\
81946	& $	1.149	\pm	0.058	$ & $	\cdots			$ &	GROND &	1.05	\\
$125931-283858$	& $	1.054	\pm	0.058	$ & $	0.121	\pm	0.036	$ &	UVOT+GROND &	1.08	\\
$125931-283858$	& $	1.241	\pm	0.022	$ & $	\cdots			$ &	UVOT+GROND &	2.33	\\
151489	& $	1.240	\pm	0.044	$ & $	\cdots			$ &	GROND &	0.21	\\
155908	& $	1.183	\pm	0.049	$ & $	\cdots			$ &	GROND &	0.76	\\
160328	& $	1.182	\pm	0.044	$ & $	\cdots			$ &	GROND &	0.10	\\
164699	& $	1.224	\pm	0.059	$ & $	\cdots			$ &	GROND &	0.23	\\
239811	& $	1.158	\pm	0.058	$ & $	\cdots			$ &	GROND &	0.84	\\
250946	& $	1.154	\pm	0.063	$ & $	\cdots			$ &	GROND &	0.53	\\
$312320-474249$	& $	1.005	\pm	0.248	$ & $	0.144	\pm	0.140	$ &	UVOT+GROND &	0.31	\\
$312320-474249$	& $	1.248	\pm	0.083	$ & $	\cdots			$ &	UVOT+GROND &	0.39	\\
329172	& $	1.185	\pm	0.056	$ & $	\cdots			$ &	GROND &	0.31	\\
415466	& $	1.133	\pm	0.061	$ & $	\cdots			$ &	GROND &	0.12	\\
$501081-1101930$	& $	1.147	\pm	0.076	$ & $	\cdots			$ &	GROND &	0.20	\\
501081	& $	1.138	\pm	0.048	$ & $	\cdots			$ &	GROND &	0.50	\\
588101	& $	1.161	\pm	0.075	$ & $	\cdots			$ &	GROND &	0.20	\\
669176	& $	1.141	\pm	0.098	$ & $	\cdots			$ &	GROND &	0.11	\\
843664	& $	1.075	\pm	0.066	$ & $	\cdots			$ &	GROND &	0.04	\\
1101930	& $	1.096	\pm	0.101	$ & $	\cdots			$ &	GROND &	0.88	\\
%\hline
%\endfoot
\hline									
\hline									
\end{longtable}
%\tablefoot{}\\
}
}

\begin{figure}
  \centering
 \includegraphics[width=\columnwidth]{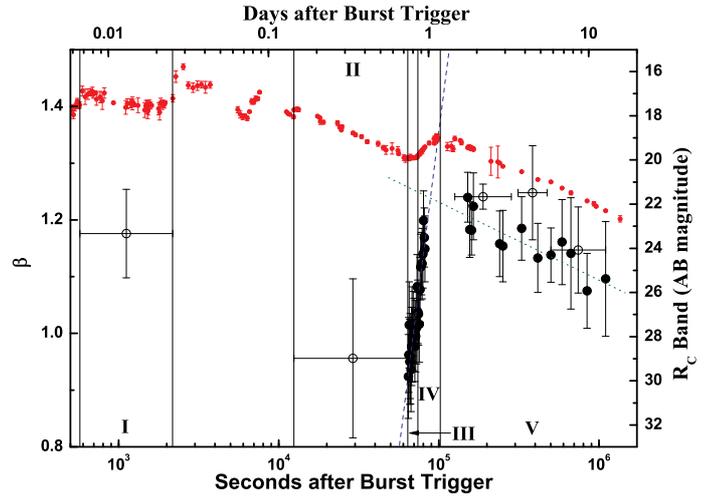}
   \caption{The evolution of the spectral slope of the optical/NIR afterglow, $\beta$. Open points are either derived from UVOT (plus TAROT, REM in the first epoch) or from joint UVOT+GROND fits. Closed points are GROND-only. See Table \ref{betatab} for the data. Red points are a composite $R_C$-band light curve of GRB 111209A (identical to the one given in Figs. \ref{KPobs} and \ref{KPzone}), added for purposes of orientation. The regions are: I. Early prompt emission. II. First afterglow-like decay. III. Plateau. IV. Rebrightening. V. Late afterglow. The plateau/rebrightening shows a marked rise of the spectral slope, from $\approx0.95$ to $\approx1.25$, where it then remains essentially constant. The two different behaviours themselves can be fitted linearly (in temporal log-space), the rising phase has a slope of $2.25\pm0.37$ (dashed blue line), while the constant phase has a slope of $-0.14\pm0.06$ (dotted green line), indicating some low-significance chromatic evolution may still be present. Data beyond those shown here show a significant departure from a power-law form due to the increasing dominance of the supernova component.}
              \label{betafig}
    \end{figure}

Multiple photometric measurements of a GRB afterglow obtained at a similar point in time (or corrected for the evolution of the afterglow) yield a spectral energy distribution (SED), a very-low resolution spectrum. The spectrum of a GRB afterglow is intrinsically a power-law due to the synchrotron-radiation nature of the emission \citep{Sari1998ApJ}, but a curvature can be induced by line-of-sight dust in the host galaxy even after the correction for Galactic extinction \citep[e.g.,][]{Fynbo2001AA,Kann2006ApJ,Kann2010ApJ}. The afterglow evolution is generally achromatic, a result that led \cite{Kann2010ApJ} to create compound light curves by shifting data of other filters to the $R_C$ band. Significant evidence for colour evolution of afterglows was quite sparse until recently, some well-known examples show that it is typically found at early times, and only if a bright afterglow leads to dense multi-wavelength sampling, which is the case for such examples as GRB 030329 \citep{Lipkin2004ApJ}, GRB 061126 \citep{Perley2008ApJ,Gomboc2008ApJ} GRB 080319B \citep{Bloom2009ApJ,Racusin2008Nature,Wozniak2009ApJ} and GRB 130427A \citep{Panaitescu2013MNRAS,Laskar2013ApJ,Vestrand2014Science,Maselli2014Science,Perley2014ApJ}.

GROND, due to its inherent simultaneous seven-colour observation mode, is predestined to study the SEDs of GRB afterglows, as each observation yields a seven-band SED with no concerns about variability. Combined with the ability to respond to GRBs within minutes, and the general collecting power of a 2.2m telescope, it is no wonder GROND has allowed the detailed study of multiple SEDs for several GRB afterglows and has allowed the discovery of significant spectral evolution in many cases, such as XRF 071031 \citep{Kruehler2009ApJ}, GRB 080129 \citep{Greiner2009ApJ2}, GRB 080413B \citep{Filgas2011AA2}, GRB 081029 \citep{Nardini2011AA}, GRB 100814A \citep{Nardini2014AA}, GRB 100621A \citep{Greiner2013AA} and GRB 091127 \citep{Filgas2011AA}, the latter even exhibiting smooth spectral evolution without any actual afterglow variability.

From the UVOT (as well as TAROT and REM during the prompt emission) data we have already derived two SEDs (Sects. \ref{fitting_prompt}, \ref{fitting_AG1}), both of which needed joint multi-colour fits over moderately long temporal baselines to yield viable results. For GROND, each ``shot'' gives us one SED.

We have a total of 46 epochs of data, and fit each with a simple power-law as well as with dust models based on Milky Way, Large and Small Magellanic Cloud dust based on the parametrization of \cite{Pei1992ApJ}. The results for the simple power-law as well as those with additional UVOT (and others) data are given in the online Table \ref{betatab}, and are plotted in Fig. \ref{betafig}. We have already remarked \citep[see also][]{Stratta2013ApJ} on the red-to-blue evolution from the prompt emission to the pre-rebrightening afterglow. The value derived from the decaying UVOT afterglow is in excellent agreement with the GROND-derived values for the beginning of the plateau phase. For the GROND data, we find a clear spectral evolution during the first day, which was already evident from the fact that the temporal slopes were different for each filter (Table \ref{tab_fit1}). We are able to linearly (in $\log(t)$-space) fit the evolution with a steep rising slope, $\alpha_{\beta,1}=2.25\pm0.37$. Starting on the second day, the afterglow SED has reached a constant value of $\beta\approx1.25$, fitting these data yields $\alpha_{\beta,2}=-0.14\pm0.06$, which is only in agreement with no further evolution at the $2\sigma$ level, therefore, some chromatic evolution may still be present but it is of low significance. Data beyond 21 days (epochs $41-46$) are not displayed as the dominating supernova component (G15,K18A) induces a strong curvature into the spectrum, so power-law fits lose their validity. We furthermore perform three fits using both UVOT and GROND data, now derived from joint power-law fits to the three afterglow segments separated by the ``steps'' (Sect. \ref{fitting_2}, see also K18A), the time spans over which these fits have been performed are also given in Table \ref{betatab}. Our SED results are for the most part in good agreement with those given by \cite{Stratta2013ApJ}, e.g., they find $\beta=1.07\pm0.15$ using GROND data published in the GCNs \citep{Kann2011GCN1}. Their spectral slope during the peak of the rebrightening, $\beta=1.0\pm0.1$, is somewhat bluer than what we expect from our $\beta$-evolution. Later SEDs, based only on UVOT data, show lots of scatter but agree within errors with our results.

The dust-model fits using GROND-only data yield discouraging results. At such low redshifts, it is not possible to distinguish between the three models using optical (and NIR) data only \citep{Kann2006ApJ}, conversely, even small fluctuations in the $g^\prime$ data (which is especially susceptible to atmospheric effects during observations like residual twilight, seeing and haze) lead to large differences in results, everything between moderate line-of-sight extinction of $A_V=0.5$ mag to negative curvature. Furthermore, the simple power-law model fits are all already statistically acceptable (excepting those during the supernova phase as mentioned above). \cite{Stratta2013ApJ} reach similar conclusions. Our joint UVOT+GROND fits, though, yield the largest wavelength span possible, and improve upon the GROND-only analysis. While the last epoch yields no evidence for dust, and the middle one no statistically significant evidence, the first of the three joint fits (where UVOT data is densest and has the best S/N) reveals curvature in the SED in the observer-frame ultraviolet. Using the SMC dust model (LMC and MW dust yield similar results), we detect additional extinction along the line of sight at the $3.4\sigma$ level: $\AV=0.121\pm0.036$ mag. This value is not in contradiction with the general result of low line-of-sight extinction, and a very typical value for GRB afterglows \citep[e.g.,][]{Kann2006ApJ,Kann2010ApJ}. This value was adopted in G15 and we also use it here where needed.

An important conclusion we can draw here is that we find no evidence for a thermally dominated spectrum such as in the case of the ``Christmas Burst'' GRB 101225A, which showed an evolving, cooling blackbody and not a typical afterglow synchrotron spectrum \citep{Thoene2011Nature}, nor do we see an extremely UV-bright blackbody spectrum as exhibited (probably) by GRB 110328A/Swift J164449.3+573451 \citep{Levan2011Science} and (definitely) by Swift J2058.4+0516 \citep{Cenko2012ApJ}. This is a strong argument against GRB 111209A being similar to these classes of explosive transients.

%------------------------------------------------------------------------------------------------------------------------------------------------------------------------------------------------------------------
%------------------------------------------------------------------------------------------------------------------------------------------------------------------------------------------------------------------

\subsection{The host galaxy}
\label{SectHost}

\begin{figure}
  \centering
 \includegraphics[width=\columnwidth]{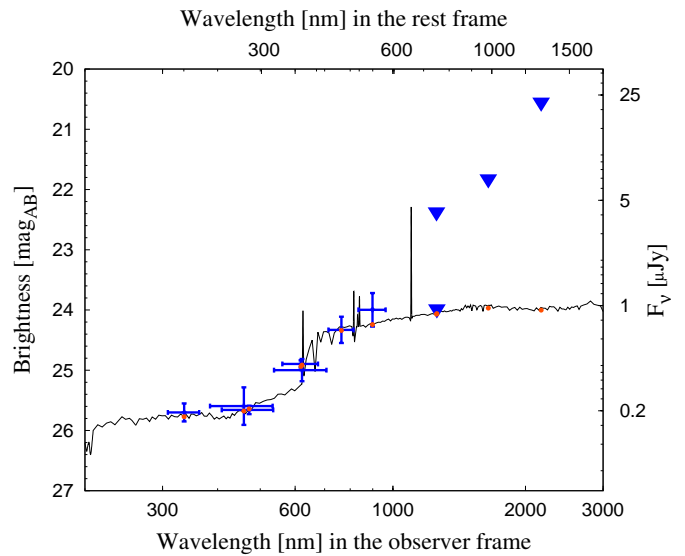}
   \caption{\texttt{LePHARE} fit to the magnitudes of the GRB 111209A host galaxy.  The detections are $F336Wg^\prime g^\prime_G r^\prime r^\prime_G i^\prime_G z^\prime_G$, the upper limits $JJ_GH_GK_{S,G}$ (the subscript G denotes GROND filters presented in this work, the non-GROND data have been taken from L14). The red $\bullet$ marks represent the photometry values as determined by the synthetic SED. The fit is acceptable, $\chi^2=1.36$. The host galaxy is very typical for a GRB host, a low-mass, low-extinction, young star-forming galaxy (see, e.g., \citealt{Schulze2014AA} for a comparison). For specific values, see Table \ref{LephareTab}.}
              \label{Lephare}
    \end{figure}

\begin{table}[t]
\caption{Fit results for the \texttt{LePHARE} host-galaxy fit.}
\label{LephareTab}
\centering                        
\begin{tabular}{l r}       
\hline\hline   
Parameter	&	Value	\\\hline   
Redshift & 0.677 (fixed) \\ \vspace{1mm}
$\chi^2$ & 1.36 \\ \vspace{1mm}
Extinction law & Calzetti \\ \vspace{1mm}
$E_{(B-V)}$ & 0.04 \\ \vspace{1mm}
$M_B$ & -17.93 \\ \vspace{1mm}
Age (Gyr) & $0.84^{+1.24}_{-0.50}$\\ \vspace{1mm}
SFR ($M_\odot$ yr$^{-1}$) & $0.22^{+1.33}_{-0.01}$\\ \vspace{1mm}
log sSFR (1/Gyr) & $0.39^{+2.52}_{-0.38}$ \\ \vspace{1mm}
$\log\left(\textnormal{Mass} (M_\odot)\right)$ & $8.89^{+0.58}_{-0.36}$\\ %\vspace{1mm}
\hline \hline
\end{tabular}
\end{table}

In the deep late-time data taken at 280 days post-trigger, we detect the host galaxy in $g^\prime r^\prime i^\prime z^\prime$ in the $3-5\sigma$ range. While our $JHK$ observations reach deeper than any of our previous ones of this field, the host remains undetected. L14 also obtained host-galaxy observations, yielding clear detections in $g^\prime r^\prime$ which are in excellent agreement with the values we obtain (and having significantly higher S/N), an additional low-S/N detection in \emph{HST} $F336W$ and Gemini $u^\prime$, as well as a much deeper limit in $J$ than the one we have achieved. The SN does not contribute significantly any more during this epoch, with expected magnitudes $g^\prime\approx28.5$ mag, $r^\prime\approx28.0$ mag, $i^\prime\approx27.5$ mag, $z^\prime\approx27.2$ mag. We find that the host galaxy is an unresolved point source, and thus must be very compact, this is confirmed by L14 who detect only marginal extension even in their high-resolution \emph{\emph{HST}} imaging. We add their data to our host-galaxy study, using the host-galaxy magnitude they derive in $F336W$, just as for the afterglow+SN fit (G15,K18A).

We use the photometric SED-fitting code \texttt{LePHARE} \citep{Arnouts1999MNRAS,Ilbert2006A&A}\footnote{http://www.cfht.hawaii.edu/$\sim$arnouts/LEPHARE} to determine host-galaxy parameters from the detections. As a cross-check, in case there is still a contribution of the SN, we also use the host-galaxy values as determined by our joint afterglow+SN fit (G15,K18A); we find the results are consistent with the previous fit, within uncertainties. Our fit result is shown in Fig. \ref{Lephare}. We find that the host galaxy is a low-mass galaxy with very low global extinction (using the \citealt{Calzetti2000ApJ} starburst-galaxy extinction law) a moderate star-formation rate (SFR) (which is in full agreement with that derived from emission line properties by \citealt{Kruhler2015AA}) but a high specific star-formation rate (sSFR); for result values, see Table \ref{LephareTab}. Due to the lack of NIR data and the low S/N of the optical data, parameters such as age, SFR, sSFR and mass are not well-determined. From the emission lines in their spectra, L14 derive a moderately high metallicity. \cite{Kruhler2015AA} derive, from the same spectra, a metallicity which is 0.3 dex lower, but in agreement within the large errors.

To determine the offset of the afterglow from the host galaxy, we use the $r^\prime$-band image taken at 5.8 days (see Fig. \ref{Field}), which has the highest detection S/N. Using the result catalogue of this image as an astrometry catalogue, we determine a mapping uncertainty between the two epochs of 0\farcs018. We estimate the purely statistical error of the afterglow/host localization via the size of the mean stellar Full-Width Half Max multiplied by the error of the afterglow/host (the afterglow, but also the host, are point sources), finding 0\farcs011 for the afterglow and 0\farcs176 for the host galaxy, the latter error dominates the total error on the offset 0\farcs177. We determine an offset of 0\farcs396 in RA and -0\farcs180 in Dec., and thus in total an offset of $0\farcs44\pm0\farcs18$. This translates to a projected offset of $3.06\pm1.25$ kpc. This is somewhat larger than the median projected offset found by \citet[][1.3 kpc]{Bloom2002AJ}\footnote{Note that these authors used a different world model than we do, but as they themselves point out, the angular distance scale over a large redshift range is quite insensitive to redshift, and therefore also to small changes in the world model.}, but still well within the distribution those authors found, which extends to 7 kpc (we caution that this sample is not unbiased and may therefore not represent the true offset distribution).

L14 use their two epochs of \emph{HST} WFC3 F336W data to determine the offset using a method analogous to our own. They derive a significantly smaller offset ($0\farcs011\pm0\farcs038$), indistinguishable from being right in the centre of the galaxy. As the \emph{HST} data have both a better S/N than our data and a much finer resolution, their offset value is likely more trustworthy than the one we have determined.

In passing, we note that G15 also derived the offset between the afterglow position and the SN position, and found that it was $<200$ pc (0 within errors), making it extremely unlikely that SN 2011kl is not caused by the same progenitor as GRB 111209A itself.

%------------------------------------------------------------------------------------------------------------------------------------------------------------------------------------------------------------------
%------------------------------------------------------------------------------------------------------------------------------------------------------------------------------------------------------------------

\subsection{The light curve of GRB 130925A}
\label{130925A}

Another ULGRB with detailed GROND observations is GRB 130925A (see Appendix \ref{ELDGRBs} for more details on this event). \cite{Greiner2014AA} have presented optical/NIR observations obtained with GROND and VLT/HAWK-I (late host-galaxy detections). These data show strong, rapid variability during the very long prompt emission phase. Furthermore, the afterglow is very strongly extinguished, the largest extinction measured so far with high significance along a line-of-sight to an afterglow. This leads to the optical transient not being detected at all in $g^\prime r^\prime$, whereas in $i^\prime z^\prime$, only the early optical emission of the prompt peak is detected above host-galaxy level. In $JHK_S$, both the early prompt emission as well as a late-time afterglow are detected. \cite{Greiner2014AA} present two SED fits of the optical NIR data. At peak, they measure an intrinsic spectral slope $\beta=1.3\pm0.4$ coupled with an extinction $\AV=5.0\pm0.7$. The later afterglow, in combination with X-ray data, yields $\beta=0.32\pm0.03$ and $\AV=6.56\pm0.76$. Both of these measurements must be treated with caution. The early optical/NIR emission is likely coupled to the largest prompt emission peak at high energies, and therefore is likely to have a different colour than the late afterglow, additionally, the intrinsic SED may not be described by a simple power-law, though of course such a deviating intrinsic spectrum can not be deduced from the data. The late SED is not based on contemporaneous data in the optical/NIR and the X-rays, additionally, the X-ray afterglow clearly stems from a different source than the optical/NIR emission \citep[see][for interpretations]{Evans2014MNRAS,Zhao2014ApJ,Piro2014ApJL}. We will therefore use both sets of values, as the true slope and extinction are likely to lie somewhere in the range.

Using our knowledge of the redshift $z=0.347$, the spectral slope and the extinction, we can use the method of \cite{Kann2006ApJ} and shift the light curve\footnote{More specifically, the light curve is given in the observer frame, but plotted as if the GRB had occurred at $z=1$.} of GRB 130925A to $z=1$. For the SMC dust model (as used by \citealt{Greiner2014AA}), we find large corrections $dRc$ to $z=1$, despite the very low redshift. The ``prompt'' values yield $dRc=-2.91^{+1.03}_{-1.04}$ mag, whereas the ``afterglow'' values give $dRc=-5.29^{+0.95}_{-0.94}$ mag. The very large error bars are due to the very large -- in absolute terms -- errors of the line-of-sight extinction. The two results are discrepant at the $1.7\sigma$ level. As stated above, the optical transient is not detected at all in the $r^\prime$ band. Since the method of \cite{Kann2006ApJ} is normalized to the $R_C$ band, we need to evaluate how the afterglow would look like in the observed $r^\prime$ band. We subtract the individual host-galaxy magnitudes from all bands, and then create an $i^\prime z^\prime JHK_S$ SED at the prompt emission peak. We find results that agree within errors with those presented in \cite{Greiner2014AA}. It is not possible to distinguish the dust model for such a low redshift \citep{Kann2006ApJ}, but Milky Way-type dust yields the smallest intrinsic spectral slope coupled with the highest extinction. Since several highly extinguished GRB afterglows have shown the presence of a prominent $\approx2200$ {\AA} bump \citep{Kruehler2008ApJ,Eliasdottir2009ApJ,Perley2011AJ,Zafar2012ApJ}, it is quite likely such a feature also exists in the spectrum of GRB 130925A, but the combination of low redshift, extremely high extinction and a bright host galaxy makes it undetectable.

The extrapolation of the extremely red ($\beta_0=4.77\pm0.11$) and curved SED at peak allows us to determine that the expected observed peak magnitude in the $r^\prime$ band would be $r^\prime\approx23$ mag. We then take the (host-galaxy subtracted) $H$ band (to which we add 118 s, to make $t_0$ identical with the trigger time of \emph{Fermi} GBM), which offers the highest data density and latest detections, and shift it downward by the difference, a total of 7 magnitudes (in Vega). This light curve is then corrected using the two $dRc$ values derived above. If one adds the large, one-magnitude error margin, the ``afterglow value''-derived shifted light curve reaches the same luminosity as the corrected afterglow of GRB 111209A (Sect. \ref{SectDiscLumin}). The ``prompt value''-derived result is among the least luminous afterglows detected so far. This may indicate that the ``afterglow value''-derived shift depicts a more realistic scenario, but there is no clear way to determine where within the range of $\approx4$ magnitudes the shifted light curve truly lies.

\section{Discussion}
\label{SectDisc}

\subsection{Energetics}
\label{Energy}

The true energy release of GRB 111209A is an important parameter both for the determination of the nature of the event, as well as for the modelling of the event with spin-down emission from a magnetar.

Building upon the correlation between intrinsic peak energy of the prompt emission $E_{p,z}$ and isotropic-equivalent energy release $E_{\rm iso}$ \citep{Amati2002AA}, \cite{Ghirlanda2004ApJ} discovered a tighter correlation between $E_{p,z}$ and $E_\gamma$, the collimation-corrected energy release (corrected for beaming, see also \citealt{Friedman2005ApJ}). The ``price'' for a tighter correlation is the addition of a new parameter, the jet-break time $t_b$, which is needed to determine the jet opening angle $\theta$ \citep{Sari1999ApJ}. Additionally, the external medium density $n_{\rm ISM}$ (or the wind parameter $A_\star$, see, e.g., \citealt{Zeh2006ApJ}, \citealt{Nava2006AA}, \citealt{Ghirlanda2006AA}) and the efficiency $\eta_\gamma$ are needed as input parameters, these are usually best-estimates, with $n\approx1-10$ cm$^{-3}$. Using the prompt energetics as derived by \cite{Golenetskii2011GCN}, we find $E_{p,z}=520\pm89$ keV, and a bolometric (restframe $1-10,000$ keV) $\log E_{\rm iso,bol}=53.83\pm0.06$ (one of the most energetic GRBs at $z<0.9$, \citealt{Perley2014ApJ}). GRB 111209A agrees well with the Amati correlation.

Our measurement of a jet-break time (G15,K18A, see Sect. \ref{fitting_3}) also allows us to place this GRB in the context of the Ghirlanda relation. Using the standard density values, we derive an opening angle of $\theta=0.15-0.19$ radian (for $n_{\rm ISM}=1\cdots10$ cm$^{-3}$). For $n_{\rm ISM}=0.1$ cm$^{-3}$ (as used by \citealt{Stratta2013ApJ}), we find $\theta=0.11$ radian, in good agreement with the value \cite{Yu2015MNRAS} employed, but significantly less than found by \cite{Stratta2013ApJ}. Both results are smaller than the value \cite{Ryan2015ApJ} derived from the X-ray afterglow, $\theta=0.34^{+0.11}_{-0.13}$ radian, but larger than the value \cite{Metzger2015MNRAS} derive, $\theta=0.05$ radian. We point out that \cite{Metzger2015MNRAS} overestimate the mean isotropic gamma-ray luminosity by a factor of 10, though; using input values from G15, it is $L_\gamma\approx4\times10^{49}$ erg s$^{-1}$. Using the correct value and the equation given in \cite{Metzger2015MNRAS}, we find $\theta\approx0.16$ radian, in perfect agreement with our result above. Note that the range we find corresponds to an opening angle of $8.6-10.9$ degrees. \cite{Prajs2017MNRAS} use a mean opening angle of 12 degrees to estimate that the ULGRB rate is roughly similar to the superluminous-supernova (SLSN) rate they derive, and that SLSNe may therefore be associated with ULGRBs in a more general manner (see K18A for more discussion on ULGRBs and SLSNe). They use an estimate as ``[t]o date, jet-breaks have not been observed in ULGRBs'' \citep{Prajs2017MNRAS}. The upper bound we derive for the opening angle assuming a typical ISM density agrees well with their estimate, implying their rate estimation does not need to be significantly changed. {\cite{Perna2018ApJ}, who study numerically whether BSG progenitors with fallback are able to launch jets and power ULGRBs, use an injection opening angle of $\theta=0.28$ radian, but note that the cocoon of the massive envelope will collimate the jet further, implying qualitative agreement with our opening angle result.}

The main result, though, is that for all these values, GRB 111209A is a strong outlier of the Ghirlanda correlation and a hyper-energetic GRB, with $\log E_\gamma[\textrm{erg}]=51.60-52.10$, a value similar to the hyper-energetic events studied by \cite{Cenko2011ApJ}. To bring this GRB into agreement with the Ghirlanda relation´\footnote{{Of course, it must be noted that the Ghirlanda relation is purely empirical, and so far has no clear physical meaning, therefore, being an outlier of this relation does not necessarily imply that any input parameters \emph{need} to be adjusted to bring it into agreement with the relation. Recent research suggests it is less tight than once assumed \citep{Wang2018ApJJet}.}}, assuming standard values otherwise, a significantly lower circumburst medium density, $n_{\rm ISM}\approx10^{-4}$ cm$^{-3}$, would be needed, leading to an opening angle of just $\theta=0.046$ radian {(2.6 degrees, in this case $\log E_\gamma[\textrm{erg}]=50.86$)}. Such a small opening angle would make it an outlier in the large sample of \cite{Ryan2015ApJ}, who find $\theta\geq0.055$ radian at 95\% confidence{, but in full agreement with the recent results of \cite{Wang2018ApJJet}, who find many early breaks\footnote{{In these cases, of course, the narrow opening angles occur in typical external medium densities, whereas for GRB 111209A, the break is late but the density extremely low.}} and a typical $\theta=(2.5\pm1.0)^\circ$}. We note here such a low circumburst density would be in agreement with the explanation \cite{Evans2014MNRAS} propose for GRB 130925A (see K18A for more discussion) and that such low circumburst densities are not unprecedented, having been found for other GRBs from reverse-shock-flare modelling \citep{Laskar2013ApJ,Laskar2016ApJ,Alexander2017ApJ}.

\citet[][henceforth GF17]{Gompertz2017ApJ} study GRB 111209A/SN 2011kl in the context of the magnetar model under the assumption that the magnetar powers the entire event, including the prompt emission and the afterglow. They present two separate lines of evidence that GRB 111209A is surrounded by a \textit{high} circumburst density medium (in contrast to our argument using the Ghirlanda relation given above), and combined with a lower limit on the jet-break time, they conclude only an extreme magnetar rotating near break-up speed can power the event (in contrast to other results, e.g., \citealt{Yu2017ApJ} find the lowest rotation rate for SN 2011kl compared with a sample of SLSNe, but as GF17 point out, these studies only deal with powering the SN {-- see also the recent results from \citealt{Wang2017ApJ3}, who use the bolometric light curve of K18A to derive magnetar-spin-down parameters, and the rotation rate they find is slower than any in the SLSN-sample of \citealt{Nicholl2017ApJ3}}).

We find {some} inconsistencies in the work of GF17 that imply that a less extreme magnetar {may} also account for GRB 111209A. {As a first note, a part of their work remarks upon the model of \cite{Metzger2015MNRAS} and that work's strongly collimated jet, but as we pointed out above, \cite{Metzger2015MNRAS} overestimate the isotropic luminosity of the GRB and therefore the collimation, this implies that the modelling of \cite{Metzger2015MNRAS} does not need to invoke extreme collimation. Alternatively, in the case of such collimation, the extremely low maximum density limit \cite{Metzger2015MNRAS} give (which GF17 contrast with a significantly higher lower limit they derive) is not valid anymore and the two results are not in conflict.}

\begin{figure}[t]
  \centering
 \includegraphics[width=\columnwidth]{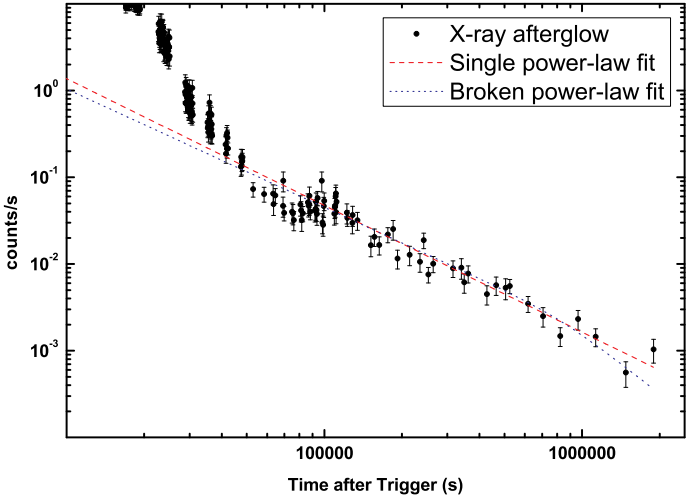}
   \caption{The late X-ray afterglow of GRB 111209A. We show two fits using all data from 109 ks onward, a singe power-law fit (red dashed) and a broken power-law fit (blue dotted), with the parameters fixed to those derived from the optical afterglow fitting. See text for more details.}
              \label{XRT}
    \end{figure}

\begin{figure}[t]
  \centering
 \includegraphics[width=\columnwidth]{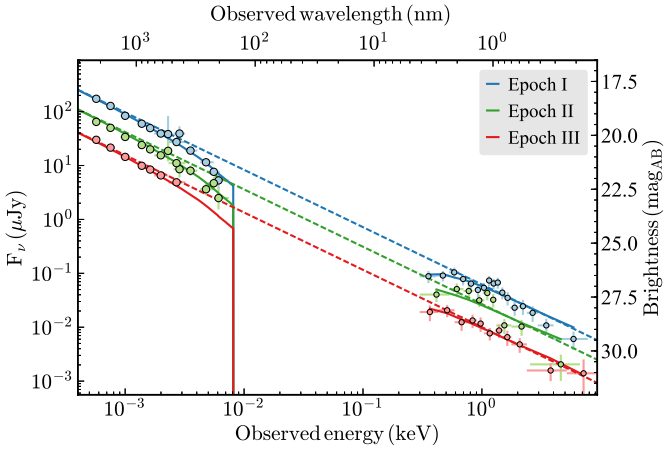}
   \caption{Optical-to-X-ray SED fits for the late afterglow. The three epochs are the ones given in Table \ref{betatab}, between the ``steps''. The SEDs are all well-fit with single power-laws, and the derived parameters are identical in all three cases. There is no significant evidence for spectral evolution or the existence of a cooling break.}
              \label{XRTsed}
    \end{figure}

\begin{table*}[t]
\tiny
\caption{Results of the optical-to-X-ray SED fits using a single power-law.}
\label{tab_XRT}
\centering                        
\begin{tabular}{c c c c c c c c}       
\hline\hline   \vspace{1mm}
Epoch & Fit statistic/dof  & Fit statistic/dof  & Fit statistic/dof  & Normalisation & Spectral slope $\beta$ & $\AV$ & NH \\
(time span in s) & ($\chi^2$, optical) & (C-stat, X-rays) & (Total) & [$10^{-5}$ keV/cm$^2$/s/keV] & & [mag] & [$10^{20}$ cm$^{-2}$] \\\hline
I (125931-283858) & 11.53/13 & 134.43/170 & 145.96/179 & 9.35 & $1.064^{+0.017}_{-0.016}$ & $0.122^{+0.024}_{-0.023}$ & $5.2^{+6.0}_{-5.2}$ \\[1mm]
II (312320-474249) & 12.43/13 & 84.83/85 & 97.26/94 & 4.05 & $1.059\pm0.031$ & $0.209^{+0.085}_{-0.075}$ & $1.8^{+9.3}_{-1.8}$ \\[1mm]
III (501081-1101930) &  1.46/7 & 76.32/105 & 77.77/108 & 1.51 & $1.060^{+0.026}_{-0.028}$ & $0.049^{+0.087}_{-0.049}$ & $0.1^{+8.5}_{-0.1}$ \\[1mm]
%\vspace{1mm}
% \\
\hline \hline
\end{tabular}
\end{table*}

The second issue is the jet-break time. For one, GF17 completely ignore the optical-afterglow fit we presented already in G15, where we derive a jet break at 9.1 days (observer frame; 5.4 days rest-frame). They instead argue that the X-ray afterglow light curve (as given on the Swift XRT repository, \citealt{Evans2007AA,Evans2009MNRAS}) shows no signs of a break until 22 days (observer frame) post-trigger. To put this statement on a statistical footing, we have fit the X-ray light curve. We use data starting 1.26 days post-burst, our first fit is with a single power-law, and our second fit is with a broken power-law, whereby we fix the jet-break time and the post-break slope to the values derived from the optical fit. We fix the sharpness of the break $n=5$. These fits can be directly compared as they have identical degrees of freedom. For the first fit, we find $\alpha_{X,1}=1.46\pm0.05$, and $\chi^2=41$ for 36 degrees of freedom; for the second fit, $\alpha_{X,1}=1.36\pm0.06$ and $\chi^2=52$ (Fig. \ref{XRT}). \cite{DePasquale2016MNRAS2} use a similar approach to derive a lower limit on the possible jet-break time of GRB 130427A, taking $\Delta\chi^2=2.7$ as a significance criterion. With $\Delta\chi^2=11$ in our case, the X-ray afterglow alone significantly rules out a break at the optical time. We note, though, that the last XRT data point lies above the best fit, whereas the next-to-last data point at 17 days lies beneath the fit.

The final X-ray detection by \emph{Swift} may have an alternate explanation. \cite{Levan2013ApJ} reported the detection of highly luminous X-ray emission associated with the SLSN SCP06F6, and \cite{Margutti2017ApJX} present not only limits on SLSNe and some further (faint) detections, but also compare them to the X-ray light curve of GRB 111209A. Their figure 1 shows that while most limits are deeper than the luminosity of the final data point of GRB 111209A ($\approx9\times10^{43}$ erg s$^{-1}$ according to \citealt{Margutti2017ApJX}, $\approx6\times10^{44}$ erg s$^{-1}$ following GF17), this last detection is significantly less luminous than the detection of SCP06F6. While no definite statement can be made, it is not implausible that SN 2011kl was accompanied by luminous X-ray radiation, which we are detecting here. {Hereby, we point to K18A and references therein for multiple lines of argumentation why SN 2011kl resembles SLSNe more than it does typical GRB-SNe. The nature of luminous X-ray radiation accompanying SLSNe is far from being clear, as is the evolution of such X-ray light curves compared to their optical counterparts. The UV/optical evolution SCP06F6 was significantly slower than that of the less luminous SN 2011kl, therefore, it may be possible that an X-ray transient associated with SN 2011kl itself (and not the GRB 111209A afterglow) would also evolve faster than the X-ray transient associated with SCP06F6, and be detectable by \emph{Swift} at the given time.} Therefore, the late X-ray light curve could consist of a breaking afterglow caused by an achromatic jet break and a rising (possibly peaking, as the detection is near the SN light-curve peak) SN X-ray component.

Bearing this in mind, we repeat the exercise excluding the very last detection, and now find $\alpha_X=1.49\pm0.06$, and $\chi^2=39$ for 35 degrees of freedom, $\alpha_X=1.41\pm0.06$ and $\chi^2=40$, respectively; it is now $\Delta\chi^2=1$, implying the X-ray data do not strongly support a jet break but certainly do not rule it out either. A free fit to the entire data (after 1.26 days) does not find a break, whereas the exclusion of the last point leads to a break being found, but the break-time and post-break slope are degenerate, anchored by only a single data point (the last one in this fit, and next-to-last one in total).

A further issue involves the broadband SED of the GRB, which GF17 argue is one of the indicators of a high circumburst density. They do use optical/NIR afterglow data from G15 to derive the SED in this regime. The values of $\beta_o$ and $A_V$ they derive are perfectly consistent with our results. Furthermore, they derive a significantly steeper slope in the X-ray regime, which we confirm by using time-sliced spectra (from 109 ks onward, beyond the plateau phase, \citealt{Gendre2013ApJ}) from the XRT repository. We derive a spectral slope difference of $\Delta\beta_{X,o}=0.43\pm0.19$, which would be indicative of a cooling break between the optical and X-ray regimes, as also stated in GF17. The problem arises from the value of the decay slope of the optical/NIR afterglow. GF17 use nine epochs of GROND data unaffected by SN 2011kl to derive a slope $\alpha_o=1.30\pm0.05$. This slope is shallower than the one we derive, $\alpha_o=1.55\pm0.01$. While GF17 do not show their fit or give a goodness-of-fit value, the value they derive points to them not taking the rebrightening episodes into account. Using their value and comparing it with the X-ray decay slope, they find $\Delta\alpha_{X,o}\approx0.25$, as expected for a cooling break in the case of a constant-density medium and slow cooling \citep[e.g.,][]{Zhang2004IJMPA}. Our measured X-ray decay slope, especially if a single power-law fit is assumed (as GF17 do), lies close to our measured optical decay slopes, which is unexpected in the case of a cooling break lying between the X-ray and the optical domain, unless the circumburst medium density $\propto\rho^{-1}$ (between a constant-density and a wind medium) and hence $\nu_c$ does not evolve and $\alpha_o=\alpha_X$. To further study this aspect, we create three NIR-to-X-ray SEDs, following the SED epochs given in Table \ref{betatab}, i.e., for each of the smoothly decaying segments of the late afterglow between the ``steps''. These SEDs are shown in Fig. \ref{XRTsed}. We fit them with single power-laws, broken power-laws and broken power-laws with $\beta_2=\beta_1+0.5$. We find no significant evidence that the SEDs can not be explained by single power-laws; in all cases {where a broken power-law is involved}, the break is found in the range $0.5-2$ keV, within the X-ray data itself. {This short, steeper segment yields a slightly improved fit, but in all cases, these improvements are not statistically significant.} The spectral slope for all three fits is identical and in excellent agreement with the intrinsic spectral slope found in the optical/NIR SED alone; the same is true for $\AV$, we give the values in Table \ref{tab_XRT}. All in all, we find no significant evidence for a cooling break or spectral evolution.

{To further study the likely location of the cooling break, we use the $\alpha-\beta$ relations. We only look at the simple models (e.g., \citealt{Zhang2004IJMPA}, their Table 1, see also \citealt{Gao2013NewARG}) and find fast cooling is ruled out in all cases. The intrinsic spectral slope we derive, $\beta=1.06$, is quite red and would usually be indicative of the regime blueward of the the cooling break $\nu_c$, and an electron distribution power-law slope $p=2.12$, and implied slope redward $\beta_{IR}=\beta_{o,X}-0.5=0.56$ which is in perfect agreement with the mean value found in \cite{Kann2006ApJ}. But taking the rather steep afterglow decay slope (as mentioned, $\alpha_o=\alpha_X$ within small errors) into account, we find the only $\alpha-\beta$ relation in full agreement with the data (i.e., $=0$ within errors) is for slow cooling, spherical expansion, a constant-density medium (``ISM''; note that GF17 consider an ISM medium ruled out in the case of the narrow-jet solution of \citealt{Metzger2015MNRAS}), and a cooling break \emph{blueward} of the X-ray regime (therefore, the marginal curvature seen in the harder X-rays may be induced by the curvature of the cooling break). This directly implies a very soft electron distribution power-law index of $p=3.12$, a rare case, but not unheard of and in agreement with the non-universality of $p$ in GRB afterglows \citep{Shen2006MNRAS}. This result therefore does not support the claim of GF17 concerning a \emph{high} circumburst density, as $\nu_c$ scales with $n^{-1}$ (e.g., \citealt{Zhang2007ApJ2}). Note that if we were to assume a  higher column density (GF17 derive lower limits in the range $10...10^4$ cm$^{-3}$ from the radio observations, see below), it implies that the event is an even stronger outlier of the Ghirlanda relation, and even more energetic, e.g., all other parameters unchanged, for $n=100$ cm$^{-3}$, we find an opening angle $\theta\approx0.26$ radian (14.9 degrees) and a collimation-corrected energy release of $\log E_\gamma[\textrm{erg}]=52.4$.}

{In the simple models we look at, this also implies $p=\alpha_{post-break}$, which would imply $\alpha_{post-break}>3$, significantly steeper than the value $\alpha_{post-break}=2.33\pm0.17$ which we find. It is therefore possible that there is indeed a contribution of SN 2011kl even in the $u$ band, and that the intrinsic afterglow decay is even steeper than what is observed when assuming SN 2011kl does not contribute in the $u$ band, which implies that SN 2011kl would be even more luminous than the result presented in G15 and K18A, further strengthening the claim that this SN is dissimilar to normal GRB-SNe.}
% Therefore, the case for a cooling break is not as clear-cut as GF17 imply, and the derived limits on the circumburst-medium density may be less significant.

{We note that for high values of the energy fraction stored in magnetic fields $\epsilon_B\gtrsim0.01...0.1$, synchrotron cooling would become important, and $\nu_c$ would lie redward of the X-rays, and the X-ray flux would become independent of density. But for one, we find from the closure relations that the cooling break lies blueward of the X-rays. Furthermore, several works have performed afterglow modelling and have shown that typically $\epsilon_B$ is very low, likely $\gtrsim10^{-4}$ \citep{Kumar2010MNRAS,Santana2014ApJ,Wang2015ApJS}. These findings strengthen our result of a cooling break blueward of the X-rays.}

Finally, GF17 use modelling of the radio data \citep{Hancock2011GCN,Hancock2012GCN} to derive further lower limits on the circumburst density, again finding moderately high values contrasting our estimation given further above {under the assumption that the Ghirlanda relation is valid}. The spectral shape of the radio detections is peculiar, which GF17 explain by interstellar scintillation. Alternatively, the anomalous spectrum may be created by the overlap of multiple components, e.g., a reverse shock and a forward shock, as in the cases of GRB 130427A \citep{Laskar2013ApJ,Perley2014ApJ} and GRB 160625B \citep{Alexander2017ApJ,Troja2017Nature1}. In such a case, scintillation would not be necessary to fully explain the deviation from the expected spectrum (though of course a lower amount of scintillation may still exist), the source size could be significantly larger and again, the limits on the circumburst medium density could be lower. As there are no further X-ray or radio observations, to our knowledge, both of our alternative explanations can not be pursued any further, though.

{\cite{Metzger2018ApJ} recently presented a study on the effects of fallback accretion on rapidly spinning proto-magnetars. They find that under certain conditions, the additional accretion can keep the magnetization within a certain regime which enables the launch of a relativistic jet for thousands of seconds, while at the same time also enabling additional energy injection into the late-time supernova, creating a viable solution to GRB 111209A/SN 2011kl. One caveat of this model is that it does not imply that the maximum extracted energy from the magnetar increases compared to the case without any additional accretion.}

{The total rotational energy budget of a magnetar is}

\begin{equation}
E_{rot}=\frac{I}{2}\Omega_0^2,\,\,\Omega_0=\frac{2\pi}{P_0},
\end{equation}

{hereby $I\approx1.3\times10^{45}\,\mathrm{g\,cm}^2$ is the moment of inertia of the neutron star (assuming a typical mass $M_{NS}=1.4M_\odot$), $\Omega_0$ is the initial angular velocity and $P_0$ is the initial spin period. For a typical magnetar near break-up rotation speed ($P_0\approx1$ ms), it is $E_{rot}\approx3\times10^{52}$ erg, though \cite{Metzger2015MNRAS} argue that neutron stars near the maximum possible mass (see \citealt{Margalit2017ApJ,Shibata2017PhRvD,Ruiz2018PhRvD,Rezzolla2018ApJ} for recent research) and with larger radii and extreme spin periods could have up to an order of magnitude higher $E_{tot}$. We note that GF17 argue that ``the majority of this [energy] is carried away by neutrinos''. Such a mechanism would be expected in Type II core-collapse SNe, but in the case of a spindown-powered central engine, such as a magnetar, the energy is released as a Poynting flux which is converted into radiation and kinetic energy. While neutrinos may be produced in the dissipation region via $p-\gamma$ interactions, the expected cross section and neutrino flux are very low \citep{Zhang2013PhRvL}. Therefore a large fraction of the energy is available to drive the GRB, afterglow and supernova.}

%{\emph{The following paragraph is work in progress and will need further input. I'm writing it non-bold-faced for now to contrast real text and comments, I'll later clean it and make it bold-faced for the revised version.}}

{Without broadband modelling, which is beyond the scope of this work (and may be hard to achieve considering how sparse the radio data is), it is not possible to pin down the true energetics of the entire event, but we can make an estimate. The collimation-corrected prompt energy release, as has been pointed out, varies strongly depending on the assumed circumburst medium density, from $\log E_\gamma[\textrm{erg}]=50.9$ for the ``Ghirlanda'' solution, $\log E_\gamma[\textrm{erg}]=51.6-52.1$ for typical density values, and up to $\log E_\gamma[\textrm{erg}]\approx52.4$ for a density 10 times higher than typical values, representative of the indications of high density that GF17 have derived.}
%{\emph{This begs question one: Is there a realistic possibility of changing some other parameter to get a narrow jet (and therefore high beaming factor) while at the same time keeping a typical or even high circumburst density?}}

{The two further components that need to be taken into account are the kinetic energy of the SN $E_{k,SN}$ and the kinetic energy of the ultrarelativistic jet $E_{k,Jet}$. For the bolometric light curve presented in G15, these authors derive $E_{k,SN}=(5.5\pm3.3)\times10^{51}$ erg. We look to the work of \cite{Wang2017ApJ3}, who applied the magnetar model to the bolometric light curve of K18A, to derive estimates of $E_{k,SN}$. We use the values they derive for models B1 and C1 (pure magnetar, and magnetar + $^{56}$Ni, respectively, both for opacity $\kappa=0.07$ cm$^2$ g$^{-1}$) and $E_{k,SN}=0.3M_{ej}v_{sc0}$, with $M_{ej}$ the ejecta mass of the SN and $v_{sc0}$ the ejecta scale velocity, and find $E_{k,SN}=\left(1.1^{+0.72}_{-0.64}\right)\times10^{52}$ erg and $E_{k,SN}=\left(1.3^{+0.72}_{-0.67}\right)\times10^{52}$ erg, respectively. We note here that depending on the asphericity of the SN explosion (which may occur along the polar axis as a result of the jet), this energy may be overestimated by a factor of two to five \citep{Mazzali2014MNRAS}.}
%{\emph{Next questions. First, Zach, I don't see anything in Wang, Cano et al. on the kinetic energy of the SN, only on the magnetar parameters. Secondly, I didn't quite follow you... How does Ek scale with luminosity? After all, the light curve shape of ``my'' bolometric SN is quite similar to the one from G15, but it is significantly brighter, this should have some influence.}}

%{\emph{I have less clue concerning the kinetic energy of the jet. I guess this is usually derived from broadband modelling, but I simply can't ask for another half year for this paper (and who knows how many more co-authors...) Zach wrote me that there is a broad range of efficiencies (0.1 - 90\%) in converting Ek to Egamma, but as I understand it, generally Ek will be of similar dimension or higher than Egamma.}}

{The broadest possible spread of energies concerns $E_{k,Jet}$. This value is connected to the energy released in gamma-rays via the efficiency $\eta_\gamma\equiv E_{\gamma,iso}/(E_{\gamma,iso}+E_{k,iso})$, hereby $E_{k,iso}$ is the isotropic kinetic energy of the fireball at the end of the prompt emission \citep{Zhang2007ApJ2}. Modelling of GRB prompt emission and afterglows has shown $\eta_\gamma$ can have a wide range of values, e.g., $\lesssim50\%$ \citep{Beniamini2015MNRAS}, and even somewhere between 0.01\% and 90\% \citep{Zhang2007ApJ2,Wang2015ApJS}, implying $E_{k,iso}=(0.1...10000)E_{\gamma,iso}$. For very low $\epsilon_B$ values (see above), radiative efficiencies are typically on the order 10\%, implying $E_{k,iso}\approx10E_{\gamma,iso}$. \cite{Lu2018ApJ} have studied a sample of GRB-SNe and, among other results, have derived efficiencies for these events. They find that GRB 111209A is a strong outlier compared to the rest of the sample, with a very high efficiency of $76\%$, implying that in contrast to all other GRB/SNe they study, the GRB energy dominates compared to the SN energy. We note that this still depends on which circumburst medium density is assumed, as we have derived a broad range of potential collimation-corrected energy releases, which are coupled to $E_{k,Jet}$ via the efficiency $\eta_\gamma$. For our range of collimation-corrected $E_\gamma$ values, this implies $\log E_{k,Jet}\approx50.4...51.9$.
%An extremely high efficiency would imply $\log E_{k,Jet}\approx49.9...51.4$, whereas very low efficiencies would yield unrealistically high $E_{k,Jet}$ values up to $<10^{56}$ erg.
}

{In total summation, we find that, assuming, as pointed out above, neutrino losses are negligible, a ``normal'' magnetar would be able to power the entire event, reasonably independent of circumburst medium density, albeit with a very short rotation period, as GF17 imply (in full agreement with \citealt{Lu2018ApJ} who find that GRB 111209A/SN 2011kl lies very close to the classical magnetar limit). Of the three components, $E_{k,SN}$ has the least insecure value and sets a lower bound (assuming no asphericity) of the total energy of the event at $\approx10^{52}$ erg which is still within the bounds of the energy a classical magnetar can deliver. A more powerful magnetar (the ``Metzger magnetar'') could power the event while not rotating near break-up velocity. For a $10^{53}$ erg magnetar, even if the total energy of the event is equipartitioned between the kinetic energy of SN 2011kl and the energy of GRB 111209A and its jets, a spin period of $\approx3$ ms is derived. Such a high-mass neutron star near the NS/BH boundary would indeed be in agreement with a high initial spin period, as shown in the models of \cite{Aguilera-Dena2018ApJ}. We therefore conclude that GRB 111209A/SN 2011kl can be powered by a magnetar central engine, and other energy sources (e.g., a Black-Hole central engine) that would be in conflict with the interpretation of the spectrum (G15, \citealt{Mazzali2016MNRAS}) are not needed.}

%{\emph{Concerning the summation of everything, my guesstimate is that for the narrow-jet case, the total will be in accordance with the typical 1.4 Msol magnetar, for typical CBM densities one might need to invoke the massive magnetar plus whatever else Metzger added in his 2015 paper to get to 1E53 erg - and the high density model will really be stretching it. Additionally, another look in GF17 revealed some discussion about how even if you have 3E52 or even 1E53 of energy, you will only be able to extract about 1/10th or so because of neutrino losses. If this is correct, then the whole thing is in real trouble...}}

%{\emph{The referee mentions that the magnetar could not be slower than 3 ms or so, not sure where that comes from... GF17 themselves, after all, make the case for a 1 ms magnetar.}}

%{\emph{Finally, the referee points to a possible discussion of further energy sources. Is there anything? I mean, while at the same time believing the spectral analysis of SN 2011kl and that it indicates magnetar powering? Zach mentioned to me the possibility of a spin-up via accretion of additional material and a time-varying braking index.}}

%------------------------------------------------------------------------------------------------------------------------------------------------------------------------------------------------------------------
%------------------------------------------------------------------------------------------------------------------------------------------------------------------------------------------------------------------

\subsection{Models for the plateau/rebrightening phase during Day 1}
\label{SectDiscModel}

We have found a remarkable behaviour during first day of GROND observations. The afterglow, instead of fading, brightens slowly, then very steeply, and does so at a {somewhat} different rate in each band, leading to a rapid colour evolution {seen especially well in the spectral slopes of the SEDs of each GROND epoch}. Suffice to say, this is not typical behaviour for a GRB afterglow, and the simplest versions of the fireball model do not explain this evolution. We must therefore turn to more advanced scenarios.

\subsubsection{A powerful energy injection?}

Several GRBs have shown very strong rebrightenings at late times (i.e., hours to days, and therefore unrelated to the prompt emission), where the afterglow did not just show a flattening of its decay, but actually increased significantly, often by more than a magnitude. Examples for such events have been seen in, e.g., GRB 970508 \citep{Djorgovski1997Nature,Castro-Tirado1998Science}, GRB 030329 \citep{Uemura2003Nature,Lipkin2004ApJ}, GRB 060206 \citep{Wozniak2006ApJ,Monfardini2006ApJ}, GRB 070125 \citep{Updike2008ApJ,Chandra2008ApJ}, GRB 080928 \citep{Rossi2011AA}, and GRB 090926A \citep{Rau2010ApJ,Cenko2011ApJ}; see also \cite{Laskar2015ApJ} for a recent comprehensive study of several further events. \cite{Yu2015MNRAS} model the rebrightening (based on data from \citealt{Stratta2013ApJ}) with such an energy injection{, and \cite{Zhang2001ApJ} even make the case that a millisecond magnetar engine would produce such a bump}. On the other hand, such energy injections do not exhibit spectral changes (stemming from a single blastwave, with non-variable microphysical parameters, see especially \citealt{Laskar2015ApJ}), therefore this interpretation seems unlikely {-- indeed, \cite{Zhang2001ApJ} specifically predict an \emph{achromatic} bump}. 

\subsubsection{A two-component jet model?}

\begin{figure*}[t]
  \centering
 \includegraphics[width=\textwidth]{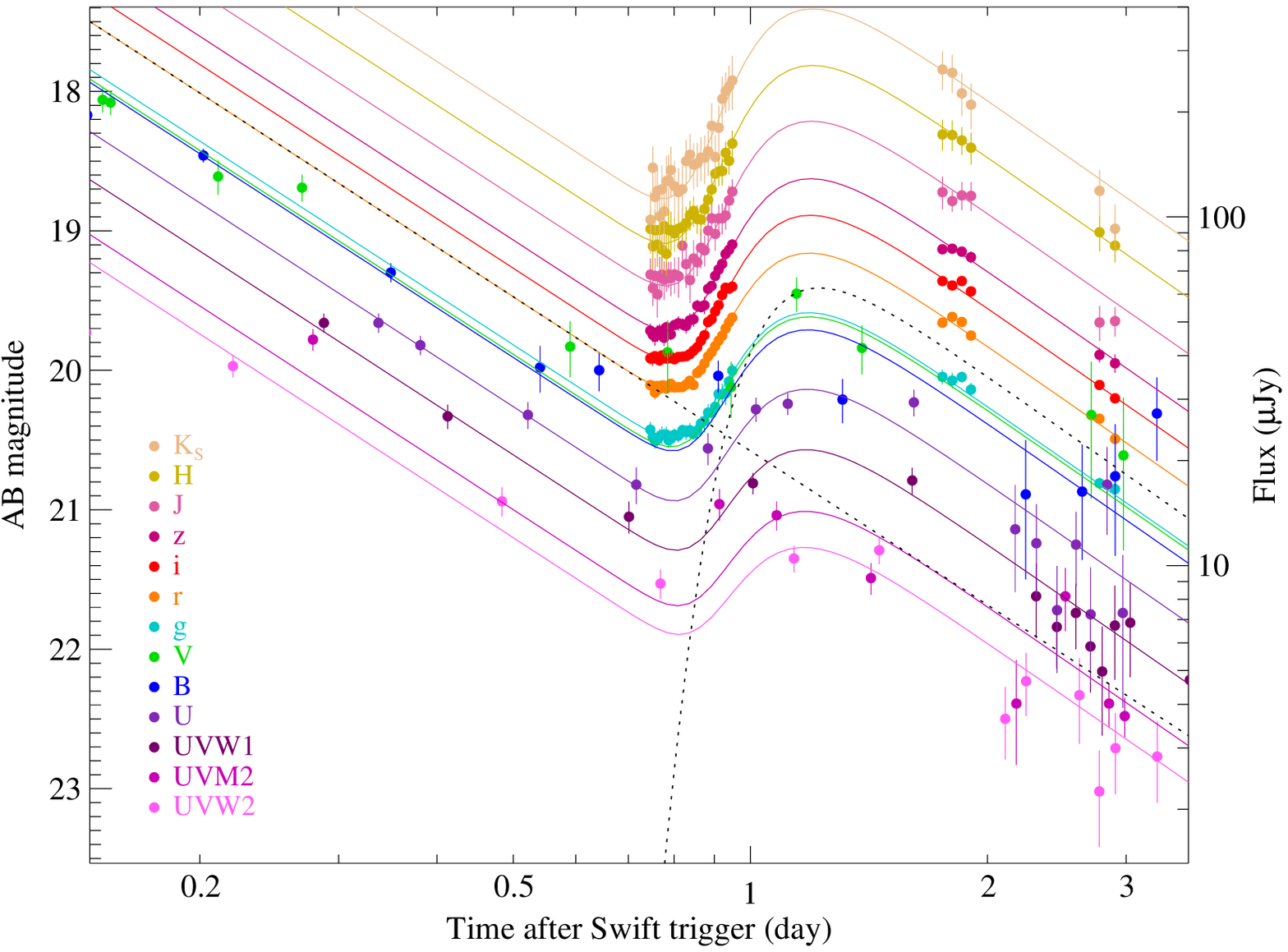}
   \caption{Decomposition of the optical/NIR afterglow of GRB 111209A across the rebrightening episode. UVOT data are the only data available before $\approx0.8$ days. With the beginning of GROND observations, a steeply rising second component becomes dominant (the two different components are shown as dotted lines for the $r^\prime$ band), leading to a colour change in the afterglow. The fit ends before the first ``step'' in the late afterglow.}
              \label{LCdecomp}
    \end{figure*}

In recent years, GROND has revealed multiple complex afterglows which exhibit rebrightenings which often rise extremely rapidly and show colour changes across the rebrightening, such as GRB 080413B \citep{Filgas2011AA2}, GRB 081029 \citep{Nardini2011AA}, GRB 091029 \citep{Filgas2012AA}, GRB 100621A \citep{Greiner2013AA} and GRB 100814A \citep{Nardini2014AA}. As with GRB 111209A, the simultaneous seven-colour capabilities of GROND allowed a detailed study of the SED evolution in those cases as well. GRB 081029 shows a similar evolution to GRB 111209A, with a steep rise in slope from $\beta\approx0.77$ to $\approx1.05$. Also, no contemporaneous rise was detected in the X-rays. GRB 080413B exhibits a switch from a very flat early SED ($\beta=0.22$), to a much steeper later one ($\beta\approx0.9$); and again there is no rebrightening seen in the X-rays\footnote{The actual rebrightening is missed in the optical/NIR as well due to lack of telescope coverage.}. GRB 100814A is similar to GRB 080413B with a transition from a once more very flat early SED ($\beta\approx0.10$), to a much steeper later one ($\beta\approx0.7$). GRB 091029 shows more mild spectral evolution but a complete decoupling of optical and X-ray evolution. GRB 100621A shows very high line-of-sight extinction which complicates the determination of spectral changes, but features the steepest rise to peak of all these events. Additionally, GRBs 081029, 100621A and 100814A show superposed variability on top of the peak, such a phenomenon can not be checked for in GRB 111209A due to the lack of dense, high signal-to-noise coverage during this time.

One explanation that may work in at least some of these cases is a two-component/double-jet model \citep[e.g.,][]{Berger2003Nature,Peng2005ApJ,Granot2006MNRAS}. In this scenario, the central engine emits a narrow, highly relativistic jet which is responsible for the prompt emission and the early afterglow, and a much wider and slower jet which produces a rebrightening when it decelerates, henceforth dominating the afterglow emission, especially in the radio regime.  If the emission regions of these two jets have different microphysical parameters, their resulting spectral energy distributions will look different, causing a spectral change. \cite{Filgas2011AA2} successfully invoked this model to explain GRB 080413B. In the case of GRB 081029, \cite{Nardini2011AA} argued that the generic two-component jet model is unable to explain the very steep rebrightening, though it may work if one assumes that the wide jet was launched at a significantly later time (and the lower Lorentz factor may suppress high-energy emission, making it undetectable because the X-ray afterglow of the narrow jet outshines it). Even in the case of GRB 100814A, the decomposition into two overlapping components yields a steep rise of the second component \citep{Nardini2014AA}.

The extremely long duration of the prompt emission of GRB 111209A makes it a prime candidate for the launch of multiple jets at different $t_0$ times with different Lorentz factors. To check if it is possible to explain the temporal and spectral evolution of the afterglow of GRB 111209A with two afterglow components with different spectral slopes, we use a multi-band fitting procedure which is able to incorporate multiple broken power-laws as well as components of different colour \citep[e.g.,][]{Perley2008ApJ,Perley2008ApJ2,Perley2014ApJ}. The result is shown in Fig. \ref{LCdecomp}. We find an acceptable ($\chi^2=319$ for 285 degrees of freedom) fit with decay slopes $\alpha_1=1.47\pm0.02$ and $\alpha_2=1.60\pm0.03$. The former value is steeper than we found by fitting the UVOT data alone, the latter value in perfect agreement with the GROND-only fit. The spectral slope changes by $\Delta\beta=0.39\pm0.03$, also in agreement with what we found by fitting the two data sets separately. What is astonishing is the rising slope of the second component, we find a value of $\alpha_r=-39\pm1$, which exceeds even the rise of GRB 100621A \citep[$\alpha_r\approx-14$,][]{Greiner2013AA}. Here, it should be mentioned that the $t_0$ we use is the \emph{Swift} trigger time, whereas the GRB clearly started several thousand seconds earlier. Setting $t_0$ to an earlier time, though, just makes the slopes steeper, although, at close to one day after the GRB, such a change would have only a minimal influence. The situation therefore is similar to the findings of \cite{Huang2006ApJ} for GRB 030329, who were unable to model the rebrightening correctly as the data was too steep for their numerical model. Our fit, of course, is entirely empirical.

One aspect we have not incorporated is a ``hidden'' jet break of the narrow jet, which, e.g., was needed to correctly fit the afterglows of GRBs 080413B and 100814A \citep{Filgas2011AA2,Nardini2014AA}. The combination of multiple small energy injections (see Sect. \ref{SectDiscEnergy} below) and the rising SN at late times does not allow us to deduce such a hidden component, but it is very likely such a break would occur significantly earlier than the late break we do detect in the multi-band afterglow+SN fitting (K18A).

A detailed modelling, especially under assumption of a strongly delayed $t_0$ and possibly in combination with a surrounding low-density void (as \citealt{Evans2014MNRAS} invoke to explain the properties of GRB 130925A) is beyond the scope of this paper, though. Here, we conclude that the rebrightening as well as the spectral slope change can in principle be accommodated by the superposition of two different afterglow components, without the need to invoke, e.g., time-variable microphysical parameters (within a single emission region).

%------------------------------------------------------------------------------------------------------------------------------------------------------------------------------------------------------------------
%------------------------------------------------------------------------------------------------------------------------------------------------------------------------------------------------------------------

\subsubsection{Energy injections into the late afterglow}
\label{SectDiscEnergy}

Our light curve fits in Sect. \ref{fitting_2} have revealed that after the purported peak and before the SN, the light curve did not decay entirely smoothly, instead, it experiences two short flattening phases, before resuming a similar decay to beforehand. For the three sections, we find $\alpha_{3.1}=1.59\pm0.03$, $\alpha_{3.2}=1.44\pm0.09$ and, from the broken power-law + SN fit, $\alpha_{late,1}=1.55\pm0.01$. These three decay slopes, while not strictly identical, are very similar to each other, additionally, we detect no more significant colour changes during this phase (Appendix \ref{SectBeta}).

This combination of factors points to the variability being due to energy injections into the afterglow, exactly the explanation we can likely rule out for the main rebrightening. Such ``refreshed shocks'' can be due to slow-moving shells in the jet catching up with the forward-shock front, or late ejection of shells \citep[e.g.,][]{Rees1998ApJ,Panaitescu1998ApJ,Kumar2000ApJ,Sari2000ApJ,Zhang2002ApJ} -- especially the latter case is attractive in consideration of the extreme duration of GRB 111209A \citep{Yu2015MNRAS}. Additional emissive components due to very long-lasting central engine activity \citep[e.g.,][]{Dai1998AA,Rees2000ApJ,Ghisellini2007ApJ} are less likely to contribute, both due to the achromatic evolution (energy injections do not change the electron index $p$ or microphysical parameters, whereas a central-engine flare would likely exhibit a different spectrum), and the ``step-like'' afterglow form (a central-engine flare would be superposed, and the later afterglow would lie on the extrapolation of the earlier decay).

Energy injections creating such ``steps'' have been proposed to explain similar features in such afterglows as GRB 030329 \citep[e.g.,][]{Granot2003Nature}, GRB 021004 \citep{Bjoernsson2004ApJ,deUgarte2005AA} and GRB 060526 \citep{Thoene2010AA}. GRB 090926A at later times also exhibited a very similar behaviour \citep{Rau2010ApJ,Cenko2011ApJ}. They may also play a role in the light curves of SLSNe \citep{Yu2016ApJL}, though the density variation model may also apply \citep{Inserra2017MNRAS}.

%------------------------------------------------------------------------------------------------------------------------------------------------------------------------------------------------------------------
%------------------------------------------------------------------------------------------------------------------------------------------------------------------------------------------------------------------

\subsection{The luminosity of the afterglows of GRB 111209A and GRB 130925A}
\label{SectDiscLumin}

In \cite{Kann2010ApJ}, we presented a large sample of \emph{Swift}-era long-GRB afterglow light curves. Comparing the afterglow of GRB 111209A to this large sample can aid us in the determination of the progenitor. Again, we add published data to the light curve (\citealt{Nysewander2011GCN,Stratta2013ApJ}, L14). At early times, as we have no overlapping observations, we assume that the $R_C$ data from TAROT are identical to our $r^\prime$ data (once they have been transformed to Vega magnitudes), then we shift the UVOT \emph{v/white} data points (these two agree very well with each other intrinsically) to the $R_C$ magnitudes, and adjust the rest of the UVOT data to this backbone. We also use UVOT data to fill the gap between the first and second day of GROND observations. At late times, we use the host-subtracted $U$ data for the post-break, SN-less afterglow. With a few exceptions, the resulting light curve shows little scatter. Furthermore, we shift the GROND $r^\prime$ magnitudes by $0.09-0.11$ magnitudes (brighter, depending on the spectral slope) to transform them to $R_C$.

We add GRB 130925A as a direct comparison. The light-curve treatment is detailed in Sect. \ref{130925A}.

%\begin{figure*}[t]
\begin{figure}[t]
  \centering
 \includegraphics[width=\columnwidth]{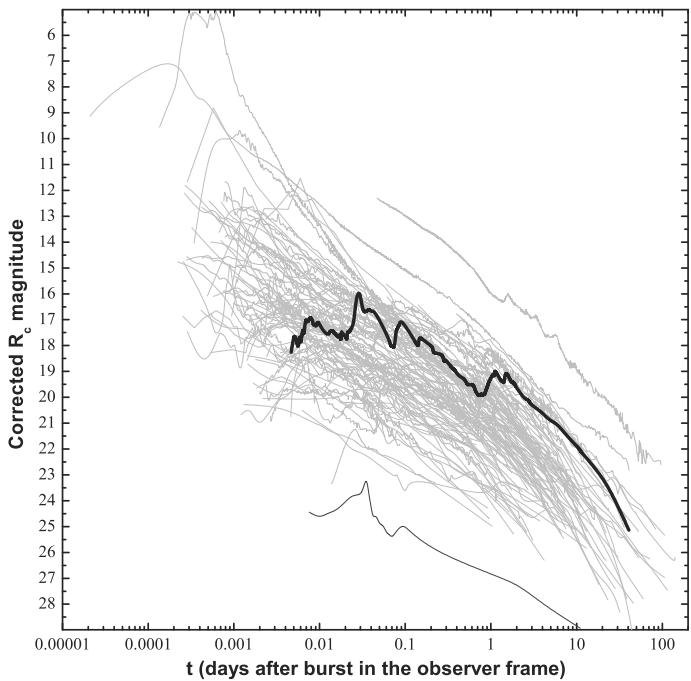}
   \caption{The $R$-band light curve of the afterglow of GRB 111209A in comparison to the long-GRB light-curve sample of \cite{Kann2010ApJ,Kann2011ApJ}. Data are corrected for Galactic extinction and, if possible, host galaxy contribution, but otherwise as observed. The strong variability of the afterglow at early times is evident. At late times, the afterglow is among the brightest observed so far. This is in strong contrast to the $R$-band afterglow of GRB 130925A (thin black curve), which is exceedingly faint due to very high host-galaxy extinction \citep{Greiner2014AA}.}
              \label{KPobs}
%    \end{figure*}
    \end{figure}

%\begin{figure*}[t]
\begin{figure}[t]
  \centering
 \includegraphics[width=\columnwidth]{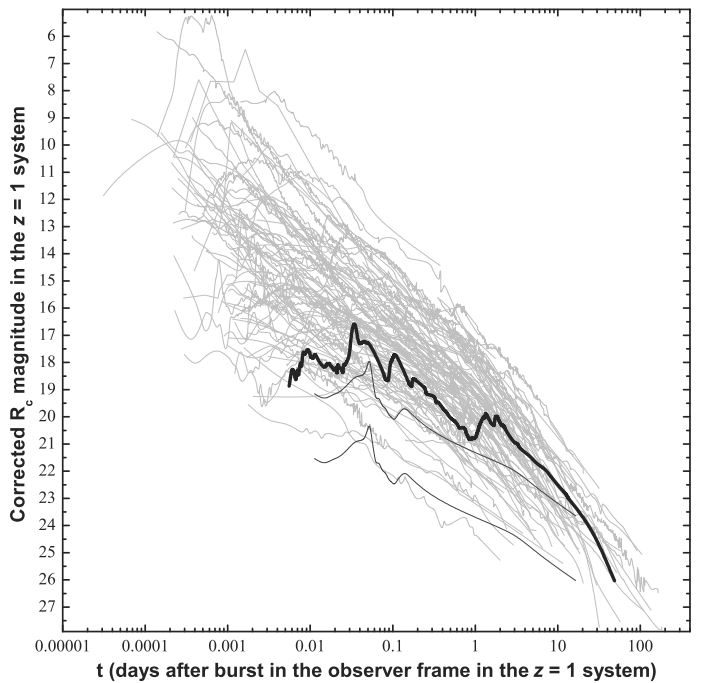}
   \caption{As Fig. \ref{KPobs}, but now in the $z=1$ system, where all light curves are directly comparable, having been corrected for all line-of-sight extinction and shifted to the same frame both in magnitude and time. The light curve of the afterglow of GRB 111209A is now seen to lie in the middle of the distribution of known afterglows, actually being among the least luminous at early times, and otherwise unremarkable from a luminosity standpoint. The afterglow of GRB 130925A is even fainter (the two different curves are based on two different extinction corrections, see text), and shows a remarkably similar evolution at early times.}
              \label{KPzone}
%    \end{figure*}
    \end{figure}

In Fig. \ref{KPobs}, we show the light curves of GRBs 111209A and 130925A, corrected for Galactic extinction and host-galaxy contribution, but otherwise as observed, in comparison to the \cite{Kann2010ApJ,Kann2011ApJ} sample. The extreme variability during the prompt emission is evident in both GRBs. For GRB 111209A, starting with the peak of the flare at $\approx2500$ s \citep{Gendre2013ApJ,Stratta2013ApJ} which roughly corresponds to the brightest peak of the prompt emission \citep{Golenetskii2011GCN}, the afterglow is at multiple times, and especially after the strong rebrightening we detected with GROND, among the brightest afterglows observed so far \citep[see also][]{Hoversten2011GCN3}. GRB 130925A, on the other hand, is extremely faint due to the very high line-of-sight extinction in the host galaxy \citep{Greiner2014AA}.

Similar to GRB 130925A (Sect. \ref{130925A}), we shift the optical light curve of GRB 111209A to $z=1$. Hereby, we are not interested in a detailed epoch-by-epoch shifting (which would probably just increase the scatter), therefore we use only two spectral slopes to shift different parts of the light curve. We use the fit from the early prompt emission ($\beta=0.63\pm0.25$, $\AV=0.25\pm0.11$) to shift data from the first two \emph{Swift} orbits (up to 0.09 days), and the joint UVOT+GROND result ($\beta=1.05\pm0.06$, $\AV=0.12\pm0.04$) for the rest of the data. For the two parts, we find $dRc=+0.61^{+0.20}_{-0.19}$, and $dRc=+0.89^{+0.06}_{-0.07}$ respectively.

After this shift, we find (Fig. \ref{KPzone}) that the afterglow of GRB 111209A lies well within the distribution of known afterglow magnitudes. Except for early times, when it actually belongs to the least luminous known afterglows, its luminosity is unremarkable (we find $R_C=20.75\pm0.07$, $M_B=-22.17\pm0.07$, $R_C=21.30\pm0.08$, $M_B=-21.62\pm0.08$ at one and four days after the trigger in the $z=1$ frame, respectively). \cite{Kann2010ApJ} compared a large number of long GRBs in terms of afterglow luminosity vs. isotropic energy release \citep[see also][]{Gehrels2008ApJ,Nysewander2009ApJ}, and the values for this GRB are not exceptional in this sense either (just as \citealt{Golenetskii2011GCN} have pointed out that the prompt emission is not really remarkable save for the duration). From the afterglow perspective, therefore, GRB 111209A is a typical long GRB, the only outstanding element being the strong variability, which is mostly linked to the extremely long prompt emission duration anyway.

The light curve of GRB 130925A is seen to resemble that of GRB 111209A quite remarkably. The actual luminosity of the afterglow is hard to determine, though (see Sect. \ref{130925A} for more details), but it is likely less luminous than that of GRB 111209A, possibly significantly more so (as also deduced by \citealt{Evans2014MNRAS} upon comparison of the X-ray afterglows after correcting for the dominating dust-echo component). The combination of high extinction and a bright host galaxy led to only sparse detections at late times; especially, no attempt was made to detect the (likely) accompanying SN.

\section{Conclusions}
\label{SectConc}

In this work, we have presented the detailed observations, obtained by GROND and UVOT, of the afterglow of the ultra-long duration GRB 111209A. Our main conclusions are the following:

\begin{itemize}
\item Excepting its extreme duration and slow variability, GRB 111209A is in agreement with usual GRBs. The prompt emission parameters are standard, and it agrees with the Amati relation \citep{Amati2002AA}, and with the Ghirlanda relation \citep{Ghirlanda2004ApJ} assuming a low circumburst medium density, though there are indications this might not be the case (GF17).
\item The afterglow shows a complicated evolution, featuring strong variability during the extremely long prompt emission phase. After a standard decay, a strong chromatic rebrightening follows, which we model with a two-component jet. The late afterglow also shows several smaller, achromatic rebrightenings, which are likely to be energy injections.
\item In general, though, the afterglow of GRB 111209A is unremarkable in comparison to other GRB afterglows. This is in contrast to the long-wavelength transients following other ultra-long duration events, which were weaker or were not in agreement with synchrotron emission.
\item In contrast to earlier work on this event, our dataset also reveals that this ultra-long duration GRB is accompanied by a very luminous supernova, which is spectrally dissimilar compared to usual GRB-SNe (G15,K18A). This SN is likely to be powered by a magnetar, and resembles SLSNe more than typical GRB-associated SNe. {We find that, although very energetic, the entire event is still in agreement with being powered by a magnetar central engine.}
\item Our data puts to rest a lot of speculation on the nature of this special event, but a completely unified model explaining both the duration as well as the SN characteristics remains out of reach for now.
\end{itemize}

In multiple ways, GRB 111209A is an extraordinary event. Here, we have shown that in terms of the afterglow, it is not completely ordinary, but clearly linked to standard GRBs and their emission mechanisms. Further dense multi-wavelength follow-up in the future of this very rare class of GRBs will yield additional insight into the breadth of high-energy gamma-ray transients and their possible link to the most luminous stellar explosions.

\acknowledgements
DAK wishes to dedicate this work to his father, R.I.P. 20. 08. 2015. You are sorely missed by so many.
We thank the referee for the very valuable report which helped improve the clarity of the paper.
DAK acknowledges Massimiliano De Pasquale, Daniele Malesani, Antonio de Ugarte Postigo, Christina C. Th\"one, Thomas Kampf, Cristiano Guidorzi, and Raffaella Margutti for interesting discussions and helpful comments.
DAK acknowledges financial support by the DFG Cluster of Excellence ``Origin and Structure of the Universe,'' from MPE, from TLS, from the Spanish research project AYA 2014-58381-P, and from Juan de la Cierva Incorporaci\'on fellowship IJCI-2015-26153.
%DAK and TK wish to thank Paul Vreeswijk for supplying us with the X-shooter spectrum of iPTF13ajg prior to publication in the WiseREP database.
We are indebted to Joe Lyman and Vicki Toy for supplying the bolometric light curves of GRB 120422A/SN 2012bz and GRB 130702A/SN 2013dx, respectively.
SK, DAK, ARossi, and ANG acknowledge support by DFG grants Kl 766/16-1 and Kl 766/16-3, SSchmidl also acknowledges the latter.
ARossi acknowledges support from the Jenaer Graduiertenakademie and by the project PRIN-INAF 2012 ``The role of dust in galaxy evolution''.
TK acknowledges support by the DFG Cluster of Excellence Origin and Structure of the Universe, and by the European Commission under the Marie Curie Intra-European Fellowship Programme.
RF acknowledges support from European Regional Development Fund-Project ``Engineering applications of microworld physics'' (No. CZ.02.1.01\/0.0\/0.0\/16\_019\/0000766).
DARK is funded by the DNRF.
FOE acknowledges funding of his Ph.D. through the DAAD, and support from FONDECYT through postdoctoral grant 3140326.
SSchulze acknowledges support from CONICYT-Chile FONDECYT 3140534, Basal-CATA PFB-06/2007, and Project IC120009 ''Millennium Institute of Astrophysics (MAS)'' of Iniciativa Cient\'ifica Milenio del Ministerio de Econom\'ia, Fomento y Turismo.
SK, SSchmidl, and ANG acknowledge support by the Th\"uringer Ministerium f\"ur Bildung, Wissenschaft und Kultur under FKZ 12010-514.
MN and PS acknowledge support by DFG grant SA 2001/2-1.
ANG, DAK, ARossi and AU are grateful for travel funding support through MPE.
Part of the funding for GROND (both hardware as well as personnel) was generously granted from the Leibniz-Prize to Prof. G. Hasinger (DFG grant HA 1850/28-1).
This work made use of data supplied by the UK Swift Science Data Centre at the University of Leicester.

\bibliographystyle{aa}

\newpage\clearpage

%\appendix
\begin{appendix}

\section{Initial observations of GRB 111209A}
\label{TheGRB}

GRB 111209A was localized as Trigger \#509336 on the 9th of December 2011 at 07:12:08 (UT times are used throughout the paper) by the \emph{Swift} satellite \citep{Gehrels2004ApJ} with its high-energy Burst Alert Telescope \citep[BAT,][]{Barthelmy2005SSRv}, which slewed immediately to follow it up with its narrow-field instruments, the X-Ray Telescope \citep[XRT,][]{Burrows2005SSRv} and the UltraViolet-Optical Telescope \citep[UVOT,][]{Roming2005SSRv}. The initial observation report already included the discovery of an X-ray and an optical/UV afterglow \citep{Hoversten2011GCN}. The initial detection by BAT was based on an image of 320 s exposure, indicating a long-lasting, low peak-flux event. At $T_0=+424$ s, \emph{Swift} triggered a second time (\#509337), a 64 s image trigger at much higher flux. Such a double trigger was only known from GRB 110709B beforehand\footnote{The TDE GRB 110328A/Swift J164449.3+573451 also caused a double trigger but was not actually a GRB \citep{Levan2011Science}.}, according to \cite{Zhang2012ApJ}. By this time, the satellite was already observing with its narrow-field instruments.

Rapid ground-based follow-up within minutes of the alert was reported from the TAROT South (\citealt{Klotz2011GCN1}, see also \citealt{Gendre2013ApJ, Stratta2013ApJ}), REM \citep[][both stationed at La Silla, Chile]{Fugazza2011GCN}, FTN \citep[][stationed on Hawaii]{Guidorzi2011GCN} and PROMPT \citep[][stationed at Cerro Tololo, Chile]{Nysewander2011GCN} telescopes, whereas no bright flashes were seen in wide-field surveys \citep{Sokolowski2011GCN, Wren2011GCN}. Continued observations revealed the GRB remained bright and highly variable at gamma-ray \citep{Palmer2011GCN}, X-ray \citep{Grupe2011GCN} and UV/optical wavelengths \citep{Klotz2011GCN2,Hoversten2011GCN3}, leading the \emph{Swift} team to declare this as a ``Burst of Interest'' \citep{Hoversten2011GCN2}. Spectroscopy by VLT X-shooter revealed the GRB to lie at $z=0.677$ (\citealt{Vreeswijk2011GCN}, note that \citealt{Im2011GCN} also report NIR spectroscopy with IRTF SpeX but do not report any line detections), which is quite close for a \emph{Swift} GRB \citep{Fynbo2009ApJS,Jakobsson2012ApJ}, and simultaneously ruled out a peculiar Galactic transient (which was not suspected anyway due to the high Galactic latitude) as well as an event at very high redshift, in which case time dilation would have strongly contributed to the duration (the low redshift was also already suspected due to the detection in all UVOT filters, \citealt{Hoversten2011GCN3}). Furthermore, after an initial non-detection \citep{Hancock2011GCN}, a bright radio afterglow was also reported by \cite{Hancock2012GCN}.

The true dimension of GRB 111209A was finally revealed by the observations of the Konus detector on the \emph{WIND} spacecraft \citep{Golenetskii2011GCN} in combination with the X-ray light curve\footnote{On the \emph{Swift} XRT repository \citep{Evans2007AA,Evans2009MNRAS} webpage at http://www.swift.ac.uk/xrt\_curves/00509336/ .}. The GRB is seen to begin $\approx5,400$ s \emph{before} the \emph{Swift} trigger time\footnote{Additionally, there is a possible precursor at $\approx-10,000$ s, but this burst has not been conclusively linked to GRB 111209A.}, and extend to $\approx10,000$ s. The XRT observations reveal that the X-ray emission remains roughly constant (with several flares superposed) up to $\approx20,000$ s, then a steep decay sets in. This is usually attributed to high-latitude emission and marks the point where the central engine ``turns off'', and thus the prompt emission ends \citep{KumarPanaitescu2000ApJ,Tagliaferri2005Nature,Barthelmy2005ApJ,Zhang2014ApJ}. Therefore, the prompt emission duration of the entire event is around 25,000 s, or seven hours (see also \citealt{Gendre2013ApJ}; \citealt{Lien2016ApJ} detect the event for $\approx18,000$ s post-trigger in BAT survey data), significantly longer than even GRB 060814B \citep[about four hours long,][see Appendix \ref{ELDGRBs}]{Pal'Shin2008AIPC}. \cite{Golenetskii2011GCN} also reported that the other prompt emission parameters, such as the spectral shape and the isotropic energy release, were within the typical distribution of GRBs (see also our own analysis in Sect. \ref{Energy}). While the peak flux of the GRB is low, the extreme duration leads to a large total fluence; the comparably low redshift, though, implies a large but by no means exceptional total energy release. The Konus-\emph{WIND} light curve\footnote{http://www.ioffe.rssi.ru/LEA/GRBs/GRB111209A/} indeed resembles that of a typical GRB, except that it is actually stretched by several orders of magnitude. Morphologically, it is therefore most similar to the first ``class'' mentioned in Appendix \ref{ELDGRBs} (and ``continuous'' in the classification of \citealt{Virgili2013ApJ}).

\section{Examples of GRBs of extreme duration and a rough phenomenological classification scheme}
\label{ELDGRBs}

So far, there is no fixed definition for the labels ``extremely long duration'' (which we shorten to EL-GRBs) and ``ultra-long'' (the special class of ULGRBs, L14). Such a definition would be arbitrary anyway, especially since $\TNT$ (defined as the time over which 90\% of the fluence is accumulated, starting after the first 5\% and ending at 95\%) is an observer-frame quantity (and also strongly detector-dependent). Numbers which could be used are 600 s (i.e., 10 minutes), and 1000 s. For actual ``ultra-long'' GRBs as presented by L14, we suggest a redshift-corrected duration of prompt emission activity \citep{Zhang2014ApJ} of at least one hour.

\begin{itemize}
\item The most ``generic'' EL-GRBs consist of multiple spaced-apart emission episodes of roughly similar intensity. \cite{Tikhomirova2005AstL} have found multiple cases in the data of the Burst And Transient Source Experiment (BATSE) on the Compton Gamma-Ray Observatory. Further individual examples are GRB 020410 \citep[][$\TNT\approx1500$ s]{Nicastro2004AA}, the IPN GRB 080407 \citep[][with $\TNT\approx2100$ s]{Pal'shin2012grb}, GRB 091024 \citep[][with $\TNT\approx1020$ s]{Gruber2011AA} which is also associated with an optical flash \citep{Virgili2013ApJ}, the dark GRB 090417B \citep[][$\TNT>2130$ s]{Holland2010ApJ}, the ``double burst'' GRB 110709B \citep[][with a total duration of $\approx1400$ s]{Zhang2012ApJ} and the similar GRB 121217A \citep[][with a total duration of $>1070$ s]{Siegel2013GCNR,Elliott2013AA}, the ``\emph{Swift} Birthday Burst'' GRB 141121A \citep[][with a total duration of $\approx1410$ s]{Golenetskii2014GCN17108,Cucchiara2015ApJ}, and, somewhat shorter, GRB 070616 \citep{Starling2008MNRAS}, which lasted $\approx600$ s.

Since GRB 111209A, {three} very similar events have been discovered. GRB 121027A features a highly variable X-ray light curve and extremely elevated emission at $>5000$ s after the trigger (\citealt{Serino2012GCN, Evans2012GCN}, see \citealt{Peng2013arXiv}, \citealt{Wu2013ApJ} and \citealt{Hou2014MNRAS} for theoretical treatments), which, combined with the redshift $z=1.773$, makes it one of the most luminous X-ray afterglows ever detected (L14, Starling et al., in preparation). L14 list this as a third example of an ultra-long duration GRB (see below for the other event).

Not included in the sample of L14, GRB 130925A \citep{Evans2014MNRAS,Zhang2014ApJ} was found to show flaring emission, both in gamma-rays \citep{Fitzpatrick2013GCN15255,Markwardt2013GCN15257,Savchenko2013GCN15259,Golenetskii2013GCN15260,Hurley2013GCN15278,Evans2014MNRAS} as well as X-rays \citep{Suzuki2013GCN15248,Evans2013GCN15254}, for a duration of over 12,000 s, starting with a short, soft precursor, followed by two large emission episodes and several further soft flares. While an initial similarity to Sw J1644+57 was remarked upon \citep{Burrows2013GCN15253}, not only due to the high-energy emission, but also due to the highly obscured optical counterpart \citep{Sudilovsky2013GCN15247,Greiner2014AA} and the very similar redshift \citep{Vreeswijk2013GCN15249,Sudilovsky2013GCN15250}, the later evolution of the X-ray afterglow as well as the offset from the host galaxy centre \citep{Tanvir2013GCN15489} point to this being another ULGRB and not a (relativistic) TDE. Similar to GRB 111209A, the early optical/NIR emission during the prompt phase shows strong variability as well as a multi-100 s offset between gamma-rays and optical light \citep{Greiner2014AA}. The prompt emission itself is much more variable than that of GRB 111209A, though, and does not look ``stretched'' \citep{Evans2014MNRAS,Greiner2014AA}. The late X-ray afterglow, for which spectral features have been claimed (\citealt{Bellm2014ApJ}, but see \citealt{Evans2014MNRAS}), is extremely soft, which has been explained as an expanding dust scattering halo \citep{Evans2014MNRAS,Zhao2014ApJ,Margutti2015ApJ}, or alternatively as thermal emission from a hot cocoon, which may be a link to BSG progenitors (\citealt{Piro2014ApJL,Basak2015ApJ}, but see \citealt{Evans2014MNRAS}). In strong contrast to the host of GRB 111209A, the host of GRB 130925A shows super-solar metallicities at multiple sites, including the explosion site \citep{Schady2015AA}. This event also features a peculiar radio afterglow with unique properties \citep{Horesh2015ApJ}. We discuss the optical/NIR afterglow light curve in more detail in Sect. \ref{130925A} and compare it to that of GRB 111209A in Sect. \ref{SectDiscLumin}.

{The newest member of the ULGRB class is also the longest so far, the ``Bastille Day'' GRB 170714B. It was localized by \emph{Swift} \citep{Dai2017GCN}, and in this case, the BAT data indicates that it was caught almost at onset \citep{Palmer2017GCN}, with detectable emission beginning just 70 s before the beginning of the image trigger. Initial X-ray observations were reported to show a lack of variability \citep{D'Avanzo2017GCN} but after further observations, \cite{Kann2017GCN} pointed out that the X-ray light curve showed clear signs of being that of an ULGRB, with a duration of at least 36 ks. GTC observations revealed a faint, very reddened afterglow \citep{deUgartePostigo2017GCN1}, and GTC spectroscopy revealed the GRB to lie at $z=0.793$ \citep{deUgartePostigo2017GCN2}. Radio follow-up was initially unsuccessful \citep{deUgartePostigo2017GCN3,Ricci2017GCN,Horesh2017GCN}, but a faint afterglow was finally detected in further observations \citep{Piro2017GCN}. Such a faint radio afterglow is a strong indicator that this is \emph{not} a relativistic TDE \citep{deUgartePostigo2017GCN3}. No detection of an accompanying SN has been reported so far. \cite{Hou2018ApJ} model the early X-ray emission with a quark-star model.}

\item ``Precursor GRBs'' exhibit a comparatively faint first emission episode \citep[which is spectrally similar to the main emission,][]{Burlon2008ApJ} followed by a ``silence'' that can last hundreds of seconds before a much more luminous emission episode occurs which contains most of the fluence of the GRB and can last for several hundred seconds further. A ``short'' example of such a GRB is GRB 061121 \citep{Page2007ApJ}, and the four well-known long cases are GRB 041219A \citep[main emission after $\approx200$ s, total duration $\approx500$ s,][]{Fenimore2004GCN, Blake2005Nature}, GRB 050820A \citep[main emission after $\approx220$ s, total duration $\approx600$ s,][]{Cenko2006ApJ}, GRB 060124 \citep[main emission after $\approx500$ s, total duration $\approx750$ s,][]{Romano2006AA}, and GRB 160625B \citep[main emission after $\approx180$ s, total duration $\approx800$ s,][]{Zhang2018Nature}{; in the latter case, the long duration and delayed main emission allowed strong polarization to be detected during the prompt emission \citep{Troja2017Nature1}}.

\item A seemingly very rare class of EL-GRBs are single-peak (Fast Rise, Exponential Decline, FRED) hard events, only two have been reported so far: GRB 971208 detected by BATSE and Konus-\emph{WIND} \citep[][duration $\approx2500$ s]{Connaughton1997IAUC}, and the extreme event GRB 060814B, which lasted for over four hours in the softest Konus-\emph{WIND} bands \citep{Pal'Shin2008AIPC}.

\item Morphologically similar but spectrally distinct are low-luminosity, low-redshift X-Ray Flashes (XRFs), which are associated with broad-lined Type Ic supernovae. Two of these have been extensively studied: XRF 060218 \citep[][duration $\approx2100$ s]{Campana2006Nature} associated with SN 2006aj \citep{Pian2006Nature, Mirabal2006ApJ,Modjaz2006ApJ,Cobb2006ApJ,Sollerman2006AA,Ferrero2006AA}, and XRF 100316D \citep[][duration $>1300$ s]{Starling2011MNRAS}, associated with SN 2010bh \citep{Chornock2010arXiv, Cano2011ApJ, Olivares2012AA, Bufano2012ApJ,Margutti2013ApJ}. The long duration of these events is likely due to additional emission components such as thermal emission from a supernova shock breakout \citep[e.g.,][]{Campana2006Nature,Starling2012MNRAS}.

\item Finally, there are high-energy transients which resemble GRBs but are possibly due to a different type of progenitor. The ``Christmas Burst'' GRB 101225A (with a duration of several thousand s, and possibly even days) has been explained as the inspiral of a neutron star into a helium star, creating a central engine similar to a GRB \citep{Thoene2011Nature}\footnote{Note that a non-cosmological origin has been proposed as well \citep{Campana2011Nature}, but this has now been firmly ruled out by the derivation of the host-galaxy redshift $z=0.847$ (L14), which is incidentally also significantly higher than the distance used in \cite{Thoene2011Nature}, $z\approx0.33$.}. L14 list this as the first ULGRB discovered, though it shows behaviour especially in the optical which is unlike any known GRB \citep{Thoene2011Nature}, in contrast to GRBs 111209A, 121027A, and 130925A. Finally, the extreme gamma-ray transients GRB 110328A/Swift J164449.3+573451 \citep{Levan2011Science,Bloom2011Science,Burrows2011Nature,Zauderer2011Nature}, Swift J2058.4+0516 \citep{Cenko2012ApJ,Pasham2015ApJ}, and Swift J1112.2-8238 \citep{Brown2015MNRAS,Brown2017MNRAS} which have been detected for many days at high energies, are very likely due to the tidal disruption of a star by a supermassive black hole, which ``turns on'' a blazar-like relativistic jet.
\end{itemize}
Note that \cite{Virgili2013ApJ} undertake a more broad classification, discerning between ``Interrupted Emission'' and ``Continuous Emission'' GRBs. The Interrupted Emission GRBs are all found in our first category. These authors also give details on many of the EL-GRBs we list above in the appendix of their work.

\end{appendix}

\end{document}